\documentclass[10pt,a4paper,openany]{book}
\usepackage{amsmath,amssymb}
\usepackage{natbib,aas_macros}
\usepackage{bm}
\usepackage[dvipdfmx]{graphicx}
\usepackage{ascmac}
\usepackage{fancyhdr}
\usepackage{fancybox}
\usepackage{color}

\usepackage{url}
\usepackage{multicol}
\usepackage{multirow}
\usepackage{threeparttable}

\definecolor{violet}{rgb}{0.4,0.0,0.4}
\definecolor{dblue}{rgb}{0.0,0.0,0.5}
\title{\bf GREX-PLUS Science Book v1}
\author{GREX-PLUS Science Team}

\begin{document}

\maketitle

\section*{Preface}
\label{sec:executivesummary}

GREX-PLUS (Galaxy Reionization EXplorer and PLanetary Universe Spectrometer) is a mission candidate for a JAXA's strategic L-class mission to be launched in the 2030s.
Its primary sciences are two-fold: galaxy formation and evolution and planetary system formation and evolution. 
The GREX-PLUS spacecraft will carry a 1.2 m primary mirror aperture telescope cooled down to 50 K.
The two science instruments will be onboard: a wide-field camera in the 2-8 $\mu$m wavelength band and a high resolution spectrometer with a wavelength resolution of 30,000 in the 10-18 $\mu$m band.
The GREX-PLUS wide-field camera aims to detect the first generation of galaxies at redshift $z>15$.
The GREX-PLUS high resolution spectrometer aims to identify the location of the water ``snow line'' in proto-planetary disks. 
Both instruments will provide unique data sets for a broad range of scientific topics including galaxy mass assembly, origin of supermassive blackholes, infrared background radiation, molecular spectroscopy in the interstellar medium, transit spectroscopy for exoplanet atmosphere, planetary atmosphere in the Solar system, and so on.

This document is the first version of a collection of scientific themes which can be achieved with GREX-PLUS.
Each section in Chapters~\ref{chap:extragalactic} and \ref{chap:galacticplanetary} is based on the presentation at the GREX-PLUS Science Workshop held on 24-25 March, 2022 at Waseda University.\footnote{\url{http://www.obsap.phys.waseda.ac.jp/grex-plus-sws-220324-25.html}}

\vspace{2cm}

\section*{Authors}

\begin{multicols}{2}

\noindent
Baba, Shunsuke (Kagoshima University) \S\ref{sec:AGNmolecularoutflow} \& \S\ref{sec:IGMmoleculargas}.

\noindent
Fujii, Yuka (NAOJ) \S\ref{sec:exoplanetatmosphere}.

\noindent
Gouda, Naoteru (NAOJ) \S\ref{sec:galaxycenter}.

\noindent
Harikane, Yuichi (University of Tokyo) \S\ref{sec:veryhighz}.

\noindent
Hirahara, Yasuhiro (Nagoya University) \S\ref{sec:interstellarmolecules}.

\noindent 
Inoue, Akio K.\ (Waseda University) \S\ref{chap:overview}, \S\ref{sec:veryhighz}, \S\ref{sec:cosmicinfraredbackground}, \& \S\ref{chap:synergywithotherprojects}.

\noindent
Kawashima, Yui (RIKEN) \S\ref{sec:exoplanetatmosphere}.

\noindent
Kodama, Tadayuki (Tohoku University) \S\ref{sec:lssandmassassembly}.

\noindent
Koyama, Yusei (NAOJ) \S\ref{sec:lssandmassassembly}.

\noindent
Kurokawa, Hiroyuki (University of Tokyo) \S\ref{sec:PS_theoreticalperspective}.

\noindent
Matsuo, Taro (Nagoya University) \S\ref{sec:exoplanetatmosphere}.

\noindent
Matsuoka, Yoshiki (Ehime University) \S\ref{sec:highzquasars}.

\noindent
Matsuura, Shuji (Kwansei Gakuin University) \S\ref{sec:cosmicinfraredbackground}.

\noindent
Mawatari, Ken (NAOJ) \S\ref{sec:lssandmassassembly}.

\noindent
Misawa, Toru (Shinshu University) \S\ref{sec:IGMmoleculargas}.

\noindent
Moriya, Takashi (NAOJ) \S\ref{sec:highzsupernovae}.

\noindent
Nagamine, Kentaro (Osaka University) \S\ref{sec:EG_theoreticalperspective}.

\noindent
Nakajima, Kimihiko (NAOJ) \S\ref{sec:empg}.

\noindent
Nomura, Hideko (NAOJ) \S\ref{sec:protoplanetarydisks}.

\noindent
Notsu, Shota (RIKEN / University of Tokyo) \S\ref{sec:protoplanetarydisks}.

\noindent
Ootsubo, Takafumi (NAOJ) \S\ref{sec:icysmallbodies}.

\noindent
Ohno, Kazumasa (NAOJ, UCSC) \S\ref{sec:exoplanetatmosphere}.

\noindent
Sagawa, Hideo (Kyoto Sangyo University) \S\ref{sec:solarsystemplanet}.

\noindent
Shimonishi, Takashi (Niigata University) \S\ref{sec:magellanicclouds}.

\noindent
Tadaki, Ken-ichi (NAOJ) \S\ref{sec:submmgalaxies}.

\noindent
Takami, Michihiro (ASIAA) \S\ref{sec:starformingregions}.

\noindent
Terai, Tsuyoshi (NAOJ) \S\ref{sec:icysmallbodies}.

\noindent
Toba, Yoshiki (NAOJ) \S\ref{sec:dustyagns}.

\noindent
Yamashita, Takuji (NAOJ) \S\ref{sec:dustyagns}.

\noindent
Yasui, Chikako (NAOJ) \S\ref{sec:starformingregions}.

\end{multicols}

\tableofcontents

\clearpage


\chapter{Overview of GREX-PLUS}
\label{chap:overview}


\section{Introduction}

GREX-PLUS (Galaxy Reionization EXplorer and PLanetary Universe Spectrometer) is a mission concept of a space telescope with a 1.2 m aperture and 50 K temperature, equipped with a wide-field camera in the 2-8 $\mu$m wavelength band and a high resolution spectrometer with a wavelength resolution of 30,000 at 10-18 $\mu$m, for a JAXA's strategic L-class mission to be launched in the 2030s (Table~\ref{tab:GPbasedesign}).
It aims to revolutionize the researches of galaxy formation and evolution and planetary system formation and evolution by achieving a high sensitivity that is impossible from the ground.

The main goals of GREX-PLUS in galaxy formation and evolution are to discover rare bright ``first galaxies'' in the earliest epoch of the Universe (cosmic age less than a few hundred million years) and to observe ``building blocks'' with one-hundredth the mass of a galaxy over 95\% of the history of the Universe (after several hundred million years in cosmic age). 
For this purpose, the GREX-PLUS wide-field camera will perform super wide-field surveys in the 2-8 $\mu$m wavelength band, which are 10--100 times deeper (i.e. more sensitive) for the same area or 100--1000 times wider for the same depth than previous surveys conducted by the Spitzer Space Telescope.
The image data obtained from the GREX-PLUS surveys will also be used to search for the most distant supernova explosions, to search for massive blackhole objects in the most distant Universe or those heavily obscured by dust, to measure infrared background radiation, and to perform a census in the Galactic plane and time-domain astronomy, and will be legacy data in the history of astronomy.

The main goals of GREX-PLUS in planetary system formation and evolution are to determine the location of water ``snowline'' in protoplanetary disks in the Galaxy and to understand the planet formation process, such as the segregation of rocky planets and gas giant planets.
Furthermore, GREX-PLUS explores planetary biosphere research to understand planetary atmospheric structures, the origin of surface ocean on rocky planets, and synthesis processes of organic materials (and eventually the origin of life), by observing organic molecules in protoplanetary disks and planetary atmospheres inside and outside of the Solar system. 
For this purpose, the GREX-PLUS high resolution spectrometer will open a new window to interstellar molecular spectroscopy with its extremely unique capability of the wavelength resolving power of 30,000 from space in the 10-18 $\mu$m wavelength band, which is referred to as the ``fingerprint region'' of molecular spectroscopy.

GREX-PLUS inherits and combines the scientific and technical heritages of SPICA \citep{2018PASA...35...30R}, a former ESA and JAXA joint mission candidate, and WISH \citep{2012SPIE.8442E..1AY}, a former JAXA's strategic L-class mission candidate.
WISH was a mission concept to expand highly successful optical wide-field surveys with the Subaru Telescope into near-infrared surveys from space.
The WISH working group (WG) under the ISAS/JAXA studied a passively cooled telescope with a 1.5 m aperture and a wide-field camera covering 1-5 $\mu$m wavelengths from 2008 to 2015.
Based on the research heritages, the Galaxy and Reionization EXplorer (G-REX) project was started in January 2020.
SPICA was a space infrared observatory project studied for about 20 years, aiming at mid- and far-infrared observations using a space telescope cooled to cryogenic temperature.
After cancellation of SPICA, we have developed the GREX-PLUS mission concept, which has improved the cooling performance and reliability of the G-REX telescope by utilizing the world's best space telescope cooler technology accumulated through the SPICA study, and has enhanced scientific functions by installing the SPICA's mid-infrared high resolution spectrograph. 
In this way, GREX-PLUS is planned to realize an important part of the highly evaluated SPICA's science.

GREX-PLUS is a project that leverages the collective efforts of the Japanese optical-infrared astronomy community by combining the cryogenic space telescope technology and planetary science in the mid-infrared band, which are developed in SPICA studies, and the world-leading super wide-field imaging surveys of distant galaxies conducted with the Subaru Telescope.
This project will be an essential stepping stone toward the future era of very large space telescopes in orbit.

The structure of this science book is ...

\begin{table}
    \label{tab:GPbasedesign}
    \begin{center}
    \caption{GREX-PLUS baseline design.}
    \begin{tabular}{|l|l|}
    \hline
    Telescope & $\phi1.2$ m, 50 K, diffraction limit at 4 $\mu$m \\
    \hline
    Wide-field camera & 1,260 arcmin$^2$ divided into 5 bands in 2-8 $\mu$m \\
    \hline
    high resolution spectrometer & Resolving power $R=30,000$ in 10-18 $\mu$m \\
    \hline
    Life time & 5 years (Goal; $+5$ or more years) \\
    \hline
    Orbit & Sun-Earth L2 or Earth trailing \\
    \hline
    Launch & 2023s by JAXA's H3 launch vehicle \\
    \hline
    \end{tabular}
    \end{center}
\end{table}

\section{Scientific Significance}
\label{sec:scientificsignificance}

\begin{figure}
 \begin{center}
  \includegraphics[width=15cm]{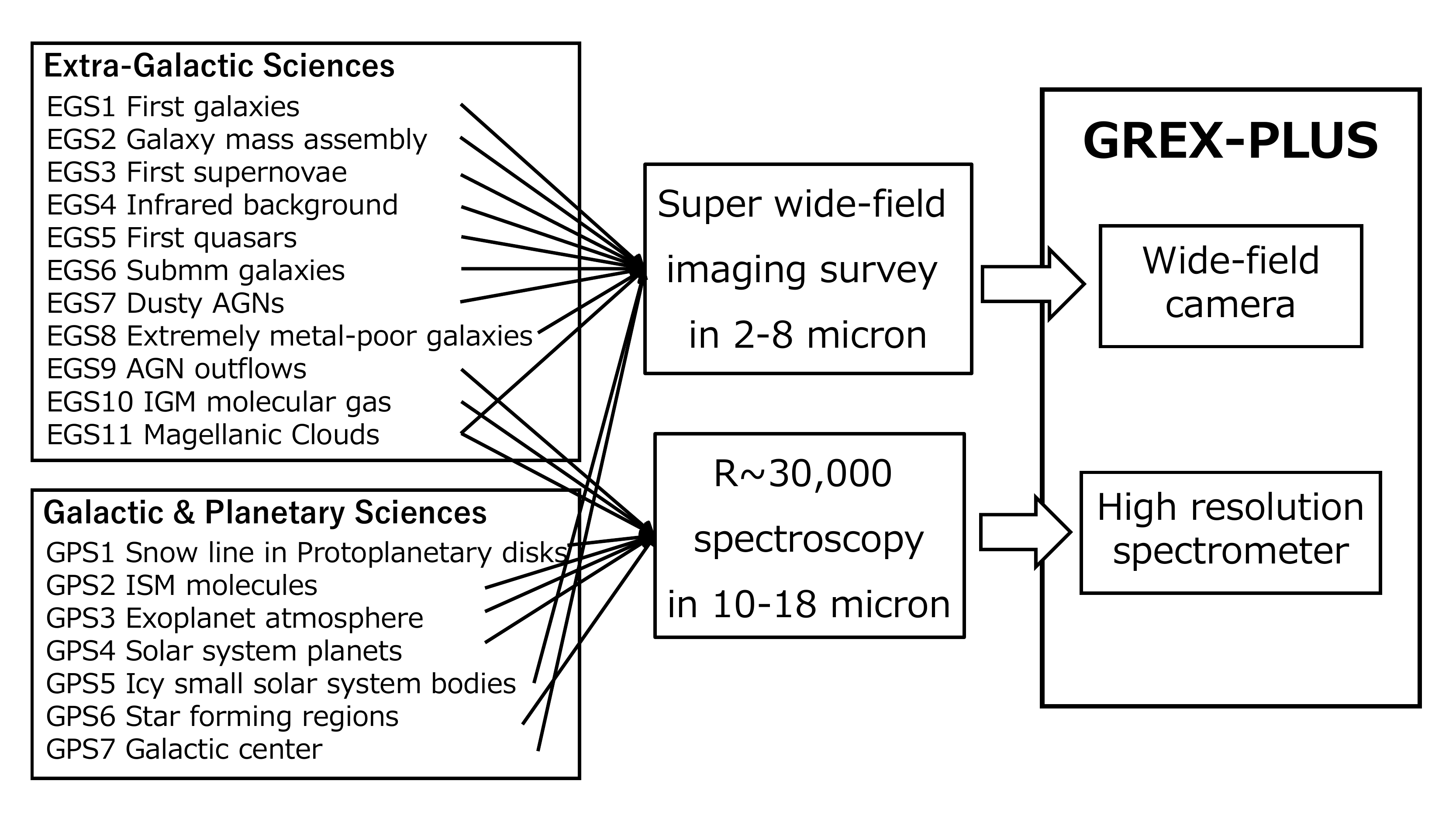}
 \end{center}
 \caption{Scientific objectives, required observations, and instruments of GREX-PLUS.
 \label{fig:SG_RG_INS}}
\end{figure}

Scientific goals in astronomy and astrophysics of ISAS/JAXA are ``to understand the origin of space and matter in the Universe'' and ``to explore the possibility of life in the Universe.''
GREX-PLUS is a project to tackle both of these goals directly. 
In this section, we describe the scientific significance of Extra-Galactic Sciences (EGS) and Galactic and Planetary Sciences (GPS) that GREX-PLUS aims to achieve, corresponding to these two major goals, respectively. 
Figure~\ref{fig:SG_RG_INS} shows the relation between the individual science cases to achieve the two major goals and the two GREX-PLUS instruments required for them. 
The individual science cases will be presented in detail in Chapters~\ref{chap:extragalactic} and \ref{chap:galacticplanetary}.

The primary goal of EGS of GREX-PLUS is to discover ``first galaxies'' and to reveal the formation processes of galaxies (EGS1; Section~\ref{sec:veryhighz}).
According to the current concordance structure formation theory based on cold dark matter, the ``first galaxies'' are thought to have formed at the cosmic age of less than 300 million years and a redshift of more than 15. 
The scientific value of proving this hypothesis by direct detection of ``first galaxies'' is extremely high. 
Such young galaxies can be detected in the ultraviolet radiation in the source rest-frame, at such a high redshift, infrared observations at wavelengths longer than 2 $\mu$m are required due to the significant redshift caused by the expansion of the Universe. 
In addition, since ``first galaxies'' are rare and faint, a sufficiently wide and deep imaging survey is required to detect them.

The secondary goal is to detect ``building blocks'' one-hundredth the mass of our Galaxy over 95\% of the history of the Universe and to determine how they merged, grew, and formed mature galaxies such as our Galaxy (EGS2; Section~\ref{sec:lssandmassassembly}). 
This is one of the ultimate goals of galactic astronomy, which began about 100 years ago at the beginning of the 20th century, and is of great scientific value. 
To achieve this goal, it is necessary to observe the optical radiation in the source rest-frame up to redshift about 8, and infrared observations at wavelengths of 2 $\mu$m or longer are again required due to the redshift. 
In addition, deep imaging with a wide field of view is also essential.

The third goal is to reveal the origin of supermassive blackholes and the galaxy-blackhole ``coevolution'' (EGS5-7). 
Galaxy-blackhole ``coevolution'' is a relatively new concept proposed about 20 years ago, in which a galaxy and a supermassive blackhole at the center of the galaxy mutually influence each other in their evolution.
It is a topic that has attracted a very high level of interest in recent years, and its high scientific value is indisputable. 
To achieve this goal, we need to search for the most distant quasars in the rest-frame ultraviolet radiation (EGS5; Section~\ref{sec:highzquasars}) and search for dusty supermassive blackholes in the rest-frame optical to near-infrared radiation (EGS7; Section~\ref{sec:dustyagns}). 
Distant galaxies bright in submillimeter wavelengths, called submm galaxies, are also a relevant galaxy population bridging quasars and dusty blackholes (EGS6; Section~\ref{sec:submmgalaxies}).
For these topics, infrared imaging observations in the 2-8 $\mu$m wavelength with a sufficient wide field-of-view and depth are required.

The discovery of most distant supernova explosions to elucidate various properties of first stars (EGS3; Section~\ref{sec:highzsupernovae}) and the discussion of signs of first stars and blackholes imprinted in the infrared background radiation (EGS4; Section~\ref{sec:cosmicinfraredbackground}) are also of high scientific value.
There is very high importance to conduct an imaging survey of distant galaxies with extremely low elemental abundances, that is a signature of primitive chemical evolutionary phases of the system (EGS8; Section~\ref{sec:empg}).
A census imaging survey of young stars in Magellanic Clouds (EGS11; Section~\ref{sec:magellanicclouds}) is also very important to understand star formation processes in lower elemental abundances than the interstellar medium in our Galaxy, which is relevant to the star formation in young galaxies in the distant Universe. 
These goals also require wide-field deep imaging surveys at wavelengths longer than 2 $\mu$m.

In addition, high resolution spectroscopy in the wavelength 10-18 $\mu$m allows us to perform various unique extragalactic sciences, including observations of the molecular gas outflow from supermassive blackholes that is essentially important to reveal feedback processes by them (EGS9; Section~\ref{sec:AGNmolecularoutflow}), observations of molecular gas in the intergalactic medium and cosmological experiments (EGS10; Section~\ref{sec:IGMmoleculargas}), and spectroscopy of massive young stars in Magellanic Clouds (EGS11; Section~\ref{sec:magellanicclouds}).

The first goal of GPS of GREX-PLUS is to identify the position of water ``snowline'' near the equatorial plane of protoplanetary disks (GPS1; Section~\ref{sec:protoplanetarydisks}).
For the theory of planetary system formation, ``snowlines'' play an important role in determining the segregation of rocky planets and gas giant planets.
It is also essential information to understand the water supply process to the surface on rocky planets, which elucidates the origin of the Earth's ocean.
Although previous observations have not succeeded in resolving the water ``snowline'' location in the equatorial plane of the disks, there is a new idea to observe a water emission line at a wavelength of 17.8 $\mu$m with a wavelength resolution of 30,000 and to velocity-resolve the Keplerian motion of water molecules, eventually identifying the ``snowline'' position.
To realize this experiment for a statistical sample of $\sim100$ protoplanetary disks, for the first time, is of great scientific value.

The second goal is to elucidate the interstellar synthesis process of organic molecules and other materials that may evolve into life eventually (GPS2; Section~\ref{sec:interstellarmolecules}). 
Understanding the evolutionary process from simple interstellar molecules to complex organic molecules that could be the origin materials of life is of extremely high scientific value as it addresses the ultimate question of the origin of life on Earth. 
For this purpose, it is necessary to realize high resolution spectroscopy with a wavelength resolution of 30,000 in the 10-18 $\mu$m wavelength band, which is also called the ``fingerprint region'' of molecular spectroscopy.

The third goal is to examine the atmospheric properties of the Solar system planets and extra-Solar planets (GPS3-4; Sections~\ref{sec:solarsystemplanet}-\ref{sec:exoplanetatmosphere}).
Since rocky planetary surfaces are expected to be the sites of life in the Universe, it is clear that understanding the atmospheric properties that determine the environments of the planetary surface has a high scientific value. 
For example, it is necessary to observe molecules that are difficult or impossible to observe from the ground, such as hydrogen molecular quadrupole emission lines at wavelengths of 12 $\mu$m and 17 $\mu$m, to obtain information that has never been obtained before. 
In this case, sufficiently high wavelength resolution is required to observe molecular emission lines in crowded wavelength bands and emission lines of rare molecules.

In addition, young stars and star forming regions in the Galaxy are also very interesting targets for GREX-PLUS (GPS6; Section~\ref{sec:starformingregions}).
Velocity resolving observations for the [Ne II] line at 12 $\mu$m with high resolution spectrometer enable us to identify the outfolw motion of gas from protoplanetary disks and to constrain the lifetime of the disks.
Imaging observations at wavelengths of 2-8 $\mu$m, especially around 3 $\mu$m water ice feature, for the Solar system provide an excellent opportunity to survey icy small bodies (GPS5; Section~\ref{sec:icysmallbodies}).
The Galactic plane including the Galaxy center is also an excellent target for GREX-PLUS imaging surveys (GPS7; Section~\ref{sec:galaxycenter}).

GREX-PLUS provides super wide-field imaging data with high sensitivity and angular resolution in the 2-8 $\mu$m wavelength band and unique mid-infrared high resolution spectroscopic data with a wavelength resolution of 30,000, all of which will eventually be released to the world as archived data and become part of the precious intellectual property of humanity. 
The data provided by both instruments will be unique and unparalleled and will be utilized in all fields of astronomy for a long time.
The new perspectives and analysis methods that we do not have today are expected to lead to various revolutionary discoveries in the future. 
This includes, of course, the discovery of unknown objects and phenomena. 
In order to promote such discoveries, we believe that developing an effective method of releasing archival data will also have a very high scientific value.

\section{GREX-PLUS Wide-Field Camera and Imaging Surveys}
\label{sec:camera}

\begin{figure}
 \begin{center}
  \includegraphics[width=7cm]{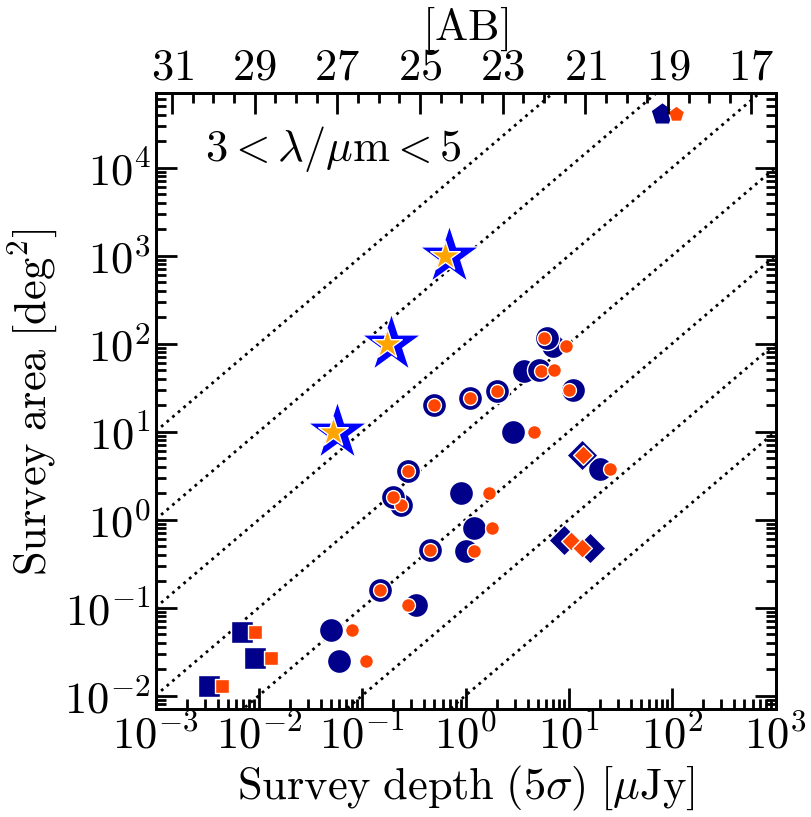}
  \includegraphics[width=7cm]{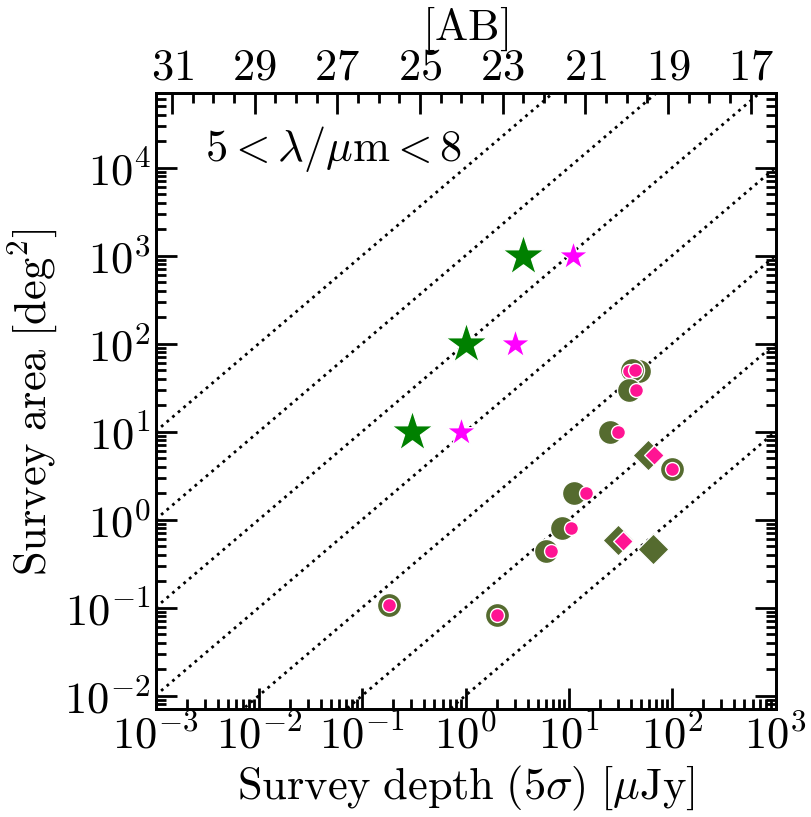}
 \end{center}
 \caption{Comparisons of survey areas and depths in the wavelength 3--5 $\mu$m bands (left) and 5--8 $\mu$m bands (right). The bluish (orange-ish) colors are $3$ (4) $\mu$m bands in the right panel, while the green-ish (magenta-ish) colors are $\sim5$ (7) $\mu$m bands in the left panel. The five-pointed-stars are the GREX-PLUS survey parameters. The circles, diamonds, pentagons, and squares are Spitzer, AKARI, WISE, and JWST surveys, respectively.
 \label{fig:surveypower}}
\end{figure}

To achieve a number of valuable sciences described briefly in the previous section and in detail in the following Chapters, GREX-PLUS will be equipped with a wide-field camera and perform super wide-field imaging surveys in the 2-8 $\mu$m wavelength band.
Table~\ref{tab:GPcamera} summarizes the performance requirements for the GREX-PLUS wide-field camera to conduct the three types of imaging surveys listed in Table~\ref{tab:GPsurveys}.
As shown in Figure~\ref{fig:surveypower}, in the 3-4 $\mu$m and 5-8 $\mu$m wavelength bands, the three surveys of GREX-PLUS are about 10--100 times deeper in similar areas and about 100--1000 times wider at similar depths than the NASA/Spitzer Space Telescope's imaging surveys achieved over its 16 years of operation.
For comparison, Figure~\ref{fig:surveypower} also shows the AKARI imaging survey in the same wavelength band and the WISE all-sky survey depth. 
The survey parameters planned for the NASA/James Webb Space Telescope (JWST) are also shown.
JWST is capable of reaching much deeper depths, but its narrow field-of-view limits the area it can cover very much.
The super wide-field surveys by the ESA/Euclid and the NASA/Nancy Grace Roman Space Telescope (hereafter referred to as Roman), scheduled in the 2020s, are limited to wavelengths shorter than 2-2.3 $\mu$m and are not shown in Figure~\ref{fig:surveypower}.

The imaging surveys with the Spitzer Space Telescope have been produced more than 2,500 scientific papers, which include, for example, photometry data of the source rest-frame optical wavelength of the most distant galaxies.
The GREX-PLUS wide-field surveys surpass the Spitzer's surveys by two to three orders of magnitude in depth and area, and completely revolutionize them.
The Spitzer Space Telescope has an angular resolution of 1.5 arcsec or worse at 4 $\mu$m wavelengths, while GREX-PLUS has an aperture about twice that of the Spitzer Space Telescope and achieves an angular resolution of 1 arcsec at 4 $\mu$m wavelength.
The angular resolution is a factor of 1.5 better, which is also a significant advantage of GREX-PLUS.
There is no other plan to realize the deep and wide imaging surveys with good angular resolution in the 2-8 $\mu$m wavelength band that the GREX-PLUS wide-field camera can achieve.
Therefore, the GREX-PLUS survey data will be an essential and fundamental legacy for almost all astronomical communities for a long time to come. 
Its scientific value will deserve the highest possible recognition.

\begin{table}
    \begin{center}
    \caption{GREX-PLUS wide-field camera performance requirements.}
    \label{tab:GPcamera}
    \begin{tabular}{|l|c|c|c|c|c|}
    \hline
     &  F232 & F303 & F397 & F520 & F680 \\
    \hline
    Central wavelength [$\mu$m] & 2.3 & 3.0 & 4.0 & 5.2 & 6.8 \\
    \hline
    Wavelength range [$\mu$m] & 2.0-2.6 & 2.6-3.4 & 3.4-4.5 & 4.5-5.9 & 5.9-7.7\\
    \hline
    Resolving power [$\lambda/\Delta\lambda$] & \multicolumn{5}{|c|}{3.7} \\
    \hline
    Pixel scale [arcsec pix$^{-1}$] & \multicolumn{5}{|c|}{0.40} \\
    \hline
    Field-of-view [arcmin$^2$] & 180 & 180 & 540 & 180 & 180 \\
    \hline
    Detector & \multicolumn{5}{|c|}{HgCdTe} \\
    \hline
    Sensitivity$^\dag$ [ABmag] & 23.8 & 23.9 & 23.4 & 22.1 & 20.9 \\
    \hline
    \end{tabular}
    \end{center}
    $^\dag$ 300 sec, $5\sigma$ for a point-source, assuming the background intensity of 0.11 MJy str$^{-1}$ at the wavelength of 3 $\mu$m (three times higher than that in the North Ecliptic Pole).\\
\end{table}

\begin{table}
    \begin{center}
    \caption{GREX-PLUS wide-field survey plan.}
    \label{tab:GPsurveys}
    \begin{tabular}{|l|p{2cm}|p{2cm}|p{2cm}|}
    \hline
     & Deep & Medium & Wide \\
    \hline
    Area [deg$^2$] & 10 & 100 & 1000 \\
    \hline
    F232 [ABmag,$5\sigma$] & 26.9 & 25.6 & 24.2 \\
    \hline
    F303 [ABmag,$5\sigma$] & 27.0 & 25.7 & 24.3 \\
    \hline
    F397 [ABmag,$5\sigma$] & 27.1 & 25.8 & 24.4 \\
    \hline
    F520 [ABmag,$5\sigma$] & 25.2 & 23.9 & 22.5 \\
    \hline
    F680 [ABmag,$5\sigma$] & 24.0 & 22.7 & 21.3 \\
    \hline
    \end{tabular}
    \end{center}
\end{table}

\section{GREX-PLUS High Resolution Spectrometer}
\label{sec:HRspectrometer}

\begin{figure}
 \begin{center}
  \includegraphics[height=8cm]{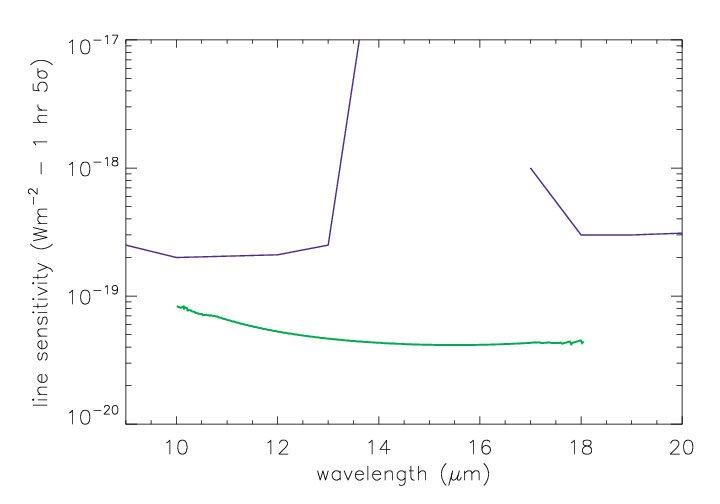}
 \end{center}
 \caption{A comparison of the line sensitivity (1 hr exposure, $5\sigma$) for unresolved narrow line emission between GREX-PLUS high resolution spectrometer (green curve) and a typical one on a ground-based 30 m-class telescope (blue curves). The gap between the wavelengths of 13.5 $\mu$m and 17 $\mu$m of the ground-based case is due to the very low transmission of the Earth's atmosphere.
 \label{fig:HRspec}}
\end{figure}

In order to realize science with high scientific value as described in Section~\ref{sec:scientificsignificance}, a high resolution spectrometer with the instrument performance shown in Table~\ref{tab:HRparameters} will be developed and installed on GREX-PLUS. 
GREX-PLUS is currently the only space telescope project with a high wavelength resolution of 30,000; JWST covers the same wavelength band, but its wavelength resolution is 3,000, which is one-tenth of GREX-PLUS.
On the other hand, ground-based large telescopes are planned to have even higher wavelength resolution. 
For example, MICHI (Mid-Infrared Camera, High-disperser, and IFU), which is being considered as a second phase instrument for the 30-m telescope TMT, will have a wavelength resolution of 60,000 to 120,000.
However, the 10-18 $\mu$m wavelength band targeted by GREX-PLUS contains many transitions of interesting molecules such as water, carbon dioxide, and ammonia, which are also abundant in the Earth's atmosphere, making ground-based observations difficult due to atmospheric absorption. 
As a result, continuous coverage of the wavelength range from the ground is impossible, and the sensitivity is limited.
As shown in Figure~\ref{fig:HRspec}, GREX-PLUS has at least several times higher emission line sensitivity than TMT/MICHI-like instrument, thanks to the great advantage of observing from outside the Earth's atmosphere. 
Another major advantage of GREX-PLUS is continuous wavelength coverage, allowing GREX-PLUS to observe wavelength bands that ground-based telescopes cannot.
Compared to JWST, which covers the same wavelength band, GREX-PLUS has 10 times higher wavelength resolution. 
For example, interesting objects can be selected in advance with JWST's medium resolution spectroscopy and decisive results can be obtained with GREX-PLUS's high resolution spectroscopy to develop very unique molecular spectroscopy. 
Specifically, there are many sciences requiring high velocity resolution, such as water ``snowlines'' in protoplanetary disks, exoplanetary atmospheres, and active galactic nuclei molecular outflows, and sciences in low temperature regions, such as detection of various organic molecules in interstellar clouds and molecules in Solar-system objects. 
No other project of this kind exists anywhere in the world, and it is easy to imagine that the scientific value of the data obtained with the GREX-PLUS high resolution spectrometer will continue to be of the highest standard for a long time.

\begin{table}
    \begin{center}
    \caption{GREX-PLUS high resolution spectrometer performance.}
    \label{tab:HRparameters}
    \begin{tabular}{|l|l|}
    \hline
    Wavelength coverage [$\mu$m] & 10-18 \\
    \hline
    Resolving power [$\lambda/\Delta\lambda$] & 30,000-34,000 \\
    \hline
    Slit size [arcsec$^2$] & $8 \times 3.5$ \\
    \hline
    Pixel scale [arcsec pix$^{-1}$] & 1.5 \\
    \hline
    Detector & Si:As, 1k$\times$1k \\
    \hline
    Continuum sensitivity$^\dag$ [mJy, $5\sigma$, 1hr] & 4.2 / 5.0 \\
    \hline
    Line sensitivity$^\dag$ [10$^{-20}$ W m$^{-2}$, $5\sigma$, 1hr] & 3.4 / 4.1 \\
    \hline
    \end{tabular}
    \end{center}
    $^\dag$ Sensitivities under the Zodiacal light intensities at the wavelength of 14 $\mu$m of 16 MJy str$^{-1}$ (low case) / 50 MJy str$^{-1}$ (high case).
\end{table}

\chapter{Extra-Galactic Sciences}
\label{chap:extragalactic}


\newcommand{\fesc}{f_{\rm esc}}
\newcommand{\Msun}{M_\odot}

\section{Theoretical Perspective}
\label{sec:EG_theoreticalperspective}

\noindent
\begin{flushright}
Kentaro Nagamine$^{1,2,3}$
\\
$^{1}$ Osaka University, $^{2}$ University of Nevada, $^{3}$ Kavli-IPMU (WPI), University of Tokyo
\end{flushright}
\vspace{0.5cm}

\subsection{Introduction}

We now live in the era of precision cosmology, where cosmological parameters are estimated with accuracies better than 10\%.  
The energy budget of the universe is dominated by dark matter ($\sim$30\%) and dark energy ($\sim$70\%) \citep{Planck18}, and the best-fit model is called the concordance $\Lambda$ cold dark matter ($\Lambda$CDM) model.  The structure formation within the $\Lambda$CDM paradigm has been studied extensively over the past few decades, and both theory and observations have made significant progress.  Astronomers are revealing the state of the early universe by studying the formation of the earliest galaxies and black holes, and we continue to push the research frontier to higher redshifts.  We were recently reminded that a new observatory could genuinely bring a breakthrough to observational astronomy, when the JWST released its magnificent new images in July 2022. 

\begin{figure}[t]
    \centering
    \includegraphics[scale=0.45]{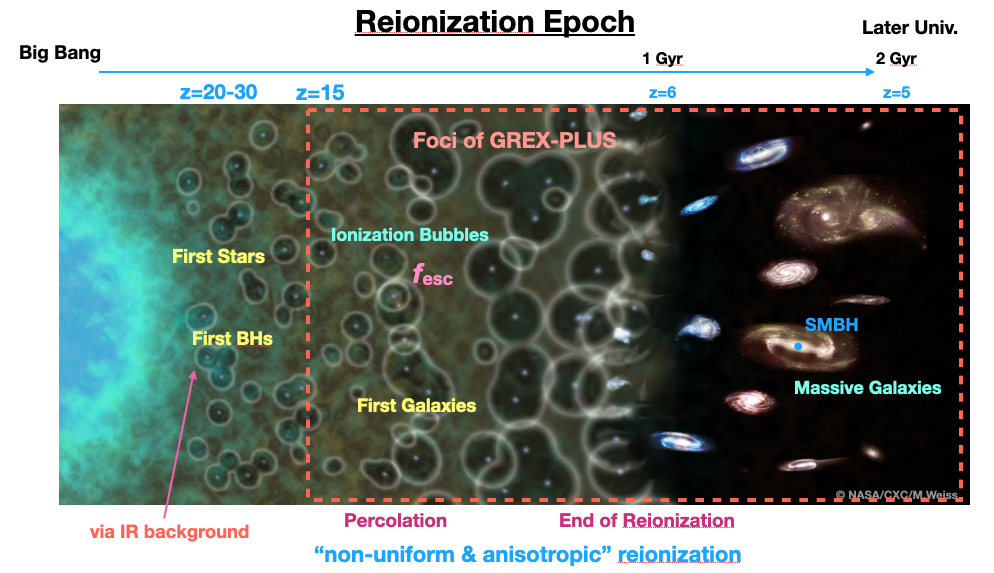}
    \caption{Structure formation in the $\Lambda$CDM cosmology and the scientific focus of GREX-PLUS.}
  \label{fig:cosmic}
\end{figure}

We show the current picture of cosmic reionization in the $\Lambda$CDM cosmology in Figure~\ref{fig:cosmic}.  
Here, the first stars and black holes form in mini-halos ($M_h \sim 10^6-10^8\,M_\odot$) at $z\approx 20-30$, followed by the formation of first galaxies at $z\approx 10-20$ in atomic-cooling halos of $M_h \sim 10^8\,M_\odot$.  
Then, the ionizing bubbles start to expand around them, gradually percolate, and the reionization completes by $z\sim 6$, as we know from the analyses of Gunn-Peterson trough in quasar spectra \citep{Fan06a}.  It is easy to imagine that cosmic reionization is a non-uniform and anisotropic process, especially in its early phase.  The critical physical parameter is the {\em escape fraction} ($\fesc$) of ionizing photons from early galaxies, as it determines the primary contributor to reionization and how it proceeds.  The balance between faint galaxies and AGNs as the source of ionizing photons is still a matter of debate, as discussed further below.

The central scientific theme of GREX-PLUS outlined in this document is ``understanding the origin of space and matter in the universe." More specific main goals in the context of galaxy formation and evolution include the following: (i) Discover rare bright first galaxies; (ii) Observe the ``building blocks" of the Milky Way-type galaxies down to $M_\star \sim 10^9 \Msun$, up to $z\sim 8$;  
and (iii) Search for the most distant supernova explosions, massive black holes hidden in dust, and measure the infrared background radiation.  
With GREX-PLUS, we will test the standard cosmology, structure formation theory, and baryon physics of galaxy formation in the earliest universe. We will provide observational evidence for the formation of the first massive galaxies in the universe. 
With these scientific goals in mind, let us look at the current status of reionization studies in detail. 

First, let us review the standard method for constructing a model of reionization history.  We start by looking at the evolution of the galaxy luminosity function. 
Observers have made significant progress in estimating the galaxy luminosity function over the past few decades, and its evolution is now constrained up to $z\sim 10$ \citep{Bouwens21}.  

\begin{figure}
    \centering
    \includegraphics[width=7cm]{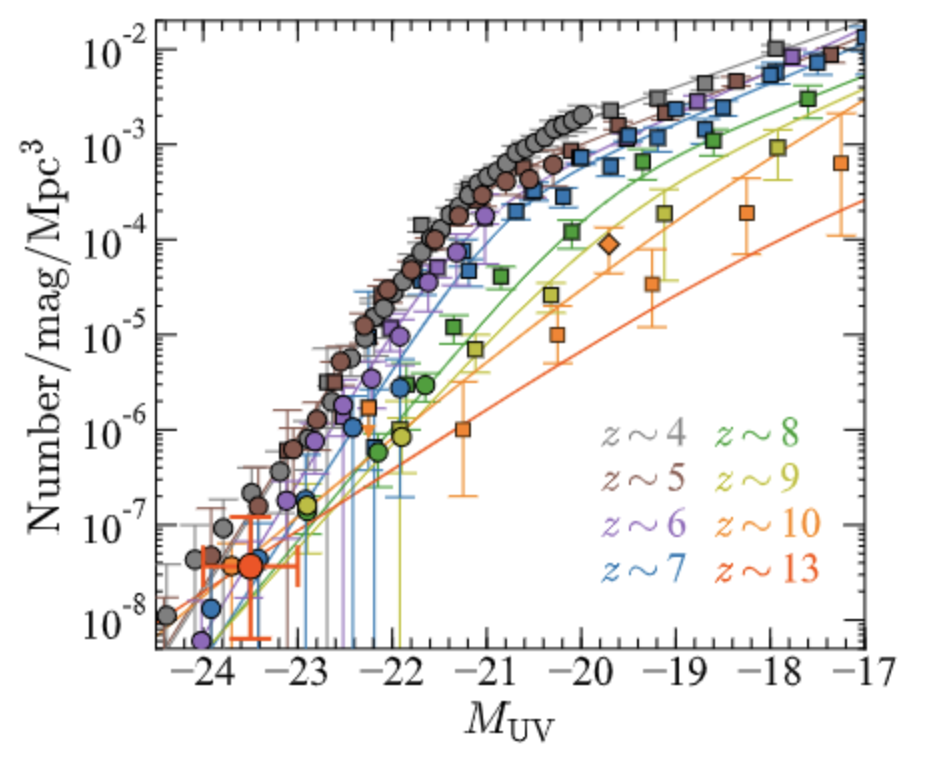}
        \includegraphics[width=8.5cm]{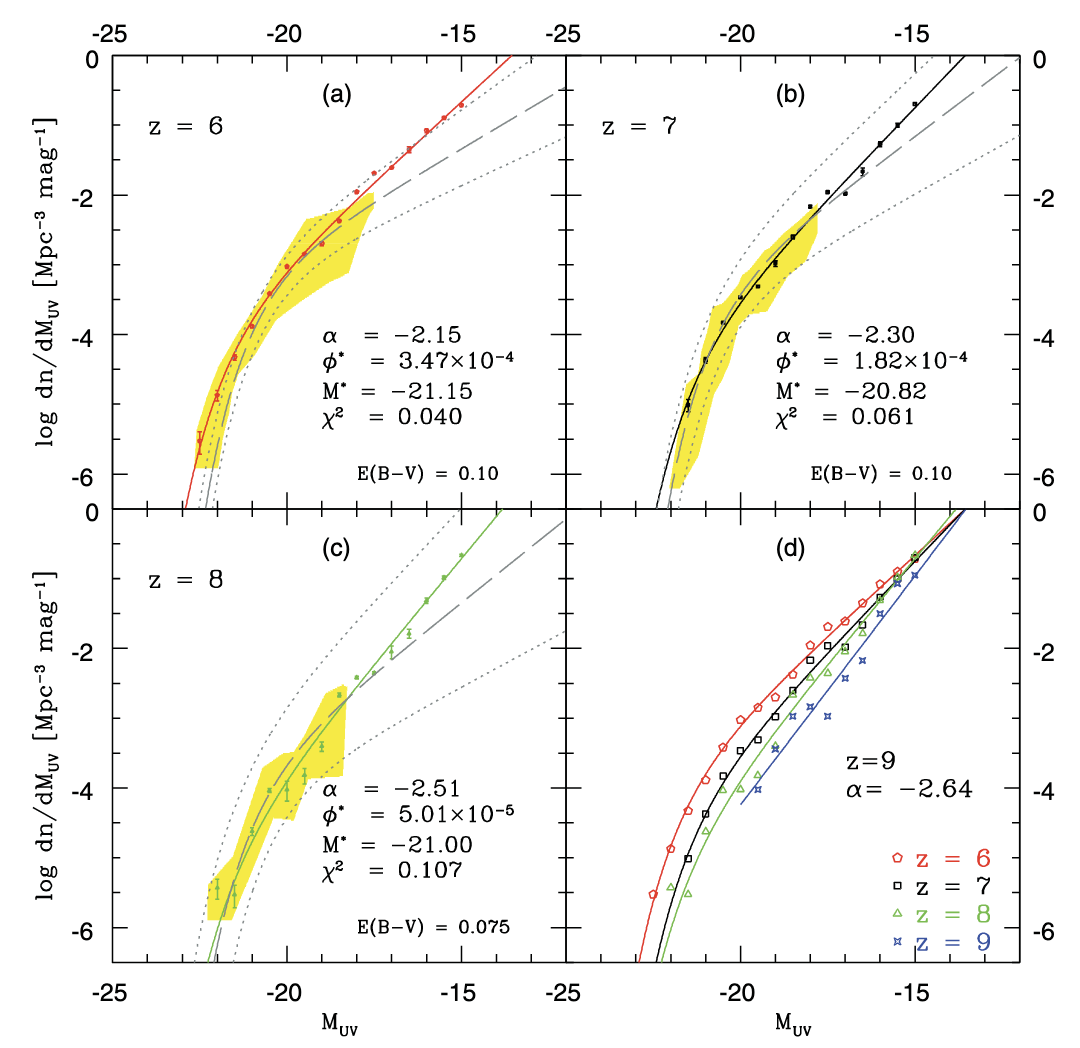}
    \caption{Examples of rest-frame UV luminosity functions of galaxies at $z=4-13$ from observations \citep[][left]{Harikane22} and cosmological hydrodynamic simulations \citep[][right]{Jaacks12b}.}
   \label{fig:LF}
\end{figure}

Most recently, \citet{Harikane22} put a constraint at the bright end of the luminosity function at $z\sim 13$ as shown in Figure~\ref{fig:LF}.  By integrating these luminosity functions, one can obtain the UV luminosity density in the universe as a function of redshift. The observed faint-end slope becomes steeper from $\alpha \sim -1.2$ ($z$=0) to $-1.6$ ($z$=3), and to even steeper slopes of $\alpha \lesssim -2.0$ at $z\gtrsim 6$, which is consistent with earlier predictions by $\Lambda$CDM cosmological hydrodynamic simulations \citep{Nag04d,Nag04e,Night06,Jaacks12b}. 

Assuming that massive stars produce UV luminosity with short lifetimes, the UV luminosity density can be converted to the cosmic star formation rate (SFR) density as a function of redshift $\rho_{\rm SFR}(z)$ \citep[e.g.,][]{Madau96,Madau14,Robertson15}.  Then the ionized volume fraction $Q_{\rm HII}$ can be computed as
\begin{equation}
    \dot{Q}_{\rm HII} = \frac{\dot{n}_{\rm ion}}{\langle n_{\rm H} \rangle} - \frac{Q_{\rm HII}}{t_{\rm rec}}, 
    \label{eq:QHII}
\end{equation}
and the production rate of the ionizing photons as 
\begin{equation}
    \dot{n}_{\rm ion}(z) = \fesc\, \xi_{\rm ion}\, \rho_{\rm SFR}(z). 
\end{equation}
Here, $\xi_{\rm ion}$ is the ionizing photon production per unit SFR. 
The $\fesc$ could depend on various quantities, e.g., SFR of galaxies, halo mass, and redshift, but it is often assumed as a constant for simplicity.

\begin{figure}
    \centering
    \includegraphics[width=8cm]{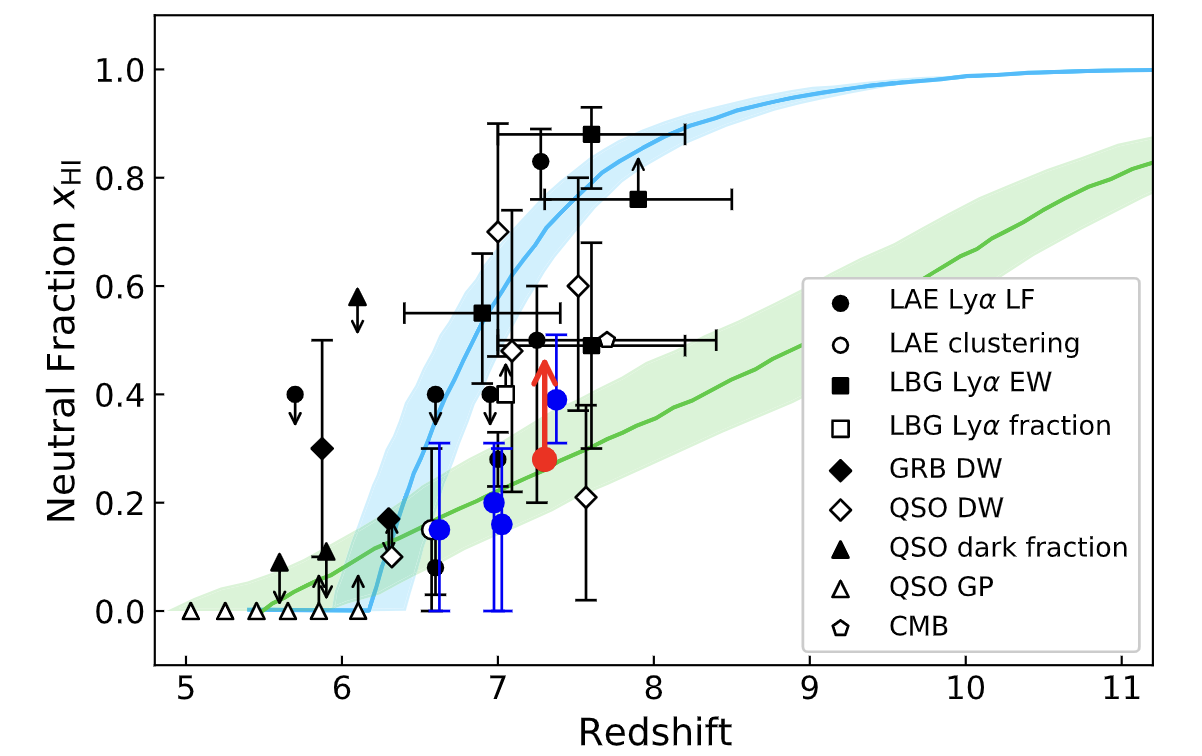}
    \caption{Volume-averaged neutral fraction as a function of redshift from \citet{Goto21}. See the main text for more details on the lines. }
   \label{fig:XHII}
\end{figure}

Once $Q_{\rm HII}(z)$ is obtained from Eq.~(\ref{eq:QHII}), then volume-averaged neutral fraction can be computed as $x_{\rm HI} = 1 - Q_{\rm HII}$ as shown in Figure~\ref{fig:XHII}. 
The green line is the ``{\em Early reionization} 
scenario" from 
\citet{Finkelstein19}, which allows higher $\fesc$ for lower mass halos and a steep faint-end slope for galaxy luminosity function; therefore the reionization is taking place earlier by lower mass galaxies.  
The blue line is the ``{\em Late reionization} scenario" from \citet{Naidu20}, which adopts constant $\fesc$ and therefore allows more contribution from more massive halos at later times, and the reionization is happening at a later time.  
The red dot in Fig.~\ref{fig:XHII} shows the lower limit obtained from the observed LAE luminosity function evolution at $z=7.3$ \citep{Goto21}.  From this figure, it is still difficult to discriminate either of the two models from current observational estimates. However, the question of ``How early did the reionization begin?" is closely related to ``When did first galaxies start to form and what are their $\fesc$?". 
Therefore the faint-end slope of early galaxies and the $\fesc$ of faint galaxies are the key parameters in the reionization study. 

\begin{figure}
    \centering
    \includegraphics[width=7cm]{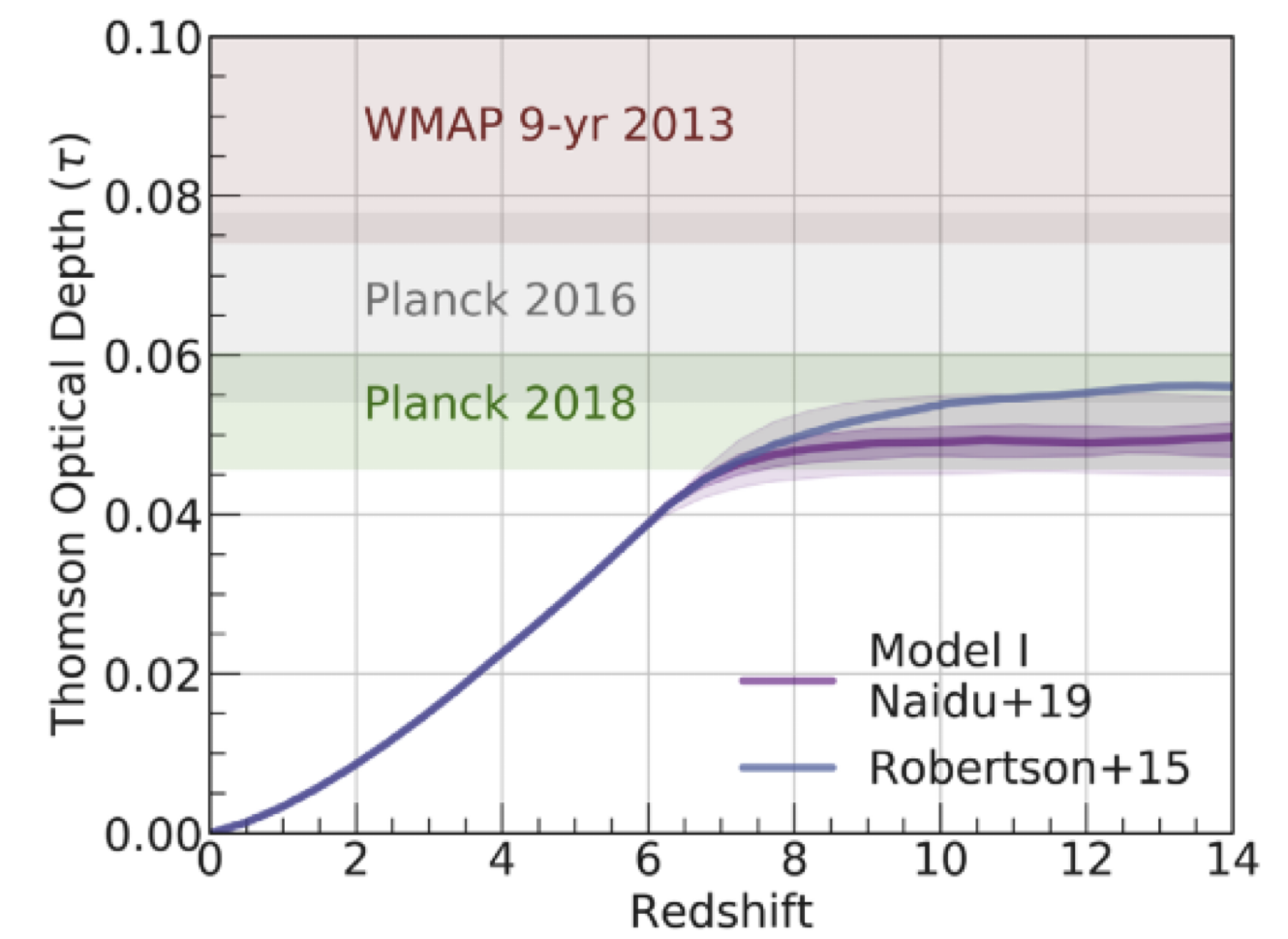}
    \caption{Thomson optical depth as a function of redshift from \citet{Naidu20}. }
   \label{fig:tau}
\end{figure}

One additional check on the cumulative Thomson optical depth up to redshift $z$ is required as follows: 
\begin{equation}
    \tau(z) = c \langle n_H \rangle \sigma_T \int_0^z f_e Q_{\rm HII}(z^\prime) \frac{(1+z^\prime)^2}{H(z^\prime)}dz^\prime, 
\end{equation}
where $c$ is the speed of light, $\sigma_T$ is the Thomson scattering cross section, $f_e$ is the number of free electrons for every hydrogen nucleus in the ionized IGM, and $H(z)$ is the Hubble parameter.  
The result is shown in Fig.~\ref{fig:tau}, together with the observational constraints from the WMAP and Planck cosmic microwave background measurement, which has come down to a lower value over the years. Both the `early' and `late' reionization scenarios should satisfy this constraint, but when they reach the final asymptotic value of $\tau_e \simeq 0.05$ would be different according to the scenario.  

Given that the parameter $\fesc$ of galaxies is the highly uncertain and most critical parameter for reionization models, it would be nice if we can predict it theoretically in the {\em ab initio} cosmological hydrodynamic simulations. 
Over the years, there have been many works trying to evaluate $\fesc$ of early galaxies and its impact on the reionization process \citep[e.g.,][]{Cen03d,Razoumov06a,Gnedin08b,Wise09,Yajima17}. 
We will not go into the details of each work here; however, it has always been difficult to simulate the effect of supernova feedback from first principles, and a subgrid model was necessary on scales below $\lesssim 100$\,pc. But there was a rough consensus from these early studies that $\fesc$ decreases with increasing halo mass with a large scatter. The galaxies in massive halos are embedded in deeper potential wells and denser gas. Therefore, the ionizing photons have difficulty getting out of the halos. 

\begin{figure}
    \centering
    \includegraphics[width=11cm]{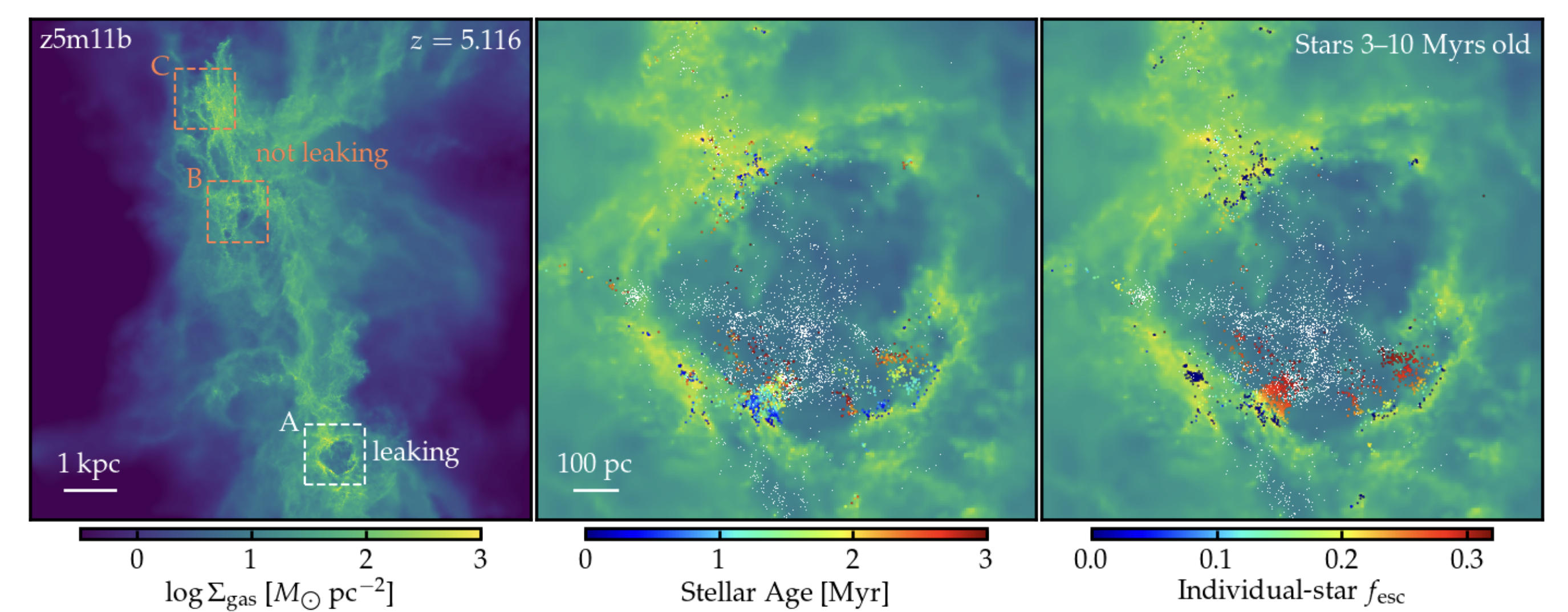}
    \caption{Zoom-in cosmological hydrodynamic simulation of high-redshift galaxies, resolving individual superbubble with different escape fractions \citep{Ma20}. The left panel shows the large-scale gaseous filament. The middle and right panels show the star particles on top of gas density in region-A of the left panel, color-coded by their stellar ages and $\fesc$, respectively.}
   \label{fig:Ma}
\end{figure}

More recently, it has become possible to push the resolution down to pc or sub-pc scales, and resolve the formation of individual superbubbles in early galaxies.  For example, \citet{Ma20} performed cosmological zoom-in simulation with GADGET-3 SPH code (FIRE-2 simulation) with pc-scale spatial resolution and a mass resolution of 100\,$M_\odot$. This high resolution allowed them to resolve each star cluster and the formation of feedback-driven superbubbles.  Figure~\ref{fig:Ma} shows that the star clusters that form in high-density regions have low $\fesc$, but those that have already moved out of the superbubble wall have higher $\fesc$.  This results in a large temporal/spatial variation in $\fesc$, with a mean around $\fesc \sim 0.1-0.2$, consistently with earlier work by \citet{Kimm14}.

\citet{Hu19} pushed the resolution further by simulating an isolated dwarf galaxy 
with a gas mass resolution of 1\,$M_\odot$ and a spatial resolution of 0.3\,pc.  This resolution also allowed him to resolve the formation of individual superbubbles of a few 100\,pc, which then break out of the disk and expand to form a galactic   
wind. The interesting point is that the resolved wind is weaker than those implemented in large-scale cosmological simulations, which requires further investigation. 
Of course, there are other studies that focused on even smaller scales (i.e., only a part of the galactic disk or ISM) and examined the impact of supernova feedback \citep{Girichidis16,Martizzi16,Gatto17,Kim18,El-badry19,Lancaster21a,Oku22}.  These high-resolution simulations reveal the need to further study the interface between supernova and surrounding ISM,  and how feedback energy propagates to larger scales in different forms (kinetic, thermal, radiation, and cosmic ray energy).   Overall, the advancement of these high-resolution simulations of supernova feedback is quite impressive on both small and galactic scales, and more progress can be expected in the next decade using more sophisticated code with more physics such as thermal conduction, magnetic fields, and cosmic rays. 

\begin{figure}
    \centering
    \includegraphics[width=9.7cm]{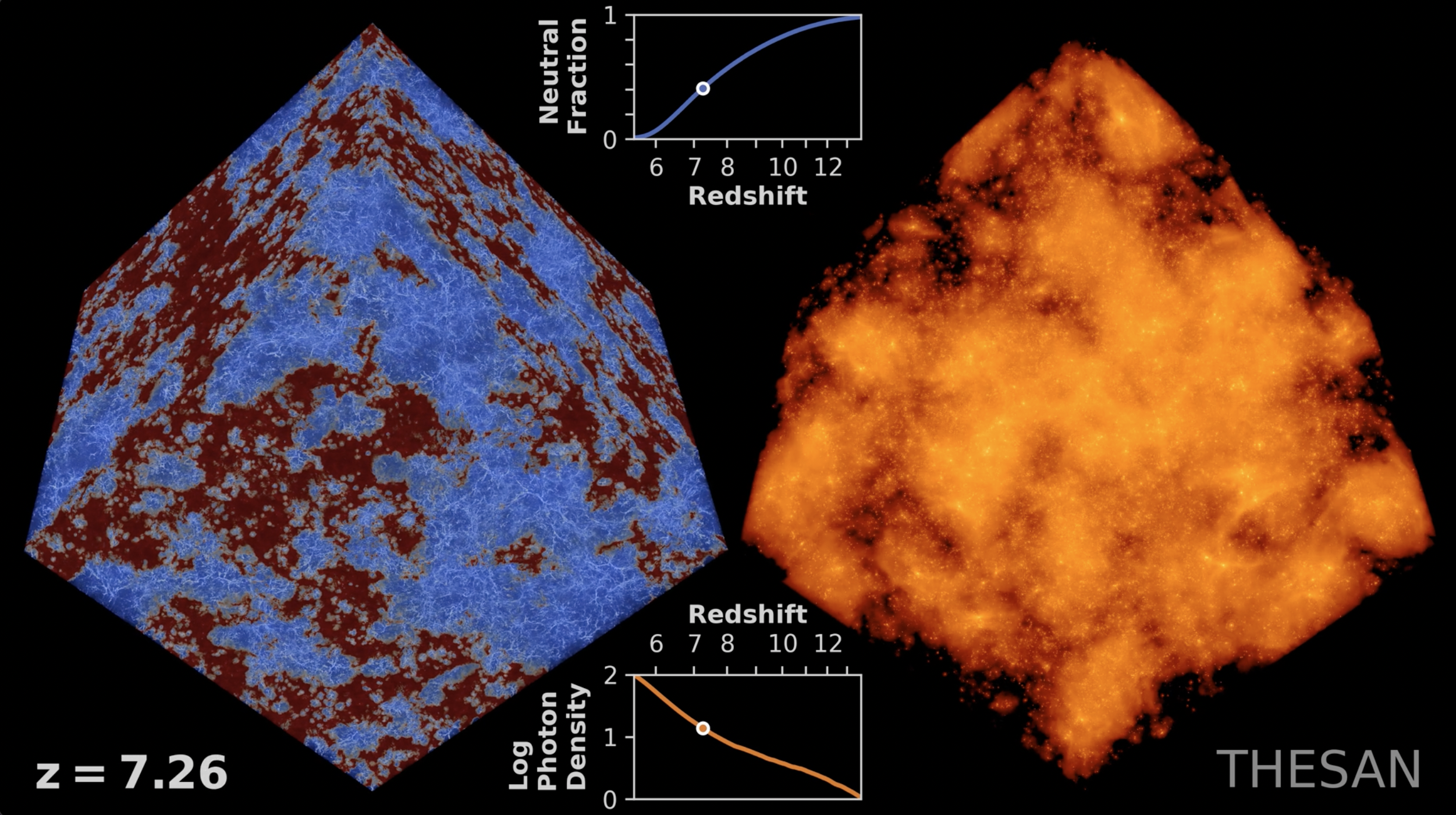}
    \caption{THESAN cosmological radiation hydrodynamic simulation. The left box shows the hydrogen neutral fraction, and the right box shows the ionizing radiation field.}
   \label{fig:THESAN}
\end{figure}

Finally, there has been significant progress on larger scales as well.  For example, in the THESAN simulations \citep{Kannan22,Garaldi22,Yeh22}, radiation hydrodynamic simulations were performed with a box size of 95.5\,cMpc using the AREPO-RT moving mesh code with on-the-fly star formation, feedback, and radiation transfer.  Previously it has been difficult to carry out such a cosmological run with all these physics simultaneously, and people mainly focused on post-processing with radiation transfer due to its heavy computational load.  The THESAN simulations inherit the subgrid models from IllustrisTNG simulation with $\fesc$ as a free parameter to match the reionization observables.  
They have succeeded in simulating a reasonable reionization history which is inbetween the `early' and `late' scenarios presented earlier in the THESAN-1 simulation with the highest spatial resolution of 10\,pc. 
The SPHINX simulations \citep{Rosdahl18,Katz21} are also radiative hydrodynamic simulations using the RAMSES-RT AMR code, with a slightly better maximum spatial resolution of $\sim$11\,pc than THESAN, but with a smaller box size of $5-10$\,cMpc. 
In their initial paper, \citet{Rosdahl18} emphasized the importance of including the effects of binary stars, which increases the total flux of ionizing radiation from the metal-poor stellar population and the escape fraction by about three times relative to only single star treatment case. 

\begin{figure}
    \centering
    \includegraphics[width=11cm]{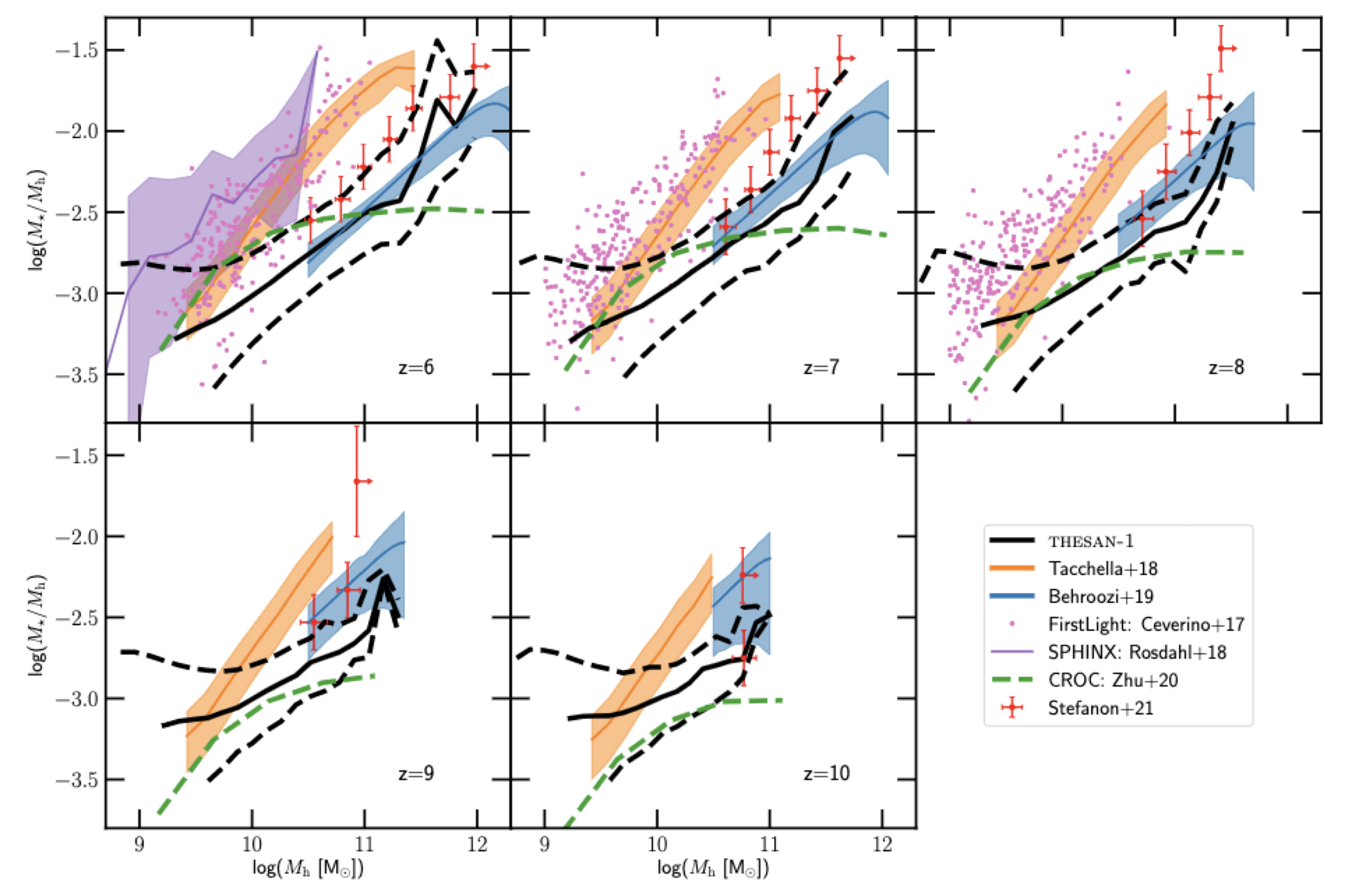}
    \caption{Comparison of SHMR from both simulations and observations \citep{Kannan22}. The blue line with a shade is from the abundance matching method \citep{Behroozi19}, and the red data points with error bars are the observational estimates of \citet{Stefanon21}. }
   \label{fig:SHMR}
\end{figure}

In addition to the CMB optical depth constraints and galaxy luminosity functions, the stellar-to-halo mass ratio (SHMR) data is also improving at $z=0-10$, as shown in Fig.~\ref{fig:SHMR}.  
This statistic is known to have a peak at $M_h \simeq 10^{12}\,M_\odot$, and a decline on both lower mass and higher mass of this mass scale, suggesting the suppression of star formation due to supernova and AGN feedback.  However, this figure shows only the lower mass side at $z\ge 6$.  One can see that there are still some discrepancies even between the abundance matching result of \citet{Behroozi19} and the observational estimate of \citet{Stefanon21}, which utilizes an abundance matching technique between the observed stellar mass function and Bolshoi dark matter simulation.  This means there remains a consistency check to be done among the latest galaxy observations and abundance matching results.
Comparison to other simulation data suggests varying results, and it is clear that we have not reached a consensus on the SHMR at high redshift yet.  It is also worth noting and interesting that \citet{Stefanon21} finds little evolution in SHMR from $z\sim 10$ to $z\sim 6$.  
See also \citet{Shuntov22} for comparisons of SHMR at lower redshifts. 


\begin{figure}
    \centering
    \includegraphics[width=6.8cm]{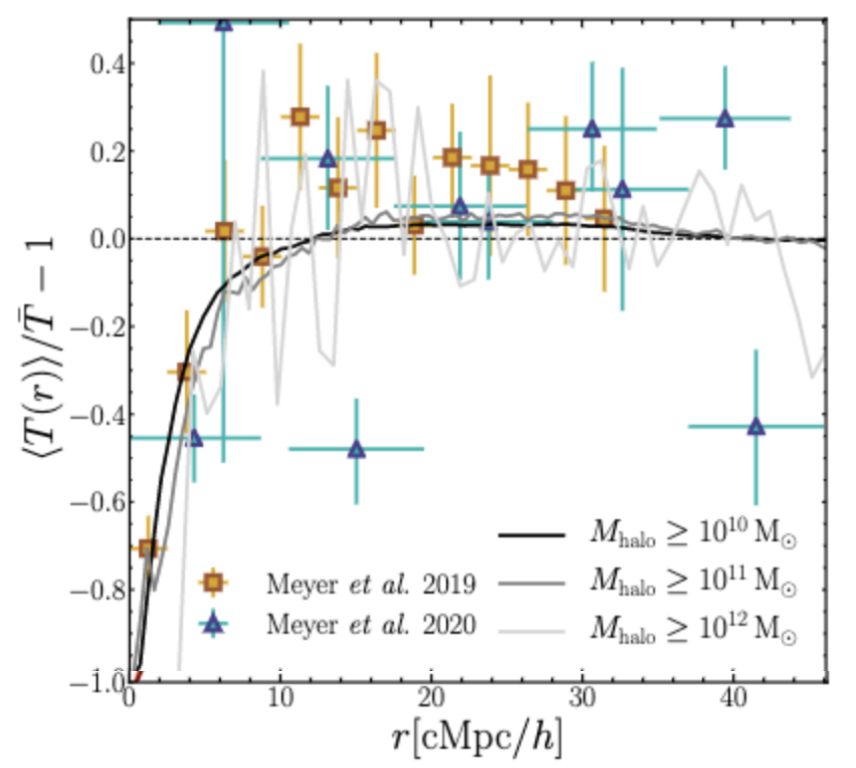}
    \caption{Average Ly$\alpha$ transmission as a function of distance from galaxies in the THESAN-1 simulation from \citet{Garaldi22} at $z=5.5$. }
   \label{fig:transmit}
\end{figure}

Finally, an even more stringent test for simulations would be the distribution of neutral hydrogen around galaxies as probed by the Ly$\alpha$ transmission (or the so-called flux decrement), a technique also called the IGM/CGM tomography \citep{Kakiichi18,Meyer19,Bosman20,Momose21a,Momose21b,Nag21,Garaldi22}. The cross-correlations between galaxies, neutral hydrogen, and metal absorption lines will give us a unique opportunity to constrain the effects of feedback and ionizing radiation field. The 2D correlation maps can also give us further information on the gas dynamics (inflow/outflow) around galaxies \citep{Turner17,Chen20}. 

In summary, the escape fraction of ionizing photons $\fesc$ remains the critical parameter for reionization studies.  Many physical processes are involved in determining $\fesc$, including star formation, radiation from massive stars, supernova feedback, and gas dynamics.  It is a significant challenge for theorists to treat all these processes self-consistently in cosmological hydrodynamic simulations, together with on-the-fly radiation transfer.  In other words, ``{\em radiation--matter coupling}" summarizes this theoretical challenge well.  Concerning the galaxies themselves, the faint-end slope of rest-frame UV luminosity function and the SHMR are the key statistics that are relevant to the reionization studies and capture the effect of feedback.  The contrast between the ``early" vs. ``late" reionization scenarios helps us to highlight and disentangle the differences in physical processes used in different reionization models. 

As a final point of this review for GREX-PLUS, (i) the minimum success of the mission is to discover the bright first galaxies; (ii) the nominal success would be to improve the constraints on the UV luminosity functions,  galaxy stellar mass functions, and SHMR at $z>6$; and (iii) the extra success would be to put a final word on ``early" vs. ``late" reionization scenarios and the average value of $\fesc$.  

\clearpage
\section{First Galaxy and Cosmic Reionization}
\label{sec:veryhighz}

\noindent
\begin{flushright}
Yuichi Harikane$^{1}$, 
Akio K. Inoue$^{2}$
\\
$^{1}$ ICRR, University of Tokyo, 
$^{2}$ Waseda University 
\end{flushright}
\vspace{0.5cm}

\subsection{Scientific background and motivation}

Observations of the highest redshift galaxies provide the strongest observational constraints on the structure formation of the universe.
In the concordance $\Lambda$-cold dark matter (CDM) structure formation model, the number density of bright massive galaxies is smaller in the earlier universe.
If there are more massive galaxies than those the theory predicts, it is either due to a lack of understanding of the baryon physics involved in galaxy formation (gas cooling and heating and star formation) or, perhaps, a flaw in the established theoretical model of the structure formation.
It is thought that these early galaxies also emitted ultraviolet (UV) photons and are responsible for cosmic reionization.
Observations of the universe at redshift $z=6-7$, when cosmic reionization completed, have progressed considerably by current telescopes, but observations of the beginning of cosmic reionization at $z>10$ have only just begun.

\begin{figure}
    \centering
    \includegraphics[width=7cm]{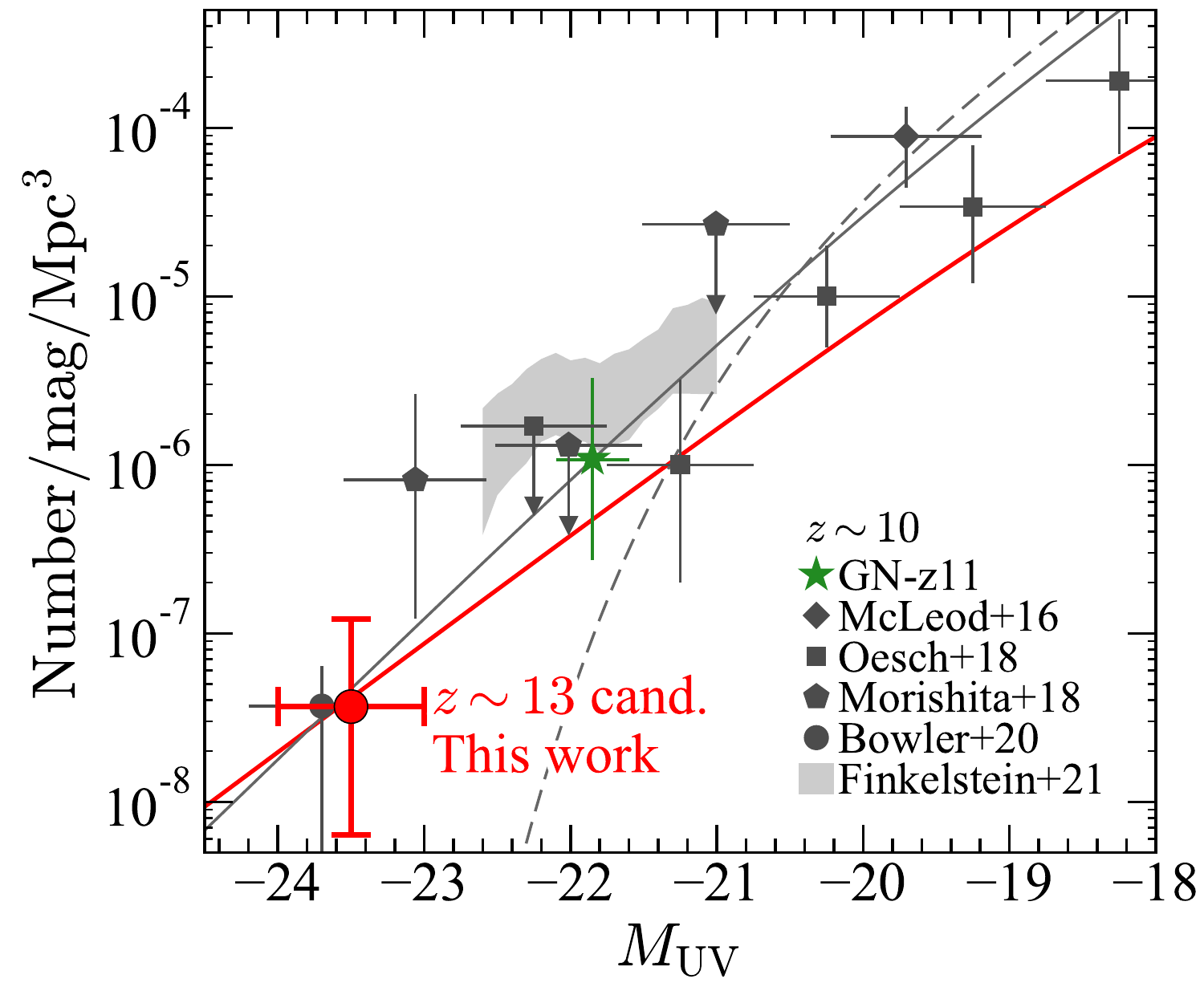}
    \caption{Rest-frame UV luminosity functions at $z\sim10$ and $z\sim13$ \citep{Harikane22}.
    The red circle shows the number density of $z\sim13$ galaxy candidates.
    The black symbols and the gray shaded region are measurements at $z\sim10$ from the literature (diamond: \citealt{2016MNRAS.459.3812M}, square: \citealt{2018ApJ...855..105O}, pentagon: \citealt{2018ApJ...867..150M}, circle: \citealt{2020MNRAS.493.2059B}, shade: \citealt{2022ApJ...928...52F}). 
    The green star is the number density of GN-z11.
    Note that the data point of \citet{2020MNRAS.493.2059B} (GN-z11) is horizontally (vertically) offset by $-0.2$ mag (+0.03 dex) for clarity.
    The gray dashed line is the Schechter function fit \citep{2016ApJ...830...67B}, whereas the gray and red solid lines are the double power-law functions at $z\sim10$ and $13$, respectively, whose parameters are determined by the extrapolation from lower redshifts in \citet{2020MNRAS.493.2059B}.}
    \label{fig:UVLF}
\end{figure}

The current records of the highest redshift galaxies spectroscopically confirmed are galaxies at $z=10.30$--$13.20$ based on the Lyman break detected with the James Webb Space Telescope (JWST) \citep{2022arXiv221204568C}, and GN-z11 at $z=10.60$ measured with detections of the Lyman break and rest-frame ultraviolet (UV) metal lines \citep{2016ApJ...819..129O,2021NatAs...5..256J,2023arXiv230207256B}.
A major surprise of GN-z11 is its remarkably high luminosity, $M_\mathrm{UV}=-21.5$ mag \citep{2023arXiv230207256B}.
Given that it is not gravitationally-lensed, GN-z11 is located in the brightest part of the rest-frame UV luminosity function.
Although the narrow field-of-view (FoV) of Hubble Space Telescope (HST)/Wide Field Camera 3 (WFC3) in the near-infrared has limited the imaging survey areas to $<1\ \mathrm{deg^2}$, several studies using HST report very luminous Lyman break galaxy (LBG) candidates at $z\sim9-10$ more frequently than the expectation from a Schechter-shape luminosity function (e.g., \citealt{2018ApJ...867..150M,2022ApJ...928...52F,2022arXiv220512980B,2022arXiv220515388L}, see also \citealt{2022ApJ...927..236R}).
More statistically robust results have come from a few square-degree near-infrared imaging surveys with Visible and Infrared Survey Telescope for Astronomy (VISTA) and UK Infrared Telescope (UKIRT) such as UltraVISTA \citep{2012A&A...544A.156M}, the UKIRT InfraRed Deep Sky Surveys (UKIDSS, \citealt{2007MNRAS.379.1599L}), and the VISTA Deep Extragalactic Observation (VIDEO) Survey \citep{2013MNRAS.428.1281J}.
These surveys have revealed that the UV luminosity functions at $z\sim9-10$ are more consistent with a double power-law than a standard Schechter function with an exponential cutoff at the bright end \citep[Figure \ref{fig:UVLF}; ][]{2017ApJ...851...43S,2019ApJ...883...99S,2020MNRAS.493.2059B}.
In addition, recently \citet{Harikane22} discovered bright galaxy candidates at $z\sim13$, whose number density is comparable to those of bright $z\sim10$ galaxies (Figure \ref{fig:UVLF}).
Previous studies also report similar number density excesses beyond the Schechter function at $z\sim4-7$ \citep{2018PASJ...70S..10O,2018ApJ...863...63S,2020MNRAS.494.1771A,2022ApJS..259...20H}, implying little evolution of the number density of bright galaxies at $z\sim4-13$ \citep{2020MNRAS.493.2059B,2022ApJS..259...20H,Harikane22}.

In addition to these observations of bright galaxies at $z\sim9-13$, several studies independently suggest the presence of star-forming galaxies in the early universe even at $z\sim15$.
A candidate for a $z\sim12$ galaxy is photometrically identified in very deep {\it HST}/WFC3 images obtained in the Hubble Ultra Deep Field 2012 (UDF12) campaign \citep{2013ApJ...763L...7E}, whose redshift is confirmed with recent JWST observations \citep{2022arXiv221204568C}.
Balmer breaks identified in $z=9-10$ galaxies indicate mature stellar populations whose age is $\sim300-500$ Myr, implying early star formation at $z\sim14-15$ (\citealt{2018Natur.557..392H}, \citealt{2021MNRAS.505.3336L}, see also \citealt{2020MNRAS.497.3440R}).
An analysis of passive galaxy candidates at $z\sim6$ reports that their stellar population is dominated by old stars with ages of $\gtrsim700$ Myr, consistent with star formation activity at $z>14$ \citep{2020ApJ...889..137M}.

\begin{figure}
    \centering
    \includegraphics[width=15cm]{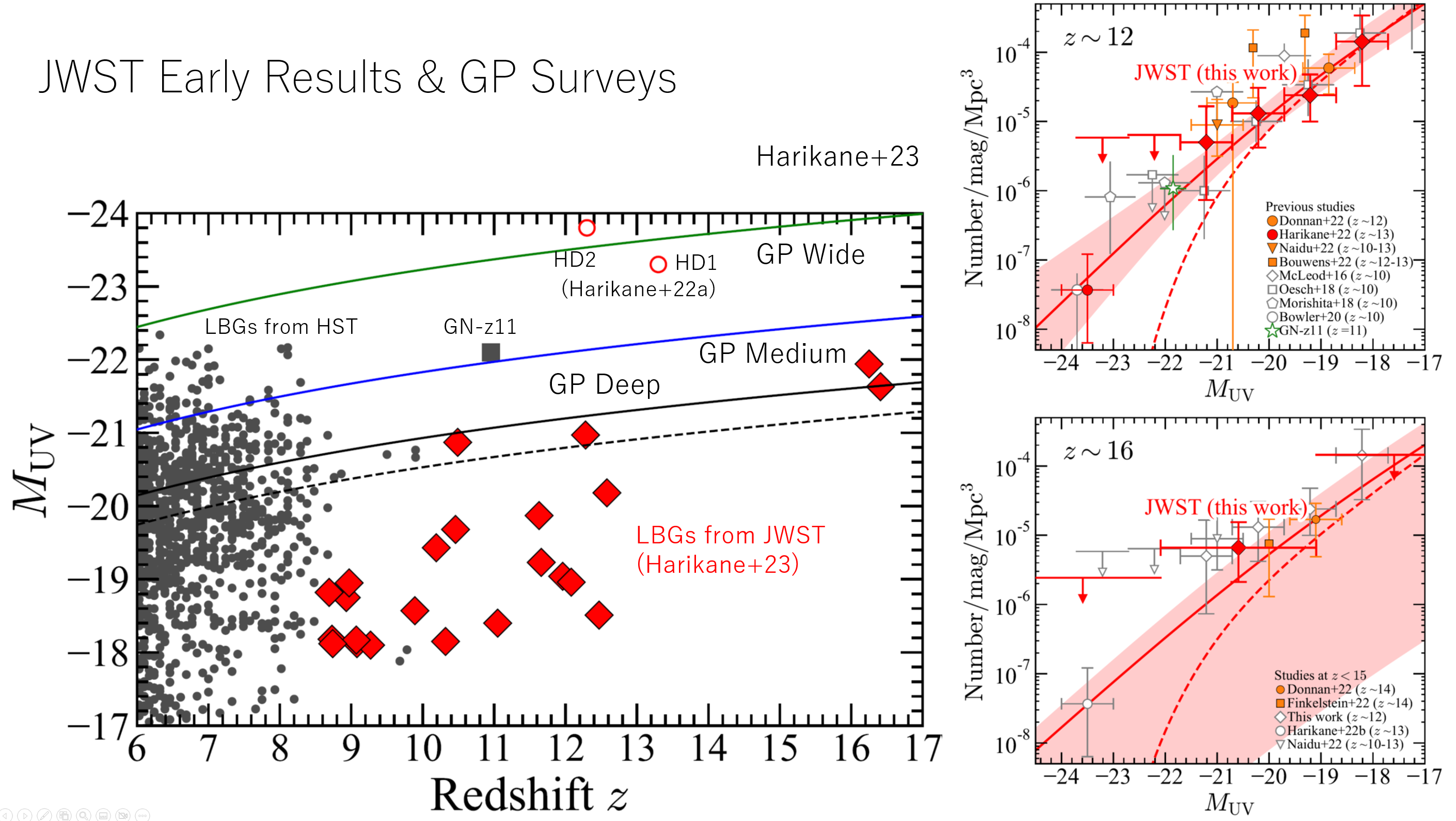}
    \caption{A summary of JWST first year survey results \citep{2023ApJS..265....5H}. The left panel shows the absolute ultraviolet magnitude of galaxies found with HST (gray) and with JWST (red). HD1 and HD2 are galaxy candidates found with ground-based near-infrared images. The black, blue, and green lines are the 5-$\sigma$ limiting magnitudes of the three GREX-PLUS imaging surveys: Deep, Medium, and Wide, respectively. The right two panels show the luminosity function of galaxies at $z\sim12$ and $\sim16$.}
    \label{fig:GPandJWST-1}
\end{figure}

Finally, the first year results of JWST again suggest the high number density of bright galaxies even at $z\sim12$--$16$ \citep[e.g., ][]{2022ApJ...940L..14N,2022ApJ...938L..15C,2022ApJ...940L..55F,2023MNRAS.518.6011D,2023ApJS..265....5H,2022arXiv221206683B} (see Fig.~\ref{fig:GPandJWST-1}).
These recent studies indicate the higher number density of bright galaxies and more active star formation in the early universe than previously thought.
Possible reasons for this high number density of bright galaxies are inefficient negative feedback in the galaxy formation process in the massive halo of the early universe, hidden AGN activity in these bright galaxies, top-heavy initial mass function, and other unknown physical process in massive galaxy formation, or perhaps a flaw in the theoretical model of the structure formation (see discussion in e.g., \citealt{2020MNRAS.493.2059B,Harikane22,2023ApJS..265....5H,2022MNRAS.514L...6P,2022ApJ...938L..10I,2022arXiv220714808M,2022arXiv220800720F,2022arXiv220807879S}).
However, the number of bright galaxy candidates is limited and the distinction between the Schechter function and the double power-law function still lacks statistical significance. 
Also, most of the galaxy candidates remain photometrically selected, and spectroscopic confirmation by ALMA or JWST is essentially needed.
Most importantly, currently-planed wide-area surveys with space telescope such as Euclid and Roman will observe only up to 2 $\mu$m, and are expected to be able to identify bright galaxies up to $z=11-13$.
To search for bright galaxies at $z\gtrsim14$, when first galaxies form, we need a wide-field survey covering 2-5 $\mu$m using the GREX-PLUS near-infrared wide-field camera.

\subsection{Required observations and expected results}

\begin{figure}
    \centering
    \includegraphics[width=15cm]{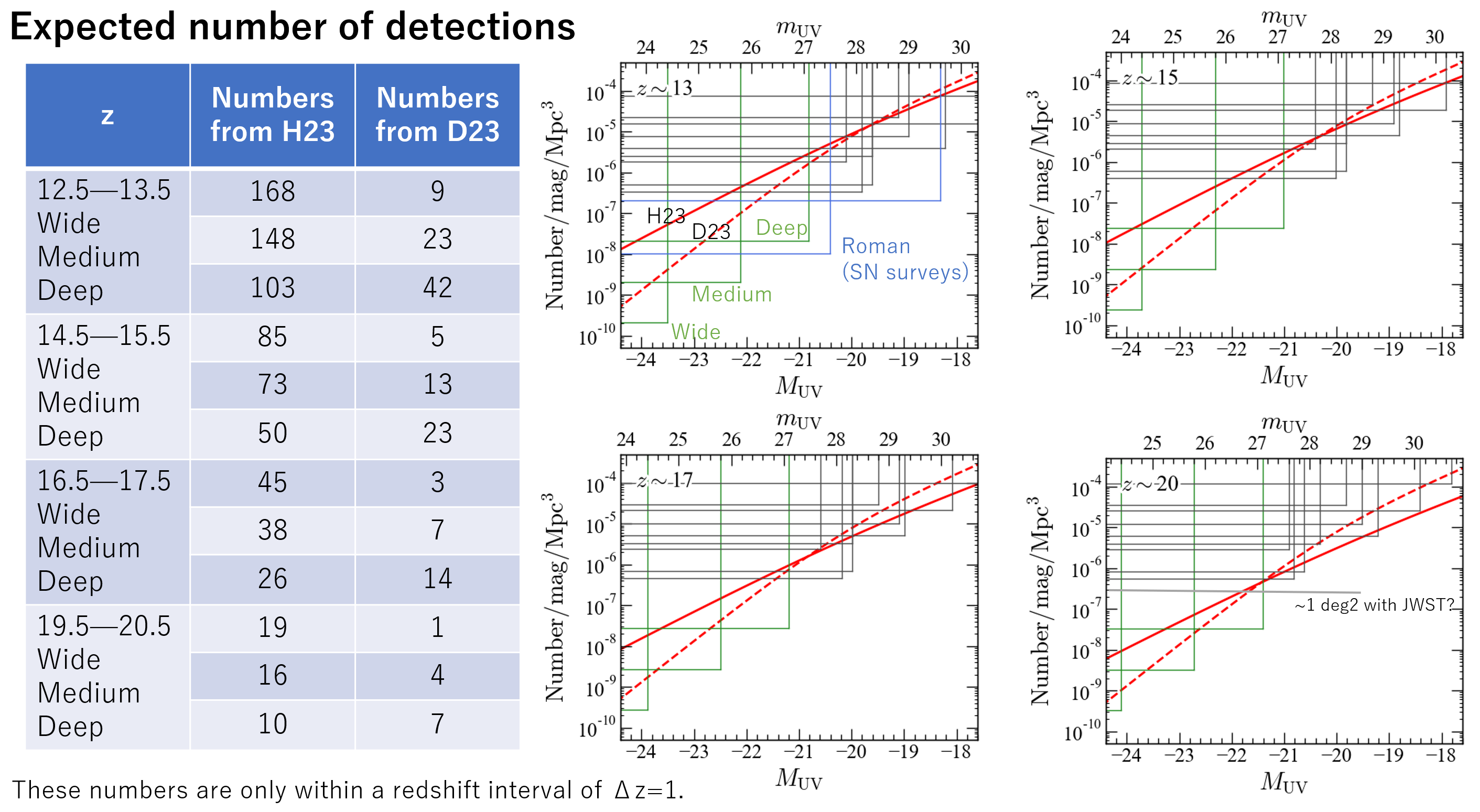}
    \caption{The expected numbers of galaxies detected in the GREX-PLUS imaging surveys (left table). In the right four panels, the red solid and dashed lines show the ultraviolet luminosity functions based on the JWST first year results and their extrapolation to $z\sim20$ (H23: \citealt{2023ApJS..265....5H} and D23: \citealt{2023MNRAS.518.6011D}). The vertical and horizontal lines in the panels show the imaging survey sensitivity and area (or survey volume), respectively. The gray lines are on-going JWST surveys, the blue lines are expected Roman surveys, and the green lines are the three GREX-PLUS surveys.}
    \label{fig:GPandJWST-2}
\end{figure}

We consider three surveys aiming to find galaxies in the early universe at $z>15$ (Fig.~\ref{fig:GPandJWST-2}; Table~\ref{tab:firstgals}) by extrapolating the observed UV luminosity functions of galaxies from $z=10$.
First, due to Ly$\alpha$ scattering by the neutral hydrogen in the intergalactic medium, photons from galaxies at $z>15$ will not be detected below the the observed wavelength of 2 $\mu$m (Ly$\alpha$ break, or Gunn-Peterson Trough; Gunn \& Peterson 1965).
Therefore, a wavelength of 2 $\mu$m or longer is required to observe galaxies at $z>15$.
In addition, in order to distinguish between Ly$\alpha$ break of $z>15$ galaxies and red colors of low redshift galaxies (e.g., passive galaxies and/or dusty galaxies), two or more photometry points are required at the rest-frame wavelengths of $0.15-0.3$ $\mu$m, longer than the wavelength of the Ly$\alpha$ break. 
GREX-PLUS is primarily designed to detect galaxies, not to spatially resolve them.
Therefore, the required spatial resolution is set to about 1 arcsec in order to separate high redshift galaxies from surrounding foreground objects and to avoid the confusion limit.
This resolution corresponds to 3-4 kpc at $z=10$ to 15, which is about the diameter of a bright high redshift galaxy \citep{2015ApJS..219...15S}.
To achieve the 1 arcsec resolution at the observed wavelength of 5 $\mu$m, an aperture of 1.2 m is required, assuming the diffraction limit.
Regarding the depth of the search, in general, we need deeper surveys with the smaller survey area to detect high redshift galaxies, while the larger area is required with the shallower depth.
Here, we consider a galaxy with magnitude of 26-27 ABmag, whose emission lines can be detected and redshifts are spectroscopically determined with a few hour observations using ALMA, JWST, or a 30-m ground-based telescope. 
Based on extrapolation of the luminosity function, a 10 (100) square-degree survey with the $5\sigma$ limiting magnitude of 27 (26) ABmag can detect tens to hundreds of galaxies with $z>15$ in the case of the double power-law function (Fig.~\ref{fig:GPandJWST-2}).
If extremely bright galaxies with $M_{\rm UV}\sim-24$ also exit in the earliest universe, they will be efficiently detected by a $\sim24$ ABmag shallow but extremely wide area survey of 1000 square-degree.
Assuming the sensitivity of a cooled telescope with a 1.2 m primary mirror aperture, a focal plane field of view of 0.25 square degrees or more is required to realize such surveys in a total observation time of less than about one year. 
The deep survey fields by the Subaru Telescope (e.g., Hyper Suprime-Cam/HSC SSP Deep fields) and by Euclid and Roman are good candidates for the survey fields by GREX-PLUS, because the optical-to-near-infrared deep images up to 2 $\mu$m obtained by such telescopes are useful to remove foreground galaxies in the search for high redshift galaxies.
Satellite orbits that can observe these fields with sufficient frequency are necessary.
There is no requirement for observation cadence, time resolution, or pointing arrival time.

\subsection{Scientific goals}

We will conduct 10 square-degree, 27 ABmag-depth (5$\sigma$, point source) and 100 square-degree, 26 ABmag-depth (5$\sigma$, point source) surveys at wavelengths of 2-5 $\mu$m and detect $>10$ galaxies with redshifts higher than those found by the JWST and Roman surveys.
This will provide stronger constraints on the physical processes of first galaxy formation and the onset of cosmic reionization.
We will also test $\Lambda$-CDM structure formation models in the early universe.

\begin{table}
    \begin{center}
    \caption{Required observational parameters.}\label{tab:firstgals}
    \begin{tabular}{|l|p{9cm}|l|}
    \hline
     & Requirement & Remarks \\
    \hline
    Wavelength & 2--5 $\mu$m & \multirow{2}{*}{$a$} \\
    \cline{1-2}
    Spatial resolution & $<1$ arcsec & \\
    \hline
    Wavelength resolution & $\lambda/\Delta \lambda>3$ & $b$ \\
    \hline
    Field of view & 10 degree$^2$, 27 ABmag ($5\sigma$, point-source) & \multirow{2}{*}{$c$}\\
    \cline{1-1}
    Sensitivity & 100 degree$^2$, 26 ABmag ($5\sigma$, point-source) & \\
    \hline
    Observing field & Fields where deep imaging data at $\lambda<2$ $\mu$m are available. & $d$ \\
    \hline
    Observing cadence & N/A & \\
    \hline
    \end{tabular}
    \end{center}
    $^a$ Primary mirror $\phi>1.2$ m is required to achieve $<1$ arcsec at $\lambda=5$ $\mu$m for the diffraction limit.\\
    $^b$ Three or more bands are required for the color selection of very high-$z$ galaxies.\\
    $^c$ A $>0.25$ degree$^2$ field-of-view of a single pointing is required from the point-source sensitivity for a $\phi=1.2$ m telescope and a supposed amount of observing time.\\
    $^d$ For example, deep fields observed with Subaru and Roman.
\end{table}

\clearpage
\section{Large Scale Structure and Galaxy Mass Assembly History}
\label{sec:lssandmassassembly}

\noindent
\begin{flushright}
Tadayuki Kodama$^{1}$, 
Yusei Koyama$^{2}$,
Ken Mawatari$^{2}$
\\
$^{1}$ Tohoku University, 
$^{2}$ NAOJ 
\end{flushright}
\vspace{0.5cm}

\subsection{Scientific background and motivation}

In the contemporary standard universe which is dominated by cold dark matter, smaller scale density fluctuations tend to have larger amplitudes in the early universe. Therefore, small low-mass objects are born first in the early universe and more massive objects are grown later by the gravitational assembly and mergers of smaller objects.
To reveal the mass assembly history of galaxies and clusters of galaxies is the fundamental approach to understand the formation and evolution of these objects, and to test the above standard theory of the universe.
If the unexpectedly large, massive galaxies are discovered in the early universe, they would force us to modify our standard views of the universe and the nature of dark matter.

In order to trace the galactic scale mass assembly history, we must observe all the way from the early building blocks of galaxies with stellar masses of 10$^9$ M$_{\odot}$ (equivalent of $\sim$ 1 per cent
of our present-day Milky Way galaxy) to the most massive galaxies with 10$^{12}$ M$_{\odot}$ (10 times larger than the MW) and sketch how the mass are grown across the cosmic times.
Mass assembly consists of two processes, star formation within individual galaxies and mergers with other galaxies. During mergers, some involve intense new star formation while others do not and just add the existing stellar masses.
These processes and their timescales are also known to depend on the surrounding environments (host halo masses and the number density of nearby galaxies), which must be closely related to the strong environmental dependence of galaxy properties (such as morphology and star formation activity) in the present-day universe.

Therefore, we aim to construct stellar mass functions (SMF) covering the full stellar mass range from the building blocks to most massive galaxies, and compare the SMF as a function of cosmic time (all the way from the early epoch to the present-day) and environment (from dense cluster cores to the general field) in detail, which will enable us to obtain the complete picture of galaxy mass assembly.

We will search for ultra massive galaxies at $z>4$ taking the advantage of our large survey volume. Existence of such massive galaxies in the early universe will put strong constraints on the hierarchical galaxy formation theory in the context of cold dark-matter dominated universe, because there is little time available for such massive galaxies to grow in the early universe. Such surveys have already been done to $z\sim4$ with ground-based NIR observations \citep{2018A&A...618A..85S}.
The higher-$z$ entention to $z\sim5$ of the survey is currently undergoing with medium-band filters at the K-band (K1,K2,K3 and K4) on SWIMS and MOIRCS which can capture Balmer-break features to $z\sim5$ (Ruby-Rush project; Kodama et al.).
We now want to extend the survey further back in time with GREX-PLUS utilizing its longer wavelength coverage and the wide field of view.

For these purposes, it is essential for us to observe distant galaxies at the rest-frame optical wavelength ($>$0.5 nm). So far, we have been largely limited to $z$=4 by ground-based observations up to 2$\mu$m (K-band). Above that wavelength (and  that redshift), ground-based observations would severely suffer from high background noise. Moreover, massive galaxies are rare on the sky and thus large area surveys are required. Space observations with GREX-PLUS is extremely unique and can play absolute key roles here because of its longer wide wavelength coverage (up to 8 $\mu$m), and its wide field of view (0.25 degree$^2$).

Moreover, we aim to conduct the clustering analyses of galaxies (and clusters as well) by utilizing the contiguous, extremely large survey data that GREX-PLUS can provide.
The clustering strength of galaxies that we can measure is a good proxy of underlying dark matter halo mass which host those galaxies. Thus we can derive stellar to halo mass ratio (SHMR) back to $z\sim8$ or so, which is also a key physical quantity that describes the galaxy formation efficiency and thus gives us crucial insights into the formation and evolution of galaxies.
In order to also investigate galaxy cluster/group scale haloes and their surrounding large scale structures, we need to cover at least 100 Mpc in comoving scale. For a statistical analysis, we request a survey over a 10 square degree area.

The most accurate stellar mass measurements can be done at the rest-frame near-infrared (NIR) wavelength (1.6--2 $\mu$m), because the mass-to-light ratio is relatively constant over various stellar populations and its scatter is at least several times smaller than the other bands.
For example, we can obtain precise stellar masses of Lyman break galaxies (LBGs) or typical star forming galaxies at $z\sim3$.
We can also calibrate the previous stellar masses measured based on rest-frame optical light with our more accurate stellar masses to be obtained.  We can thus improve the conversion equation to derive stellar masses for various galaxies with various star formation histories,

\subsection{Required observations and expected results}

\begin{figure}
    \begin{center}
      \includegraphics[width=8cm]{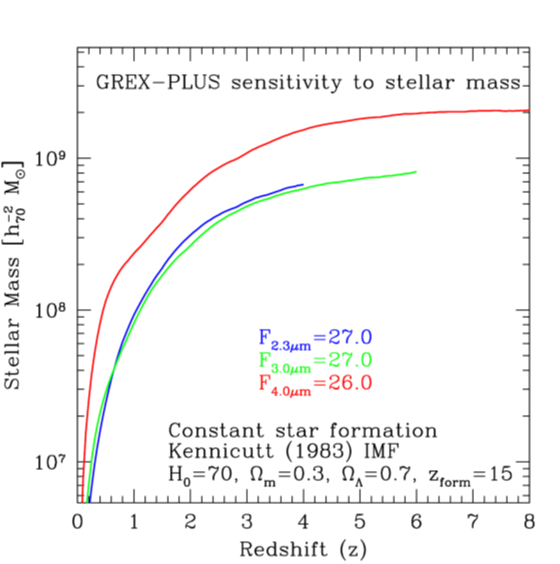}
    \end{center}
    \caption{Limiting stellar mass of galaxies as a function of redshift with GREX-PLUS. If we reach 26--27 AB magnitudes at F232 (2.3$\mu$m), F303 (3.0$\mu$m), and F397 (4.0$\mu$m) as specified in the figure, we will be able to trace the stellar mass assembly history back to $z\sim8$ and down to the building blocks of 10$^9$ M$_{\odot}$ or 1/100 of our Milky Way galaxy.}\label{fig:masslim}
\end{figure}

Figure~\ref{fig:masslim} shows the limiting stellar mass as a function of redshift that can be detected at 5$\sigma$ with GREX-PLUS.
In order to detect building blocks of galaxies with 10$^9$ M$_{\odot}$/yr back to $z\sim8$, we must reach down to 27 AB magnitudes, well matched to our planned observations.
By combining with Subaru/HSC optical imaging data, we can cover the wavelength range of 0.4--5 (and up to 8 for bright galaxies) $\mu$m, and thus can measure stellar masses precisely from the spectral energy distribution (SED) of galaxies.

We can also conduct a statistical number count of close neighbour galaxies utilizing the relatively high spatial resolution of GREX-PLUS in order to investigate the relative contribution of galaxy mergers to the mass assembly history.
The required spatial resolution for this purpose is 8 kpc or 1 arcsec so that we can resolve neighbouring galaxies beyond the cosmic noon epoch ($z>2$).

We aim to construct a large statistical sample of galaxies across the cosmic age and environment with a large area survey. We also need to map the growth history of large scale structures back to $z\sim8$.
The 100 square-degree is a sufficiently large area that corresponds to the comoving volume which can include the progenitors of 100 present-day rich galaxy clusters like Coma. This will enable us to conduct the statistical and quantitative analyses.

We can separate the two major processes of mass assembly, namely star formation and mergers, and thus can track the mass assembly history of galaxies all the way from the building blocks in the early universe to the mature galaxies in the present-day universe. We will also investigate its dependence on galaxy morphology and surrounding environment.
We will also trace the mass assembly history of large scale structures by observing the full range of environments, from rich cluster cores to the surrounding large filamentary structures.

For the precise stellar mass measurements based on rest-frame NIR light, we aim to observe LBGs at $z\sim3$ at 5--8 $\mu$m.
The typical stellar mass of LBGs based on the rest-frame optical light is around 10$^{10-11}$ M$_{\odot}$ (e.g., \citealt{2010MNRAS.401.1521M}), which corresponds to 25--24 AB magnitudes in the rest-frame NIR, if the mass-to-light ratio calibrated at the nearby universe is used \citep{2003ApJS..149..289B}.
A 10 square-degree survey can construct a sample of more than several 1000 LBGs, statistically sufficient for the above analysis.

\subsection{Scientific goals}

\begin{itemize}
\item Detect building blocks of galaxies down to 10$^9$ M$_{\odot}$ back to $z\sim8$, and construct stellar mass functions.
\item Find massive galaxies at $z>4$, and put constraints on the hierarchical galaxy formation theories.
\item Find 100 progenitors within each $\Delta$$z$=1 slice,
and trace the mass assembly history of galaxy clusters and large scale structures.
\item Observe 1000 LBGs at $z\sim3$ and measure precise stellar masses based on rest-frame NIR light, and calibrate the conventional stellar mass measurement method based on the rest-frame optical light.
\end{itemize}

\begin{table}[ht]
    \label{tab:lssandmass}
    \begin{center}
    \caption{Required observational parameters.}
    \begin{tabular}{|l|p{9cm}|l|}
    \hline
     & Requirement & Remarks \\
    \hline
    Wavelength & 2--8 $\mu$m & \multirow{2}{*}{$a$} \\
    \cline{1-2}
    Spatial resolution & $<1$ arcsec & \\
    \hline
    Wavelength resolution & $\lambda/\Delta \lambda>3$ & $b$ \\
    \hline
    Field of view & 2--5 $\mu$m: 10 degree$^2$, 27 ABmag ($5\sigma$, point-source) & \multirow{2}{*}{$c$}\\
    \cline{1-1}
    Sensitivity & 5--8 $\mu$m 10 degree$^2$, 24--25 ABmag ($5\sigma$, point-source) & \\
    \hline
    Observing field & Fields where deep imaging data at $\lambda<2$ $\mu$m are available. & $d$ \\
    \hline
    Observing cadence & N/A & \\
    \hline
    \end{tabular}
    \end{center}
    $^a$ Primary mirror $\phi>1.2$ m is required to achieve $<1$ arcsec at $\lambda=5$ $\mu$m for the diffraction limit.\\
    $^b$ Three or more bands are required for the color selection of very high-$z$ galaxies.\\
    $^c$ A $>0.25$ degree$^2$ field-of-view of a single pointing is required from the point-source sensitivity for a $\phi=1.2$ m telescope and a supposed amount of observing time.\\
    $^d$ For example, deep fields observed with Subaru and Roman.
\end{table}

\clearpage
\section{High Redshift Supernovae}
\label{sec:highzsupernovae}

\noindent
\begin{flushright}
Takashi Moriya$^{1}$
\\
$^{1}$ NAOJ 
\end{flushright}
\vspace{0.5cm}

\subsection{Scientific background and motivation}
Massive stars play essential roles in the evolution of the early universe. A significant fraction of the first stars in the universe are predicted to be massive stars that explode as supernovae \citep[e.g.,][]{2015MNRAS.448..568H}. Supernovae provide energy and elements to the surrounding media and drive the evolution of the early universe. Furthermore, massive stars are considered to be a major source of high-energy photons contributed to the cosmic reionization. In order to know the exact contribution of massive stars to the cosmic reionization, it is important to characterize the properties of massive stars in the early universe. However, it is difficult to observe individual massive stars in the epoch of reionization ($z>6$) even with JWST unless they are lensed \citep{2022ApJ...940L...1W}. Fortunately, some supernovae from massive stars such as superluminous supernovae are known to be extremely luminous. These luminous supernovae can be observed even if they appear at $z>6$. In addition, massive stars in the early universe may evolve and explode differently from those in the local universe. For instance, metallicity is lower in the early universe and thus mass loss from massive stars is also smaller. Smaller mass loss keeps mass of massive stars higher throughout their evolution. A representative example of supernovae that may only exist in the early universe is pair-instability supernovae which are theorized explosions of massive stars that keep their core mass more than 60~M$_\odot$ until the time of their explosion \citep{2002ApJ...567..532H}. Some pair-instability supernovae are predicted to be extremely luminous and they can be observed even if they appear at $z>6$ \citep{2011ApJ...734..102K}. Another example of possible luminous supernovae that only exist in the early universe is the explosions of massive stars above 10,000~M$_\odot$. Such extremely massive stars are mainly suggested to be a source of supermassive black holes, but some of them may explode during explosive helium burning and become luminous supernovae that can be observed at very high redshifts \citep{2021MNRAS.503.1206M}.

In order to discover supernovae at $z>6$, it is required to perform wide-field deep near-infrared transient surveys. Roman will be able to conduct such transient surveys and discover supernovae up to $z\simeq 7$ \citep{2022ApJ...925..211M}. The maximum redshift which Roman can reach is mainly determined by the maximum wavelength that Roman can observe ($\sim 2~\mu\mathrm{m}$). In order to discover supernovae from the epochs when the cosmic reionization is ongoing ($z>7$), it is unavoidable to conduct deep and wide transient surveys at the longer wavelengths. Therefore, we propose to conduct high-redshift supernova surveys by the near-infrared wide-field camera on GREX-PLUS.

\subsection{Required observations and expected results}
\begin{figure}
    \centering
    \includegraphics[width=0.49\columnwidth]{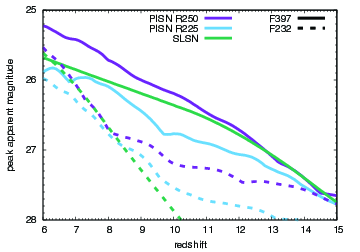}
    \includegraphics[width=0.49\columnwidth]{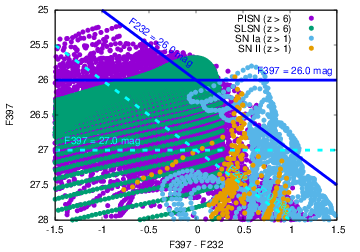}
    \caption{
    \textit{Left:} Expected peak brightness of pair-instability supernovae (PISN) and superluminous supernovae (SLSN) at high redshifts in the F232 (around $2~\mathrm{\mu m}$) and F397 (around $4~\mathrm{\mu m}$) bands. For pair-instability supernovae, we show two brightest models (R250 and R225). By conducting supernova surveys that reach deeper than 26~AB~mag per epoch, we can discover supernovae that are more distant ($z>7$) than those can be found by Roman. \textit{Right:} Supernova color-magnitude diagram with the F397 and F232 bands. We show pair-instability supernovae at $z>6$, superluminous supernovae at $z>6$, Type~Ia supernovae at $z>1$ \citep{2007ApJ...663.1187H}, and Type~II supernovae at $z>1$ (the Nugent template from https://c3.lbl.gov/nugent/nugent\_templates.html). We can find that the high-redshift supernovae can be separated from low-redshift ones through color.
    }
    \label{fig:moriya}
\end{figure}

Here, we investigate the observational strategy to discover pair-instability supernovae and superluminous supernovae at $z>7$. First, we estimate their brightness around $2~\mathrm{\mu m}$ (F232 band) and around $4~\mathrm{\mu m}$ (F397 band) by taking pair-instability supernova models by \citet{2011ApJ...734..102K} and superluminous supernova templates by \citet{2022ApJ...925..211M}. In order to discover supernovae at $z>7$, it is required to reach at least 26~AB~mag (Fig.~\ref{fig:moriya}). The two-band survey allows us to distinguish high-redshift supernovae and low-redshift supernovae even with a single epoch observation (Fig.~\ref{fig:moriya}). Such a distinction would be important to efficiently discover rare high-redshift supernovae. It also implies that we do not need to rely only on light curves to discover them and high-cadence observations are not necessarily required.

Based on the estimated light-curve properties, we conducted supernova survey simulations to estimate the expected numbers of high-redshift supernova discoveries by GREX-PLUS. We assume two-band surveys with the F232 and F397 bands. We assume the survey field of $20~\mathrm{deg^2}$ with the limiting magnitude of 26.0~AB~mag. We assume that the survey period is 5 or 6 years and the same field is observed repeatedly during the survey period. We count the number of supernovae that become brighter than the assumed limiting magnitude at least once in either band.

The number estimates are based on the cosmic star formation history of \citet{2015ApJ...802L..19R}. For pair-instability supernovae, we assume the Salpeter initial mass function to estimate their event rate. The stellar mass range is set to $0.1-500~M_\odot$ and the pair-instability supernova mass range is $150-300~M_\odot$. The superluminous supernova rate is estimated by extrapolating the local event rate ($30~\mathrm{Gpc^{-3}~yr^{-1}}$, \citealt{2013MNRAS.431..912Q}) based on the cosmic star formation history. In the case of pair-instability supernovae, we also adopted the event rate where the initial mass function is assumed to be flat above $100~M_\odot$ to show the possible case of a top heavy initial mass function. We assumed the cadences of 0.5, 1.0 and 2.0~years. The survey period is set to 5~years for the 0.5 and 1.0~year cadence surveys, and it is set to 6~years for the 2.0~year cadence survey. A longer survey period is preferred to increase the discovery numbers.

Table~\ref{tab:highredshiftsupernovanumber} summarizes our simulation results. The numbers in the parentheses are those for the case of the top-heavy initial mass function. In the $20~\mathrm{deg^2}$ survey simulations with the limiting magnitude of 26.0~AB~mag, we expect to discover around 20 pair-instability supernovae and around 4 superluminous supernovae at $z>7$. The discovery numbers will increase if massive stars are preferentially formed at high redshifts.

Supernova surveys require repeated observations of the survey field. As previously discussed, the survey cadence does not need to be short in order to discover and identify high-redshift supernovae. Indeed, our survey simulations show that the expected discovery numbers do not change significantly when we assume the survey cadences of 0.5, 1.0, and 2.0~years. However, it is preferred to have one year cadence so that we can confirm the appearance and disappearance of high-redshift supernova candidates.

For the confirmation of high-redshift supernovae, spectroscopic observations of high-redshift supernova candidates would be ultimately required. The best candidates should be spectroscopically followed by JWST or TMT and confirm their redshifts and supernova nature. Even if we fail to get supernova spectra, it is possible to obtain their distance by observing their host galaxies if they are bright enough. The required observational parameters are summarized in Table~\ref{tab:highredshiftsupernovarequirement}.

\subsection{Scientific goals}
By conducting supernova surveys with two bands in $2-5~\mathrm{\mu m}$, we discover supernovae that are at higher redshifts ($z>7$) than those discovered by Roman. The two-layer transient surveys are proposed: $40~\mathrm{deg^2}$ survey with 26~AB~mag per epoch and $1~\mathrm{deg^2}$ survey with 27~AB~mag per epoch. The supernova surveys should be continued for the duration of the GREX-PLUS survey period (5~years) with the one-year cadence, i.e., the survey field needs to be observed every year for 5~years. The high-redshift supernovae from such surveys will enable us to reveal the supernova rates at $z>7$. The supernova rates would allow us to estimate massive star formation rates at $z>7$ and constrain the initial mass function in the early universe, as well as the contribution of massive stars to the cosmic reionization. In addition, we will systematically investigate the properties of supernovae such as pair-instability supernovae that are expected to exist mainly in the early universe.

\begin{table}
    \begin{center}
    \caption{Expected numbers of high-redshift supernova discoveries by GREX-PLUS. The numbers in the parentheses show the results for the top-heavy initial mass function.}
    \label{tab:highredshiftsupernovanumber}
    \begin{tabular}{|c|c|c|c|c|c|c|c|c|c|c|}
    \hline
     \multicolumn{11}{|c|}{$20~\mathrm{deg^2}$, 26~AB~mag per epoch} \\
    \hline
     & \multicolumn{5}{|c|}{Pair-instability supernovae} & \multicolumn{5}{|c|}{Superluminous supernovae} \\
    \hline
    Cadence & $z>6$ & $z>7$ & $z>8$ & $z>9$ & $z>10$ & $z>6$ & $z>7$ & $z>8$ & $z>9$ & $z>10$ \\
    \hline
    0.5~yr & 75 (541) & 25 (186) & 4.0 (29) & 0 (0) & 0 (0) & 12 & 5.5 & 2.8 & 1.3 & 0.6 \\
    \hline
    1.0~yr & 70 (506) & 22 (164) & 2.4 (18) & 0 (0) & 0 (0) & 9.0 & 4.4 & 2.1 & 1.0 & 0.5 \\
    \hline
    2.0~yr & 81 (591) & 24 (178) & 2.2 (16) & 0 (0) & 0 (0) & 6.5 & 3.2 & 1.5 & 0.7 & 0.3 \\
    \hline
    \end{tabular}
    \end{center}
\end{table}

\begin{table}
    \begin{center}
    \caption{Required observational parameters.}
    \label{tab:highredshiftsupernovarequirement}
    \begin{tabular}{|l|p{9cm}|l|}
    \hline
     & Requirement & Remarks \\
    \hline
    Wavelength & 2--5 $\mu$m &  \\
    \hline
    Spatial resolution & $<1$ arcsec & \\
    \hline
    Wavelength resolution & $\lambda/\Delta \lambda \geq2$ & $a$ \\
    \hline
    Field of view & 20 deg$^2$, $> 26$ AB~mag ($5\sigma$, point-source) per epoch & \multirow{2}{*}{}\\
    \cline{1-1}
    Sensitivity &  & \\
    \hline
    Observing field & Fields where deep imaging data at $\lambda<2$ $\mu$m are available. & $b$ \\
    \hline
    Observing cadence & once per year & $c$ \\
    \hline
    \end{tabular}
    \end{center}
    $^a$ Having two band information at around $2~\mathrm{\mu m}$ and $5~\mathrm{\mu m}$ helps distinguishing high-redshift supernovae from low-redshift ones by color.\\
    $^b$ Host galaxy information is useful to identify high-redshift supernova candidates. \\
    $^c$ The supernova survey should be continued for 5~years.   
\end{table}

\clearpage
\section{Cosmic Infrared Background}
\label{sec:cosmicinfraredbackground}

\noindent
\begin{flushright}
Shuji Matsuura$^{1}$, Akio K.\ Inoue$^{2}$
\\
$^{1}$ Kwansei Gakuin University \\
$^{2}$ Waseda University
\end{flushright}
\vspace{0.5cm}

\subsection{Scientific background and motivation}

\begin{figure}
    \centering
    \includegraphics[width=0.6\columnwidth]{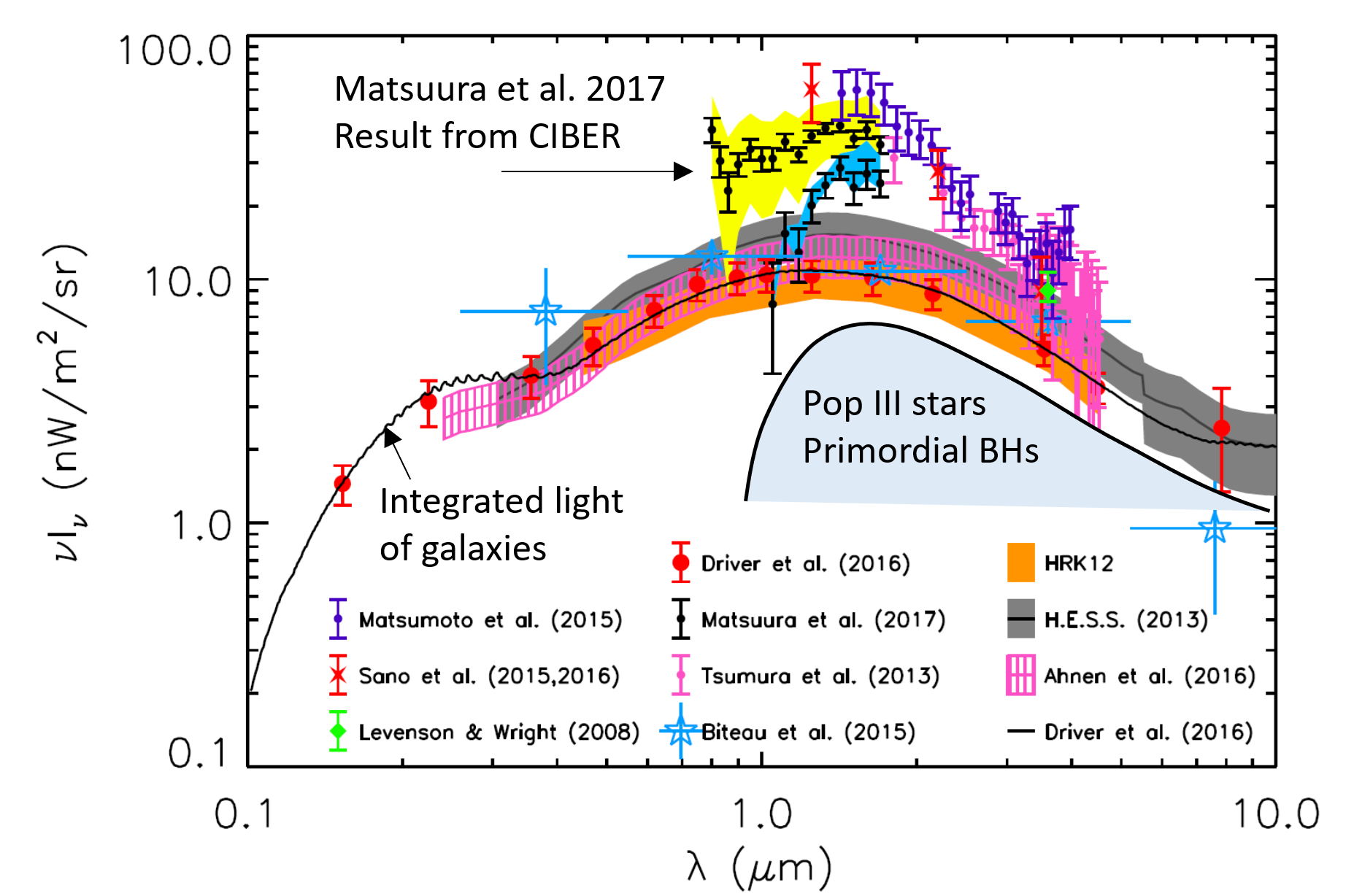}    
    \caption{
    \textit{Left:} A summary of background radiation observations. The solid black line and surrounding data show galaxy-integrated light. The gray shaded band is the upper limit to the background radiation from high-energy gamma-ray observations. Data points are the results of direct measurements of the background radiation and are several times higher than the galaxy-integrated light and the upper limit from the gamma-ray observations (adapted from \citealt{2018RvMP...90b5006K}).}
    \label{fig:EBL1}
\end{figure}

The first objects in the universe (``first stars'' or ``Population III stars'') are expected to have formed around redshift 30.
They are thought to be massive stars with masses ranging from 10 to 1,000 solar mass centered at about 100 solar mass \citep{2014ApJ...781...60H,2015MNRAS.448..568H}.
Some of the first stars will eventually become blackholes. 
There may also be blackholes created by direct gravitational collapse of primordial density fluctuations \citep{1971MNRAS.152...75H}.
We refer to all of these blackholes formed in the early universe as primordial blackholes in this section. 
The first stars and primordial blackholes are too faint to detect individually.
However, their integrated light may be detectable as the cosmic infrared background (CIB) radiation.
It is known that the spectra of objects before the epoch of cosmic reionization (around redshift 6) have a break at shorter wavelengths than their Ly$\alpha$ emission line, which is called the Ly$\alpha$ break.
This characteristic spectral break may be inprinted in the CIB spectrum (Fig.~\ref{fig:EBL1}, Left).
The CIB intensity measured by the rocket experiment, Cosmic Infrared Background Experiment (CIBER), is several times higher than the integrated light of galaxies, and it is very important to understand its origin \citep{2017ApJ...839....7M}.
However, to measure the intensity of the CIB precisely, it is necessary to subtract the zodiacal light, which is more than 10 times brighter than the CIB, from the data.
Fortunately, the CIB and zodiacal light have different spectra, which are useful for subtraction, but this is a major source of systematic errors because the zodiacal light dominates the measured signal.
On the other hand, there is an indirect measurement of the CIB from the observation of ultra-high energy gamma rays.
Because gamma rays are limited in their mean free path by collisions and pair production with CIB photons, the maximum distance that gamma rays can reach the Earth is determined by the intensity of the CIB.
If we can observe gamma rays from distant objects, therefore, we can constrain the intensity of the CIB.
The CIB intensity estimated by this indirect method is consistent with the integrated light of galaxies and not with the direct CIB measurements \citep{2013A&A...550A...4H}.
However, we note that there are uncertainties caused by the assumed spectral shapes of gamma ray and the CIB.
On another hand, the spatial fluctuation of the CIB has been detected by the analysis of the images, where individual galaxies are masked, obtained with Spitzer Space Telescope \citep{2005Natur.438...45K} and the AKARI satellite \citep{2011ApJ...742..124M}.
This indicates the existence of a CIB component that has a different origin from the integrated light of galaxies.
In addition, a spatial correlation of fluctuations in the CIB and X-ray background radiation has been reported \citep{2013ApJ...769...68C}.
This is an interesting hint to consider the gas accretion onto blackholes as the origin of the CIB.
Another interpretation is that the CIB is the accumulation of the stellar light of galactic halos \citep{2012Natur.490..514C}.
In summary, the origin of the CIB is still under debate and to solve this question remains an important scientific theme.

In June 2021, the first flight of CIBER-2, a greatly improved version of the CIBER rocket experiment, was completed, and the data analysis is in progress.
CIBER-2 has six wavelength bands covering from optical to 2 $\mu$m and tries to discriminate the contributions of primordial sources and galactic halo stars to the background fluctuations based on the spectra shape.
However, it is important to observe at wavelengths above 2 $\mu$m to observe the CIB originating from the first stars and primordial blackholes.
GREX-PLUS is equipped with a wide-field camera that covers the wavelength range from 2--8 $\mu$m to perform a wide-field imaging survey. 
Here, we propose to measure the intensity and fluctuation of the CIB using this camera.

\begin{figure}
    \centering
    \includegraphics[width=0.6\columnwidth]{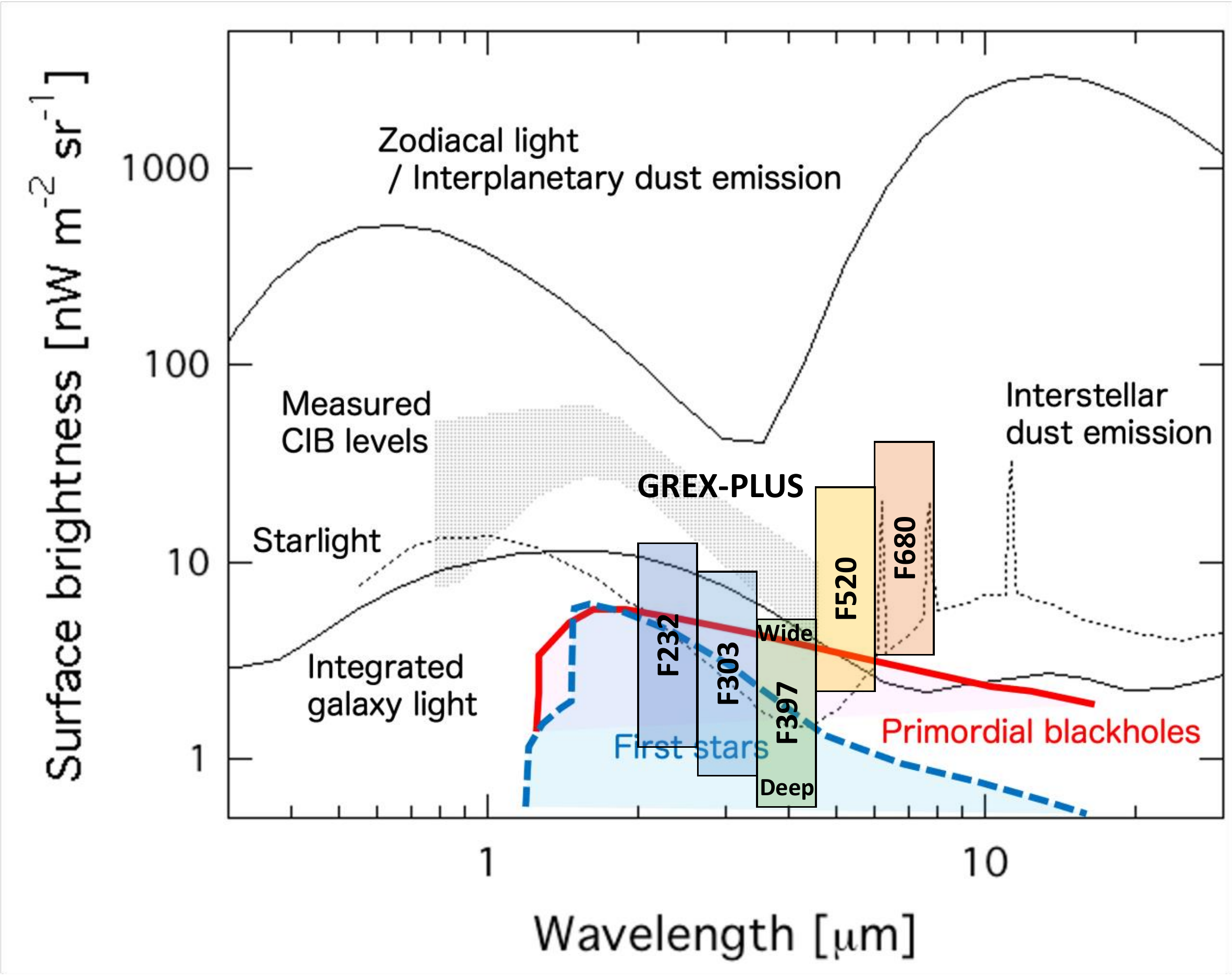}    
    \caption{
    A comparison of contributions from direct observations of cosmic infrared background (CIB), zodiacal light, Galactic interstellar dust emission, integrated galaxy light, and integrated radiation from first stars and primordial blackholes. The expected sensitivity of the five bands of the GREX-PLUS imaging surveys is shown by the vertical bands. The lower and upper ends of the bands are sensitivities of the Deep and Wide surveys, respectively (see Table~\ref{tab:GPsurveys}). This is the sensitivity when aiming to detect correlations on the smallest angular scale. For large angle fluctuations and absolute intensity measurements, the sensitivity can be greatly improved by combining many pixels.
    }
    \label{fig:EBL2}
\end{figure}

\subsection{Required observations and expected results}

We aim detecting spatial fluctuations in the CIB beyond the degree scale, where the effects of zodiacal light, the Galactic radiation, and emission from foreground galaxies are considered to be small.
For this purpose, it is necessary to observe a region of several tens of square degrees or more.
As shown in Figure~\ref{fig:EBL2}, each component of the CIB has a different spectrum, so multicolor observations are needed to separate and extract them.
The most interesting contributions from the first stars and primordial blackholes to the CIB is expected to be about 1--10 nW m$^{-2}$ sr$^{-1}$ at an optimistic estimate, and the sensitivity of this level or better is needed.

We have examined the ability to constrain the CIB intensity by GREX-PLUS Deep, Medium, and Wide imaging surveys that will perform near- and mid-infrared imaging in more than 10 square degree areas.
In Figure~\ref{fig:EBL2}, we show the CIB intensity limits converted from the $5\sigma$ point-source sensitivities for each band, assuming the diffraction limited imaging capability with a 1.2m aperture telescope.
This is the CIB sensitivity on the smallest angular scale.
For larger angular scales, a significant improvement in sensitivity can be expected by stacking many pixels after point source removal.
The sensitivity in F397, which is the highest, is sufficiently high to detect the expected intensity of the accumulation of the first stars and primordial blackholes.
Other bands are also sensitive enough to detect each component of the CIB and to separate and extract them.

The observing fields should include the fields previously observed with Spitzer and AKARI to compare the results.
Since the dominant foreground contribution comes from zodiacal light, it is also useful to evaluate its seasonal variation to better handle the zodiacal light subtraction.
Therefore, it is desirable to include monitoring observations near the North Ecliptic Pole (NEP), which can be easily and frequently observed from the satellite orbit supposed for GREX-PLUS.

\subsection{Scientific goals}

To measure the spatial fluctuation up to a degree-scale of the CIB as an accumulation of primordial objects which are not resolved individually.
To identify the origin of the CIB by extracting the contribution from primordial objects from the CIB measurements based on the auto-correlation spectrum of the measured fluctuations, cross-correlation spectrum among different wavelengths, and the spectral energy distribution.
To constrain theoretical models of primordial objects based on the multi-wavelength cross-correlation analysis including X-ray background radiation.

\begin{table}
    \label{tab:CIB}
    \begin{center}
    \caption{Required observational parameters.}
    \begin{tabular}{|l|p{9cm}|l|}
    \hline
     & Requirement & Remarks \\
    \hline
    Wavelength & 2--8 $\mu$m & \\
    \hline
    Spatial resolution & $<1$--$2$ arcsec & $a$ \\
    \hline
    Wavelength resolution & $\lambda/\Delta \lambda>3$ & \\
    \hline
    Field of view & For more than several 10 square degrees fields, & \multirow{2}{*}{}\\
    \cline{1-1}
    Sensitivity & 1--10 nW m$^{-2}$ sr$^{-1}$ & \\
    \hline
    Observing field & It is desirable to include the fields observed with Spitzer and AKARI, for example, the North Ecliptic Pole (NEP). & \\
    \hline
    Observing cadence & It is desirable to include monitoring observations in NEP to examine seasonal variation of zodiacal light. & \\
    \hline
    \end{tabular}
    \end{center}
    $^a$ Need to avoid confusion limit in each band.\\
\end{table}

\clearpage
\section{High Redshift Quasars}
\label{sec:highzquasars}

\noindent
\begin{flushright}
Yoshiki Matsuoka$^{1}$
\\
$^{1}$ Ehime University
\end{flushright}
\vspace{0.5cm}

\subsection{Scientific background and motivation}

Through extensive observations and theoretical works carried out over the past 30 years, we have come to realize that supermassive black holes (SMBHs), with masses ranging from $M_{\rm BH} \sim 10^6~{\rm M}_\odot$ to $10^{10}~{\rm M}_\odot$, are ubiquitous in the observable universe. Almost all galaxies with evolved bulges have an SMBH at their nuclei, at least in the local universe, and the SMBH mass is found to be roughly 0.1 \% of the bulge mass with a relatively small scatter. Such strong mass correlation may indicate that SMBHs have evolved with the host galaxies under close interactions, via yet unknown processes. Physical drivers behind this ``co-evolution" of galaxies and SMBHs have been a subject of extensive studies, from both theoretical and observational perspectives \citep{2013ARA&A..51..511K}. 
This research field is also attracting attention, not only from the academy but also from the general public, thanks to the recent groundbreaking observations such as the gravitational wave detection from black hole mergers and imaging of SMBH shadows with the Event Horizon Telescope. 

In the mean time, it is still not understood how such SMBHs are formed in the hierarchical structure formation. The strongest constraints come from observations of quasars in the early universe -- the most distant quasar currently known has the redshift $z$ = 7.64 and the mass $M_{\rm BH} \sim 10^9~{\rm M}_\odot$. Even if we assumed the classical maximum efficiency (the Eddington limit) of mass accretion, the progenitor mass would exceed $M_{\rm BH} \sim 10^4~{\rm M}_\odot$, which is difficult to achieve with the standard seeding models via Population III stars \citep[e.g.,][]{2021ApJ...907L...1W}. More exotic formation paths, such as the direct collapse of primordial gas clouds, have been proposed and actively discussed. In order to disentangle various theoretical scenarios, it is imperative to observe those high-$z$ quasars in greater details to look for signatures of the preceding formation epochs. At the same time, we need to push to even higher redshifts at $z \ge 8$ and measure both statistical and individual properties, including the luminosity function and clustering of quasars, the stellar and gaseous contents of the host galaxies, etc. As such, a wide-field near-IR survey with GREX-PLUS will provide crucial pieces of information on when, where, and how the first quasars have emerged, and transitioned to SMBHs known in the low-$z$ universe.

\subsection{Required observations and expected results}

\begin{figure}
 \centering
 \includegraphics[scale=0.3]{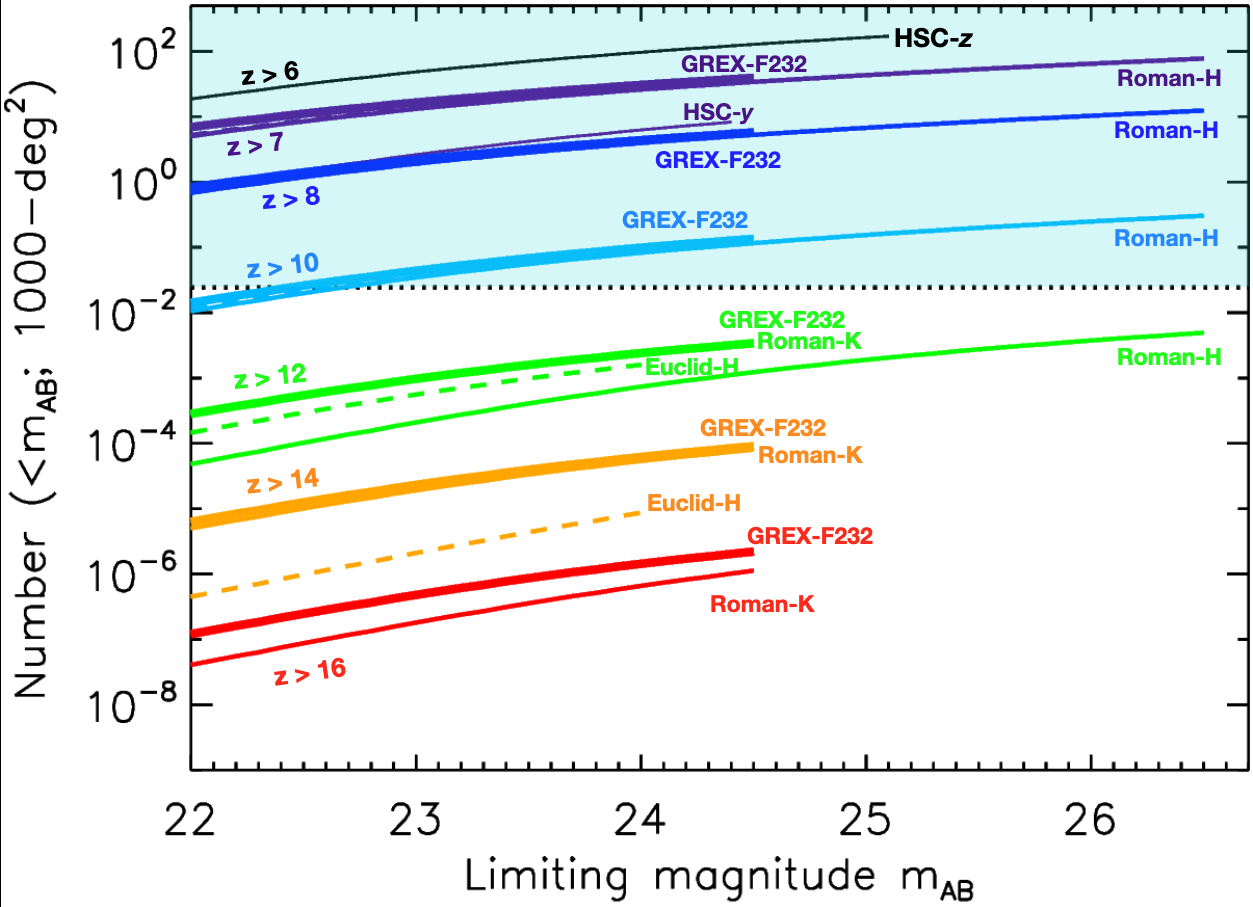}
 \caption{Expected surface density of quasars (per 1000 deg$^2$) for several cases of lower redshift cut, as a function of limiting magnitude. Four missions are considered here, i.e., Hyper Suprime-Cam (HSC) SSP survey (thin lines), {\it Euclid} (dashed lines; overlapping with other lines at $z > 7 - 10$), {\it Roman Space Telescope} (medium lines), and GREX-PLUS (thick lines).
 The light-blue shading represents the surface densities providing $>$1 quasars in the whole sky; in other words, no quasar is expected on average in the lower part of this plot.}
 \label{fig:highzquasars}
\end{figure}

Figure \ref{fig:highzquasars} presents the expected surface density of quasars (per 1000 deg$^2$) for several cases of redshift. Here we assumed the quasar luminosity function measured at $z = 6$ extrapolated to higher redshifts, using the empirical relation of the number density $\phi \propto 10^{kz}$ with $k = -0.7$ \citep{2018ApJ...869..150M}. While this is the best possible estimates we can make at the moment, we caution that there is a very large uncertainty involved in the extrapolation. The present estimates suggest that {\it Euclid} is able to find quasars up to $z \sim 10$ with the planned 15,000 deg$^2$ survey down to $J \sim 24$ AB mag. The {\it Roman Space Telescope} plans to observe 2000 deg$^2$ down to $H \sim 26.5$ AB mag in the High Latitude Survey, which will also allow for detecting quasars up to $z \sim 10$. The additional $K$-band coverage could extend the highest redshift to $z \sim 15$, but the surface density may be too low to expect even a single quasar there. 

GREX-PLUS has an advantage over {\it Euclid} and {\it Roman} at the highest redshifts. However, Figure \ref{fig:highzquasars} indicates that the situation is similar to the {\it Roman} $K$-band;  the expected number of quasars is only $\le 10^{-5}$ at $<24.5$ AB mag over 1000 deg$^2$, thus a chance to find a quasar at the relevant redshifts may be rather small. On the other hand, even non-detection would provide a valuable piece of information on the evolution of quasar number densities. We are carrying out a high-$z$ quasar survey with the imaging data taken by the HSC SSP project, and have so far discovered $\sim$100 quasars at $z \sim 6$ over $\sim$1000 deg$^2$, with the spectroscopic completeness nearly 100 \% down to $\sim$24.0 AB mag. No quasar over a similar area at $z \sim 15$ would lead to a conclusion that the number densities evolve as $k < -0.4$ $(\phi \propto 10^{kz})$; in comparison, the measurements to date suggest $k = -0.5$ at $3 < z < 6$ and $k = -0.7$ at $5 < z < 6$ \citep{2016ApJ...833..222J}. For reference, a very flat slope of the evolution, $k = -0.1$ over $6 < z < 15$, would allow for discovery of a few quasars at $z \sim 15$ by probing 1000 deg$^2$ down to 25 AB mag (though we are not aware of any existing models that predict such slow evolution). Measurements of the declining number density toward high redshifts will provide useful constraints on the theoretical models of early quasar evolution, such as the seeding mechanisms of the first supermassive black holes and their obscured fraction. In the actual survey, it is desirable to target the fields with deep images in the shorter wavelengths from, e.g., Subaru and/or {\it Roman} observations, and use two GREX-PLUS filters to cover the redshifted Ly$\alpha$ line and continuum emission. We can use a classical two-color diagram to pick up candidates via dropout selection, and identify their spectroscopic nature with large ground-based telescopes.

Finally, let us consider how far we can go to hunt for the first quasars, from a pure sensitivity perspective. The current record holders of quasar redshifts are present at $z \sim 7.5$, and have black holes masses $M_{\rm BH} \sim 10^9~{\rm M}_\odot$. Assuming the Eddington-limit accretion, the progenitor masses would be $M_{\rm BH} \sim 10^{5-6}~{\rm M}_\odot$ at $z \sim 15$ and $M_{\rm BH} \sim 10^{4-5}~{\rm M}_\odot$ at $z \sim 20$ -- within the mass range predicted by seeding scenarios via direct collapse black holes. The corresponding brightness in the near-IR are 29 -- 31 AB mag at $z \sim 15$, assuming no dust extinction. If their positions are fixed by preceding observations, such sources could be an interesting targets for very deep exposures with GREX-PLUS.

\subsection{Scientific goals}
With a 1000-deg$^2$ class survey at 2 -- 4 $\mu$m, we aim to obtain an upper limit on the rate of number density decline from $z \sim 6$ to 15. If the decline was much milder than currently measured at $z \le 6$, then we might be able to find a few quasars at $z \sim 15$. Comparison of the survey results with theoretical models will provide useful constraints on the properties of the earliest quasars, such as the formation of the first supermassive black holes and nuclear obscuration.

\begin{table}
    \label{tab:highzq}
    \begin{center}
    \caption{Required observational parameters.}
    \begin{tabular}{|l|p{9cm}|l|}
    \hline
     & Requirement & Remarks \\
    \hline
    Wavelength & 2--4 $\mu$m & $a$ \\
    \hline
    Spatial resolution & $<1$ arcsec & \\
    \hline
    Wavelength resolution & $\lambda/\Delta \lambda>3$ & $b$ \\
    \hline
    Field of view & $>$1000 deg$^2$& \\
    \hline
    Sensitivity & 25 ABmag ($5\sigma$, point-source) & \\
    \hline
    Observing field & Fields where deep imaging data at $\lambda<2$ $\mu$m are available. & $c$ \\
    \hline
    Observing cadence & N/A & \\
    \hline
    \end{tabular}
    \end{center}
    $^a$ We need to capture redshifted Ly $\alpha$ (2.3 $\mu$m at $z = 15$) and continuum emission at the longer wavelength.\\
    $^b$ At least three bands are required for the color selection (can be reduced to two bands if the bluest band comes from preceding [e.g., {\it Roman}] observations).\\
    $^c$ For example, deep fields observed with Subaru and/or {\it Roman}.
\end{table}

\clearpage
\section{Dusty Star-Forming Galaxies}
\label{sec:submmgalaxies}

\noindent
\begin{flushright}
Ken-ichi Tadaki$^{1}$
\\
$^{1}$ Hokkai-Gakuen University
\end{flushright}
\vspace{0.5cm}

\subsection{Scientific background and motivation}

\begin{figure}[tbh]
\centering
\includegraphics[width=.55\linewidth]{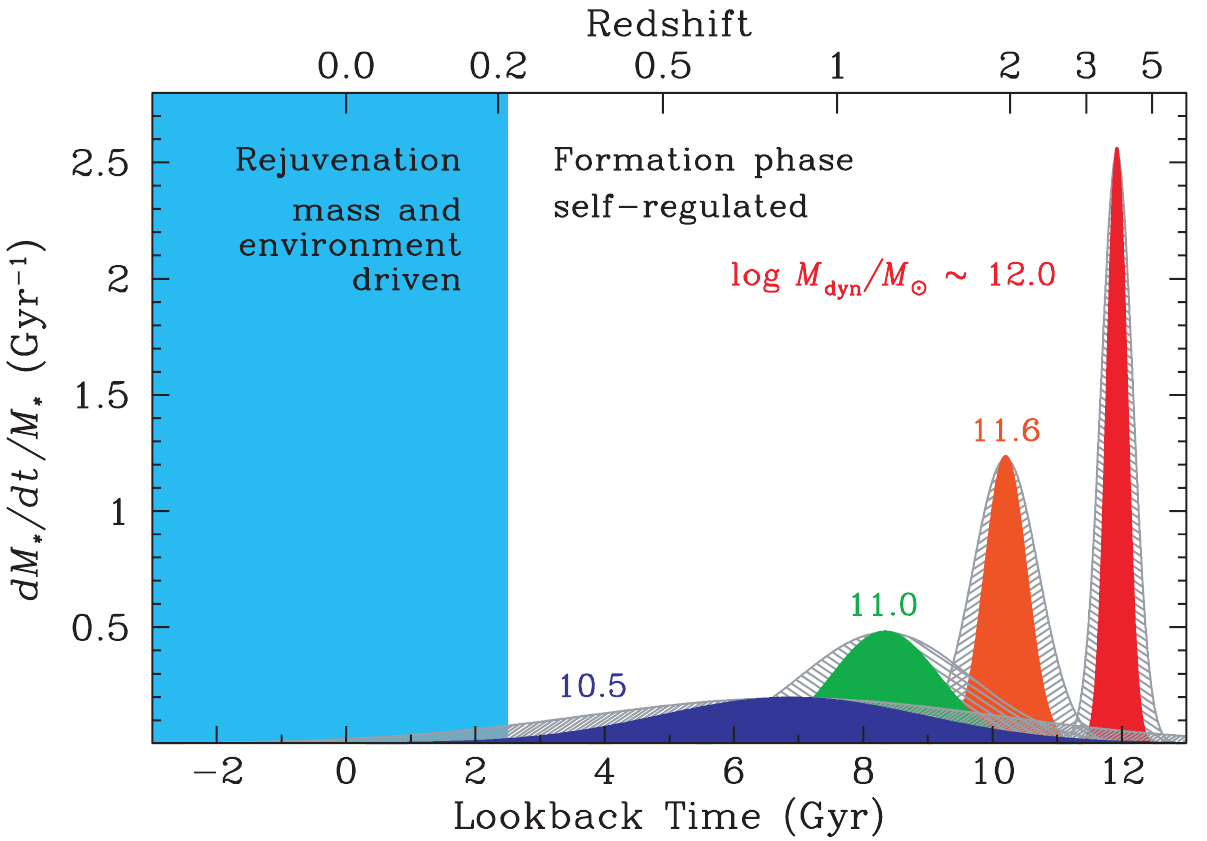}\ \ 
\caption{Star formation histories inferred from optical spectroscopic observations of nearby early-type galaxies \citep{2010MNRAS.404.1775T}. The results suggest that massive galaxies with a dynamical mass of $\log(M_\mathrm{dyn}/{\rm M}_\odot)\sim12$ experienced a short burst of star formation at $z\sim4$. 
The figure is taken from Figure 9 of \cite{2010MNRAS.404.1775T}; 'Environment and self-regulation in galaxy formation'.
\label{fig:Thomas+2010}
}
\end{figure}

The star formation history of nearby early-type galaxies indicates that more massive galaxies formed earlier in the universe and on shorter timescales (Figure \ref{fig:Thomas+2010}). 
Understanding the formation of the most massive galaxies also leads to understanding the evolution of the earliest galaxies such as bright Lyman break galaxies with $z=11-13$ \citep{Harikane22}.
Recent near-infrared spectroscopic observations of massive quiescent galaxies at $z=3-4$ support that they were formed by intense star formation activity at $z>4$ (SFR$\sim$300 ${\rm M}_\odot$ yr$^{-1}$, timescale of several hundred million years).
Although the current record redshift of spectroscopically confirmed massive quiescent galaxies is $z=4.01$ \citep{2019ApJ...885L..34T}, JWST will allow us to break this record within the next one or two years.
Dusty star-forming galaxies (DSFGs) at $z>6$ are strong candidates for the progenitors of these quiescent galaxies at the highest redshift as they can be associated with intense star formation.

While ALMA observations have identified many DSFGs at $z=4-6$, there are only three at $z>6$: SPT0311-58 at $z=6.9$ \citep{2018Natur.553...51M}, HFLS3 at $z=6.3$ \citep{2013Natur.496..329R}, G09-83808 at $z=6.0$ \citep{2018NatAs...2...56Z}.
Two of them, SPT0311-58 and HFLS3, are exceptionally bright sources that exceed 15 mJy (SFR$>$1000 ${\rm M}_\odot$ yr$^{-1}$) at 870 $\mu$m, but G09-83808 is intrinsically about 4 mJy at 870 $\mu$m (SFR$\sim$400 M$_\odot$ yr$^{-1}$) due to strong gravitational lensing effect ($\mu$=8--9), making it a more common star-forming galaxy in this era. 
The theoretical model predicts that the surface number density of 4 mJy sources at $z=6-7$ is as low as $\sim0.3$ deg$^{-2}$, and is expected to increase to $\sim$2 deg$^{-2}$ if 2 mJy sources are included \citep{2020ApJ...891..135P}.
However, these number densities have been poorly constrained by observations. 
Finding DSFGs at $z>7$ as well as making a statistical sample of DSFGs at $z=4-7$ are important for understanding when and how the most massive galaxies are formed in the early Universe.
As for a definition of DSFGs, we will mainly focus on galaxies that exceed 2~mJy (SFR$\sim$200 ${\rm M}_\odot$ yr$^{-1}$) at 870~$\mu$m.


\subsection{Required observations and expected results}

DSFG are commonly massive with a stellar mass of $\log(M_\star/{\rm M}_\odot)>10.5$ \citep{2022ApJ...933..242Z}, resulting in bright continuum emission at long wavelengths ($\sim$25 AB mag at 4 $\mu$m).
However, they are faint at short wavelengths ($<2~\mu$m) due to strong dust extinction and sometimes are completely dark even in deep HST 1.6 $\mu$m images \citep{2019Natur.572..211W}.
Therefore, DSFGs can be characterized by a red color at 2--4 $\mu$m.
JWST observations have revealed that a red source at 2--4 $\mu$m is detected in the NOEMA 1.1 mm and JCMT 850 $\mu$m maps ($S_{870}\sim2$ mJy), and is a DSFG at $z=5-6$ \citep{2022arXiv220801816Z}.
This demonstrates that wide-field imaging at 2--4 $\mu$m is a powerful approach for making a large sample of DSFGs at $4-7$ and for discovering DSFGs at $z>7$.
Roman is not ideal for this purpose because it can not cover wavelengths of $>2~\mu$m where DSFGs become bright.
Coordinated surveys between Roman ($\sim$27 AB mag at 2 $\mu$m) and GREX-PLUS ($\sim$25 AB mag at 4 $\mu$m) are effective for identifying DSFGs at $z>4$.

A red color at $2-4~\mu$m can be also explained by Balmer break with old stellar populations as well as strong dust extinction. 
Photometry with the GREX-PLUS/F520 filter is useful for breaking the degeneracy between age and dust attenuation because DSFGs at $z>6$ are even brighter at 5 $\mu$m, corresponding to the rest-frame $i$-band.
Moreover, we will be able to accurately estimate the stellar mass of DSFGs at $z>6$ by including $\sim5~\mu$m continuum data as it is less affected by dust extinction.
\cite{2020ApJ...889..137M} find that a half of red sources at $2-4~\mu$m are detected in ALMA 870 $\mu$m continuum maps, supporting that they are DSFGs.
If deep 5 $\mu$m data is not available, making ALMA follow-up observations of 870 $\mu$m continuum emission in red sources is an alternative option to distinguish between DSFGs and quiescent galaxies.
Once red sources at $2-4~\mu$m are found to be DSFGs at $z>4$, we will make ALMA follow-up observations of [C~{\sc ii}] emission line to obtain the spectroscopic redshift.
As [C~{\sc ii}] emission line is bright ($\sim0.75$ Jy km s$^{-1}$ for 2 mJy sources; \citealt{2018NatAs...2...56Z}) in DSFGs even at $z>6$, only 4 minutes integration with ALMA is required for 5$\sigma$ detection.
In the near future, it is planned to at least double the bandwidth of ALMA, allowing for more efficient spectral scans.
Nevertheless, we currently have only three DSFGs at $z>6$ due to a lack of photometrically selected samples and can not expand this research further.
Wide field surveys with GREX-PLUS will provide us with a large sample of DSFGs.


\subsection{Scientific goals}

The minimum goal is to identify $>100$ DSFGs at $z>6$ to measure the number densities as a function of redshift.
This requires a wide survey over $\sim$50 deg$^2$ with a depth of 27 ABmag at $2~\mu$m and 25 ABmag at $4~\mu$m, provided that the theoretical model shown in Figure \ref{fig:Thomas+2010} is assumed.
We also aim to estimate a stellar mass and a halo mass of DSFGs through SED and clustering analysis, respectively.
Comparing the number density, stellar mass, and halo mass with those of massive quiescent galaxies at similar redshifts, we will be able to understand an evolutionary connection from DSFGs to quiescent galaxies.
The next goal is to discover DSFGs at $z>7$, answering the question of when the most massive galaxies first formed.

\begin{table}[tbh]
    \label{tab:smgrequirement}
    \begin{center}
    \caption{Required observational parameters.}
    \begin{tabular}{|l|p{9cm}|l|}
    \hline
     & Requirement & Remarks \\
    \hline
    Wavelength & 2--5 $\mu$m &  \\
    \hline
    Spatial resolution & $<2$ arcsec & \\
    \hline
    Wavelength resolution & N/A &  \\
    \hline
    Field of view & 50 degree$^2$, 27 ABmag at $2~\mu$m ($5\sigma$, point-source) & \multirow{3}{*}{$a$}\\

    Sensitivity & 50 degree$^2$, 25 ABmag at $4~\mu$m ($5\sigma$, point-source) & \\

    \hline
    Observing field & Fields where deep imaging data at $\lambda<2$ $\mu$m are available. &  \\
    \hline
    Observing cadence & N/A & \\
    \hline
    \end{tabular}
    \end{center}
    $^a$ When deep $2~\mu$m data is available from Roman survey, we require $4.4~\mu$m and $8~\mu$m data for GREX-PLUS survey.\\
\end{table}

\clearpage
\section{Dust Obscured Active Galactic Nuclei}
\label{sec:dustyagns}

\noindent
\begin{flushright}
Yoshiki Toba$^{1}$,
Takuji Yamashita$^{1}$ 
\\
$^{1}$ NAOJ 
\end{flushright}
\vspace{0.5cm}

\subsection{Scientific background and motivation}
\begin{figure}
\centering
\includegraphics[width=0.9\textwidth]{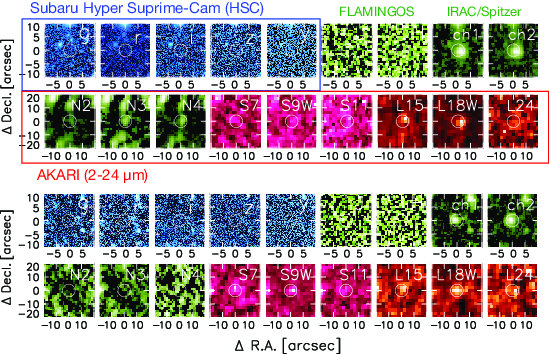}
\caption{Examples of multi-wavelength images ($g$, $r$, $i$, $z$, $y$, $J$, $H$, ch1, ch2, N2, N3, N4, S7, S9W, S11, L15, L18W, and L24, from top left to bottom right) for optically dark IR galaxies reported in \cite{2020ApJ...899...35T}. R.A. and decl. are relative coordinates with respect to objects in the AKARI NEP-wide catalog \citep{2012A&A...548A..29K}. White circles in the images also correspond to the coordinates of the AKARI NEP-wide catalog.}
\label{image}
\end{figure}

Dust-obscured active galactic nuclei (AGN) are thought to correspond to the maximum growth phase of both galaxies and supermassive black holes (SMBHs) in the context of their co-evolution.
This makes them one of the ideal laboratories to investigate the physical mechanism behind the co-evolution and the AGN feedback \citep[e.g.,][]{2010MNRAS.407.1701N,2015PASJ...67...86T,2018MNRAS.478.3056B,2022ApJ...936..118Y}.
Infrared (IR) satellites such as AKARI, Spitzer, and WISE have improved our understanding of dust-obscured AGN at a wide range of redshifts \citep[e.g.,][]{2018ARA&A..56..625H}. 
In particular, optically ``dark'' IR galaxies (that are faint or invisible in the optical and/or near-IR but bright in the IR and/or sub-millimeter) have recently been reported in several deep fields \citep[e.g.,][]{2019Natur.572..211W,2020A&A...640L...8U,2020ApJ...899...35T,2021Natur.597..489F}, and the advent of the JWST \citep{2006SSRv..123..485G} is boosting the number of discoveries for those objects \citep[e.g.,][]{2022arXiv220714733B,2022arXiv221100045P}.
Those optically dark IR galaxies are located at $1 < z < 7$ that are undetectable even by deep optical imaging surveys such as HST and Subaru HSC, meaning that those have been missed by precious optical surveys (Figure \ref{image}).
How many optically dark IR galaxies are there in the universe?
How much do they contribute to the cosmic star formation and supermassive black hole mass accretion history?
Estimating the statistical properties (e.g., luminosity function and auto-correlation function) and physical properties (e.g., black hole mass accretion rate and star formation rate) of optically dark IR galaxies will provide an avenue to answer these questions.

\subsection{Required observations and expected results}
\begin{figure}
\centering
\includegraphics[width=0.8\textwidth]{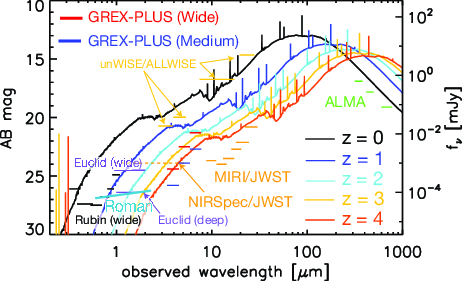}
\caption{Typical SEDs of optically dark IR galaxies found in the AKARI NEP \citep{2020ApJ...899...35T} as a function of redshift up to $z = 4$. The detection limits of ongoing and forthcoming missions are over-plotted.}
\label{SED}
\end{figure}

By the time GREX-PLUS becomes available, it is expected that the combination of Euclid \citep[e.g.,][]{2022A&A...662A.112E} and Roman \citep[e.g.,][]{2019arXiv190205569A} with HSC and Rubin will provide an insight into the properties of optically dark IR galaxies, especially in relatively nearby ($z < 2$) sources. 
Hence, we propose a systematic search for optically dark IR galaxies at $z > 3$ by taking advantage of GREX-PLUS, a wide-area survey (100--1000 deg$^2$) with unique mid-IR bands (3.5--8 $\mu$m), which cannot be realized by Euclid and Roman.
According to a typical spectral energy distribution (SED) of optically dark IR galaxies found in the AKARI North Ecliptic Pole (NEP) survey \citep{2020ApJ...899...35T}, We can expect that a few tens of objects at $z > 3$ are expected to be discovered especially by the GREX-PLUS medium survey with 100 deg$^2$ (Figure \ref{SED}).
Those high-$z$ objects will not be able to be discovered even by Euclid and Roman due to their extremely large extinction.
We estimated the expected number of optically dark IR galaxies including such a ``Roman-dropout'' galaxies discovered with GREX-PLUS based on their surface density and photometric redshifts reported in \cite{2020ApJ...899...35T}, which is summarized in Table \ref{tab:num}.
In particular, given the fact that the number of optically dark galaxies with $z > 4$ has been limited to a few dozen so far, it is possible that only GREX-PLUS can construct a statistically robust sample.

\begin{table}[htb]
    \begin{center}
    \caption{Expected number of optically dark IR galaxies discovered by GREX-PLUS.}
    \label{tab:num}
    \begin{tabular}{ccrr}
    \hline\hline
     \multirow{2}{*}{Redshift} & \multirow{2}{*}{Expected surface density [deg$^{-2}$]} & \multicolumn{2}{c}{Expected number} \\
     \cline{3-4}
        &   &   Medium [100 deg$^2$]    &   Wide [1,000 deg$^2$] \\
        \hline
        $1 < z < 2$ &   5.6 &   560    &   5600    \\
        $2 < z < 3$ &   1.5 &    150    &    1500    \\
        $3 < z < 4$ &   2.1 &    210    &   ---      \\
        $z > 4$     &   0.2 &     20    &   ---        \\     
    \hline
    \end{tabular}
    \end{center} 
\end{table}
What are the properties of these ``Roman-dropout'' IR galaxies? 
We performed a physical parameter search from a SED point of view to satiety criteria of Roman dropouts by using CIGALE \citep[Code Investigating GALaxy Emission;][]{2019A&A...622A.103B} that is also a useful SED template generation tool.
We found that most of the object stratifying criteria of Roman dropouts are $3 < z < 5$, $A_{\rm V}$ (ISM) $> 10-100$, and $L_{\rm IR}$(AGN)/$L_{\rm IR} > 0.8$, which may be very dusty AGN. 
This is observably verifiable through follow-up observations with JWST (if it is still operating in the 2030s) and ALMA.
We will address the aforementioned question by calculating various physical quantities such as spectroscopic redshifts, star formation rate, and black hole mass.

Radio detection of the optically dark IR galaxies will be examined to measure radio loudness and to search for a harbored radio-loud AGN in the center of the optically dark IR galaxies. The high radio-loudness itself is strong evidence for AGNs. A high-$z$ radio-loud optically-dark IR source is recently discovered \citep{2022MNRAS.512.4248E} and implies an abundance beyond expectation from the optically bright quasar population. The radio galaxies at $z>1$ show a dusty nature compared to the local radio galaxies \citep{2019ApJS..243...15T, 2020AJ....160...60Y}. We will explore for radio loudness of the optically dark IR galaxies identified by GREX-PLUS with the next-generation radio surveys by SKA and ngVLA.

\subsection{Scientific goals}
We propose a systematic search for optically dark IR galaxies, including duty AGN (that are not detectable by Euclid, Roman, and Rubin) with a GREX-PLUS wide and medium survey (Table \ref{tab:dustyagn}).
The survey area and 5$\sigma$ sensitivity in 3.5--8 $\mu$m for point sources in the wide and medium surveys are 1000 deg$^2$ and 21--24 AB mag and 100 deg$^2$ and 23--26 AB mag, respectively.
It would be ideal if our targets could be separated based on the proper motion, since galactic sources such as C-rich stars may be contaminated.
We will address the growth history of SMBHs and their host galaxies as a function of redshift by investigating the statistical properties (e.g., luminosity function and correlation function) and physical properties (SMBH mass accretion rate and star formation rate) of optically dark IR galaxies found in this project. 
We also aim to determine the contribution of dust-obscured AGN to cosmic star formation rate density, mass accretion rate density, and structure formation of the universe.
\begin{table}
    \begin{center}
    \caption{Required observational parameters.}
    \label{tab:dustyagn}
    \begin{tabular}{|l|p{9cm}|l|}
    \hline
     & Requirement & Remarks \\
    \hline
    Wavelength & 3.5--8 $\mu$m & \multirow{2}{*}{$a$} \\
    \cline{1-2}
    Spatial resolution & $<2$ arcsec & \\
    \hline
    Wavelength resolution & $\lambda/\Delta \lambda>3$ & $b$ \\
    \hline
    Field of view & 1000 deg$^2$, 20--22 AB mag ($5\sigma$, point-source) & \multirow{2}{*}{$c$}\\
    \cline{1-1}
    Sensitivity & 100 deg$^2$, 23--24 AB mag ($5\sigma$, point-source) & \\
    \hline
    Observing field & Fields excluding the galactic plane & $d$ \\
    \hline
    Observing cadence & $>$ 2 & $e$ \\
    \hline
    \end{tabular}
    \end{center}
    $^a$ F397, F520, and F680. Diffraction limit at primary mirror $\phi>$ 1.2 m and $\lambda$=10 $\mu$m.\\
    $^b$ Three or more bands are required for the SED analysis.\\
    $^c$ A $>0.25$ degree$^2$ field-of-view of a single pointing is required from the point-source sensitivity for a $\phi=1.2$ m telescope and a supposed amount of observing time.\\
    $^d$ In particular, the overlap area with Rubin, Euclid, and Roman is preferable.\\
    $^e$ It would be nice to be able to measure the proper motion to distinguish from C-rich stars. 
\end{table}

\clearpage
\section{AGN Molecular Outflow}
\label{sec:AGNmolecularoutflow}

\noindent
\begin{flushright}
Shunsuke Baba$^{1}$
\\
$^{1}$ Kagoshima University 
\end{flushright}
\vspace{0.5cm}

\subsection{Scientific background and motivation}

Gas outflows from active galactic nuclei (AGNs) have been proposed as a mechanism regulating the co-evolution of supermassive black holes and their host galaxies.
Such outflows have been observed in multiple phases and velocity scales, from fast ionized gas to slow molecular gas, but it is the molecular outflow that accounts for the majority of the mass.
Besides, it is also in the molecular phase that stars form from gas.
Therefore, understanding the kinematics of molecular outflows is essential for assessing the AGN feedback effects on the host galaxy.
Observations of nearby ultra-luminous infrared galaxies (ULIRGs, $L_\mathrm{IR}>10^{12}\,{\rm L}_\odot$) have shown that the kinetic power and moment rate of molecular outflows correlate with AGN luminosity, indicating that AGNs are the dominant source driving the outflow \citep{2014A&A...562A..21C}, although there may be a large scatter in the coupling efficiency and moment boost \citep{2019MNRAS.483.4586F}.
Investigating what type of molecular gas outflows were driven by AGNs in distant galaxies at the epoch of more active star formation will help us understand how present-day massive galaxies were formed.
The first spatial scale at which an AGN can directly affect the molecular gas is the region corresponding to the molecular torus that surrounds the nucleus ($\lesssim$10\,pc).
In this subsection, we propose a method to observe the kinematics of the molecular gas in this region.

\subsection{Required observations and expected results}

\begin{figure}[htbp]
\centering
\includegraphics[width=\textwidth]{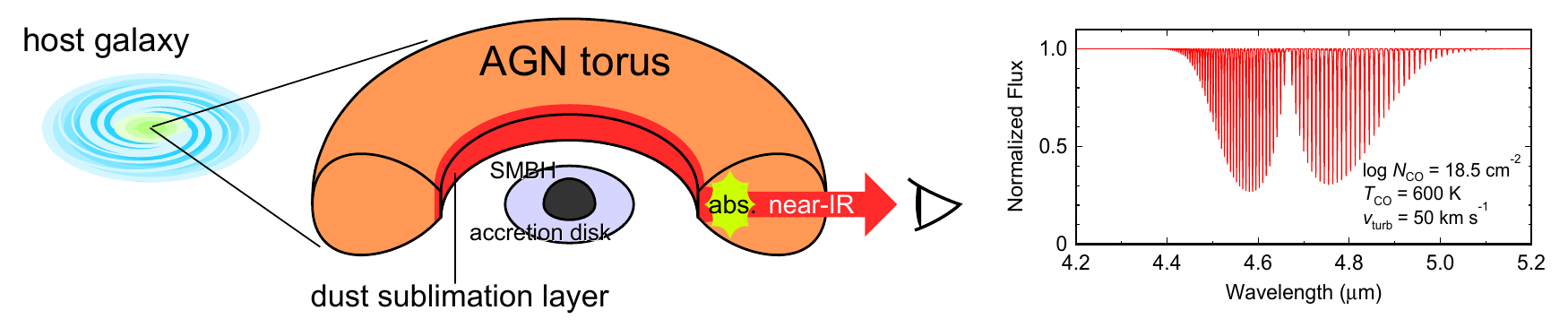}
\caption{
Left: schematic of the observation of an AGN torus region using near-infrared absorption lines.
The thermal emission from the dust sublimation layer is the dominant continuum emission source over the galaxy, and by spectroscopy of absorption lines to it, an effectively high spatial resolution can be achieved.
Right: example of a CO vibrational rotational absorption spectrum (column density $N(\mathrm{CO})=10^{18.5}\,\mathrm{cm^{-2}}$, temperature 600\,K, velocity width 50\,km\,s$^{-1}$).
Lines of different rotational levels appear at intervals of $\sim$0.01\,$\mu$m around the rest wavelength of 4.67\,$\mu$m.
}
\label{fig:Baba_torus}
\end{figure}

The biggest obstacle to observing the vicinity of an AGN is the physical smallness of the region itself.
Its angular size is only a few hundred milli-arcseconds, even for the nearest AGNs.
This hinders imaging observations of distant AGNs.
This problem can be solved by spectroscopic observation of absorption lines in the rest near-infrared region.
Figure \ref{fig:Baba_torus} illustrates the concept of this method: the dominant near-infrared continuum in an AGN host galaxy is thermal radiation from dust heated by the nucleus, especially from a compact region where the dust is sublimating.
With absorption line spectroscopy, the effective spatial resolution is determined by the size of the background source, and the vicinity of the AGN can be selectively observed with minimal influence from the host galaxy.
In the near-infrared region, a useful absorption line band is the vibrational rotation transition of CO ($v=1\leftarrow0$, $\Delta J=\pm1$, band center $\sim$4.67\,$\mu$m, line spacing $\sim$0.01\,$\mu$m).
An example spectrum is shown in Figure \ref{fig:Baba_torus}.
Dozens of lines belonging to different rotational excitation levels can be observed at once, which helps us to effectively constrain the gas column density, temperature, and line-of-sight velocity.
This is an advantage that CO pure rotational emission lines in the sub-millimeter region do not have.

Previous observations of CO ro-vibrational absorption lines have been performed for nearby AGNs;
\citet{2013PASJ...65....5S} and \citet{2021ApJ...921..141O} observed the ULIRG IRAS 08572$+$3915 NW, which has a buried AGN, with the Infrared Camera and Spectrograph (IRCS) of the Subaru telescope at wavelength resolution $R\sim10,000$ ($\Delta v\sim 30\,\mathrm{km\,s^{-1}}$) and detected an outflow component at $-160\,\mathrm{km\,s^{-1}}$ from each line of CO.
This component is found to be high temperature (several 100\,K) and large column density ($N(\mathrm{CO})\sim5\times10^{18}\,\mathrm{cm^{-2}}$) from its level population diagram, suggesting that it is in the vicinity of the AGN, such that X-rays heat it.
Thus, CO absorption at rest-frame near-infrared wavelengths is a unique probe sensitive to outflows near AGNs.
Unfortunately, ground-based observations with such high spectral resolution have been carried out for only a few nearby objects, let alone distant ones, due to poor atmospheric transmittance.
There are some examples of space-based observations using ISO, Spitzer, and AKARI \citep{2004A&A...426L...5L,2004ApJS..154..184S,2018ApJ...852...83B,2022ApJ...928..184B}, but in these cases, individual rotational lines are not resolved due to poor spectral resolution.
Even JWST, which is now just starting science operation, has a maximum dispersion of $R\sim3,000$, which is not sufficient to resolve the velocity of each line.

Observing outflows through CO ro-vibrational absorption lines for distant AGNs, which is not possible with existing telescopes, is a science that will be pioneered for the first time with GREX-PLUS.
If the wavelength resolution of the high-dispersion spectrograph is better than $R=10,000$ ($\Delta v\sim 30\,\mathrm{km\,s^{-1}}$), the velocity decomposition of CO lines can be adequately achieved.
If the wavelength range of 10--18\,$\mu$m is covered, the CO band (rest 4.67\,$\mu$m) can be observed at redshifts beyond $z=1.2$, which is an important era in galaxy evolution.
One concern is that the sensitivity of the spectrograph is a bit challenging.
The current nominal sensitivity to continuum emission is 5\,mJy at the 5$\sigma$ level for a one-hour integration.
Let us use IRAS 00397$-$1312 ($z=0.26$, $L_\mathrm{IR}=1\times10^{13}\,{\rm L}_\odot$) as an example of a luminous AGN that shows CO absorption.
If this galaxy were at $z=1.2$, the S/N reached in a 3-hour integration would be 2.7 at rest-frame 5\,$\mu$m.
Although multiple-line stacking as described below can be employed to obtain velocity profiles, it is essential to first find a promising bright target in the hyper-luminous infrared galaxy (HyLIRG, $L_\mathrm{IR}>10^{13}\,{\rm L}_\odot$) class.

\begin{figure}[htbp]
\centering
\includegraphics[width=\textwidth]{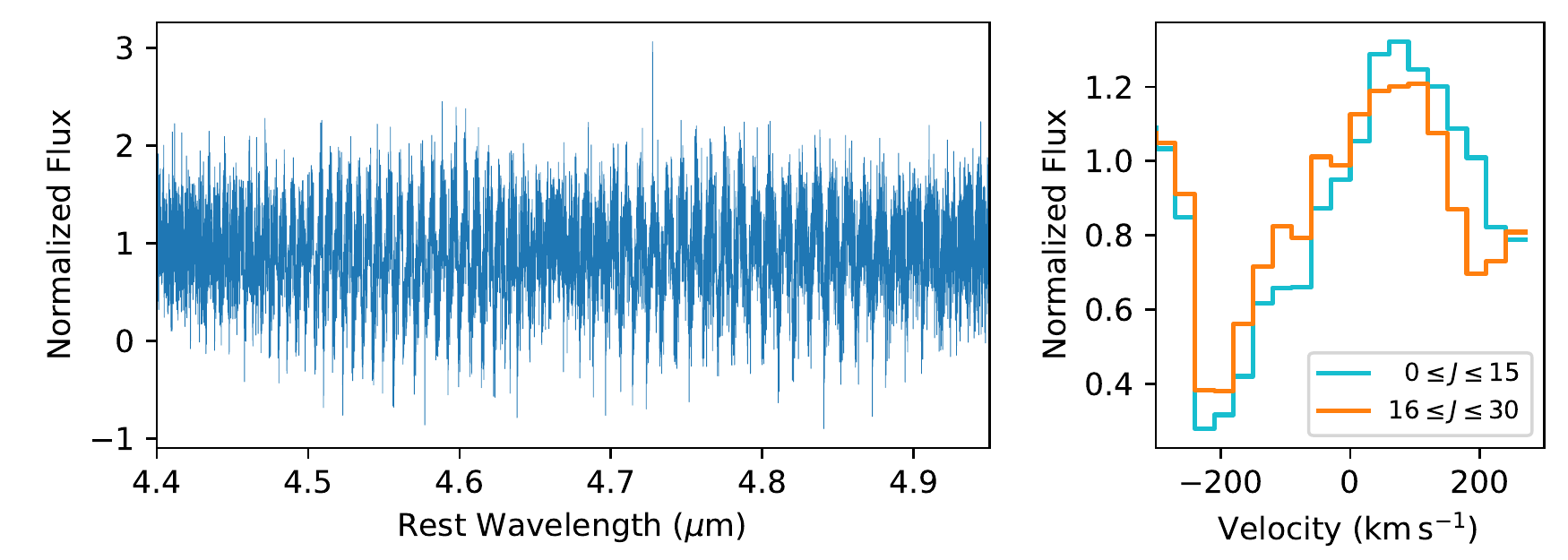}
\caption{
Left: expected spectrum of the outflow described in this text observed from the same line of sight as the propagation direction with S/N=2.7 for the continuum level.
Right: velocity profiles for the lines of different rotational levels $J$ averaged for each low and high excitation regime.
The binning is applied with a velocity width corresponding to the required resolution (30\,km\,s$^{-1}$), rather than the nominal wavelength resolution of the high-dispersion spectrometer.
}
\label{fig:Baba_spec}
\end{figure}

Now, let us assume an outflow with a certain mass rate as an example and show what kind of spectrum is expected.
Here, it is assumed that the velocity decelerates from 200\,km\,s$^{-1}$ to 50\,\,km\,s$^{-1}$ and the temperature decreases from 800\,K to 600\,K during the propagation process, and the gas density is $10^6$\,cm$^{-3}$ at the base, the opening angle is 90\,deg, and the CO to H$_2$ abundance ratio is $10^{-5}$.
The resulting mass outflow rate is $30\,{\rm M}_\odot\,\mathrm{yr}^{-1}$.
Figure \ref{fig:Baba_spec} (left) shows a spectrum of the CO absorption caused by this outflow expected when S/N=2.7 for the continuum level.
The line of sight of the observation is assumed to be in the same direction as the outflow.
As this figure shows, it is difficult to analyze the individual lines separately.
However, since the band contains many lines of different rotational excitation levels, the velocity profile can be determined well by averaging (stacking) them.
Figure \ref{fig:Baba_spec} (right) shows the averaged velocity profiles at low ($0\leq J\leq15$) and high ($15\leq J\leq30$) excitation levels.
Binning is applied with a velocity width of 30\,km\,s$^{-1}$.
In both cases, a P-Cygni profile consisting of absorption and emission, which is characteristic of outflows, is observed.
It can also be seen that the line width of absorption is narrower at high excitation than at low excitation.
This indicates that the outflow is slowing down as the gas moves from the center, where it is warm, to the outward, where it is cold (i.e., decelerating outflow).
Thus, even with the current nominal sensitivity, the overall velocity profile of the gas can be studied by line stacking, and if the temperature change associated with propagation is significant, velocity changes of an outflow can be also discussed.
The required wavelength resolution, wavelength coverage, and sensitivity are summarized in Table \ref{tab:Baba_requirements}.

\subsection{Scientific goals}

By performing pointed high-dispersion spectroscopy of CO ro-vibrational absorption lines for HyLIRG-class AGNs at $z>1.2$, we detect molecular gas outflows with an order of magnitude higher wavelength resolution than existing space telescopes.
This will provide a more detailed examination of the AGN feedback during the era of high galactic activities.

\begin{table}[h]
    \begin{center}
    \caption{Required observational parameters.}
    \label{tab:Baba_requirements}
    \begin{tabular}{|l|p{9cm}|l|}
    \hline
     & Requirement & Remarks \\
    \hline
    Wavelength & 10--18 $\mu$m &  \\
    \hline
    Spatial resolution & N/A &  \\
    \hline
    Wavelength resolution & $\lambda/\Delta \lambda>10,000$ &  \\
    \hline
    Field of view & N/A &  \\
    \hline
    Sensitivity & 5 mJy (for continuum, 1hr $5\sigma$) &  \\
    \hline
    Observing field & N/A &  \\
    \hline
    Observing cadence & N/A &  \\
    \hline
    \end{tabular}
    \end{center}
\end{table}

\clearpage
\section{Molecular Gas in the Intergalactic Medium}
\label{sec:IGMmoleculargas}

\noindent
\begin{flushright}
Toru Misawa$^{1}$, Shunsuke Baba$^{2}$
\\
$^{1}$ Shinshu University, 
$^{2}$ Kagoshima University 
\end{flushright}
\vspace{0.5cm}

\subsection{Scientific background and motivation}

Neutral hydrogen absorbers with very large column densities of log($N_{\rm HI}$/cm$^{-2}$) $\geq$ 20.3 provide strong absorption features (Damped Ly$\alpha$ systems; DLAs) in spectra of background quasars \citep{1986ApJS...61..249W}. In DLAs, hydrogen is mainly neutral, while it is almost ionized in all other class of quasar absorption systems with log($N_{\rm HI}$/cm$^{-2}$) $\lesssim$ 20. The neutrality of the hydrogen gas in DLAs is important, since such cold neutral clouds could be precursors of molecular clouds, corresponding to the birth place of stars in young galaxies \citep{2003ApJ...587..278W}. Indeed, a clear chemical evolution is found in DLAs; the metallicity increases by a factor of $\sim$50--100 from $z$ = 5 to today \citep{2018A&A...611A..76D}.

Since the first detections in 1960s, more than 35,000 DLAs have been discovered so far \citep{2020ApJS..250....8L}. \citet{2014A&A...566A..24N} recently discovered $\sim$100 extremely strong DLAs (ESDLAs) with log($N_{\rm HI}$/cm$^{-2}$) $\geq$ 21.7 in quasar spectra from the Baryon Oscillation Spectroscopic Survey (BOSS) of the Sloan Digital Sky Survey (SDSS), and confirmed that their column density frequency distribution becomes steeper than what is seen at lower column densities as \citet{2005ApJ...635..123P} had already pointed out, which could be due to a conversion of neutral hydrogen atoms to molecular hydrogen (H$_2$), dust obscuration of the background quasars, and/or a low covering factor (i.e., a small size) of the absorbers. Based on a statistical analysis of the column density distribution function, ESDLAs are supposed to have column densities of up to log($N_{\rm HI}$/cm$^{-2}$) $\sim$ 22.5 \citep{2021MNRAS.507..704H}.
In such high column density absorbers, the presence of molecular hydrogen (H$_2$) has been theoretically expected \citep{2004ApJ...609..667S}. The molecular fraction, $f$(H$_2$) = 2N(H$_2$)/[2N(H$_2$)+N(HI)], actually increases according to the total hydrogen column density in the ISM toward Galactic stars \citep{1977ApJ...216..291S}. Extragalactic molecular hydrogen have also been discovered in high column density absorbers at high redshift like ESDLSs through the detection of Lyman- and Werner-band absorption lines \citep{2014A&A...566A..24N, 2019MNRAS.490.2668B}.

\begin{figure}[ht]
   \begin{center}
   \includegraphics[width=100mm,angle=270]{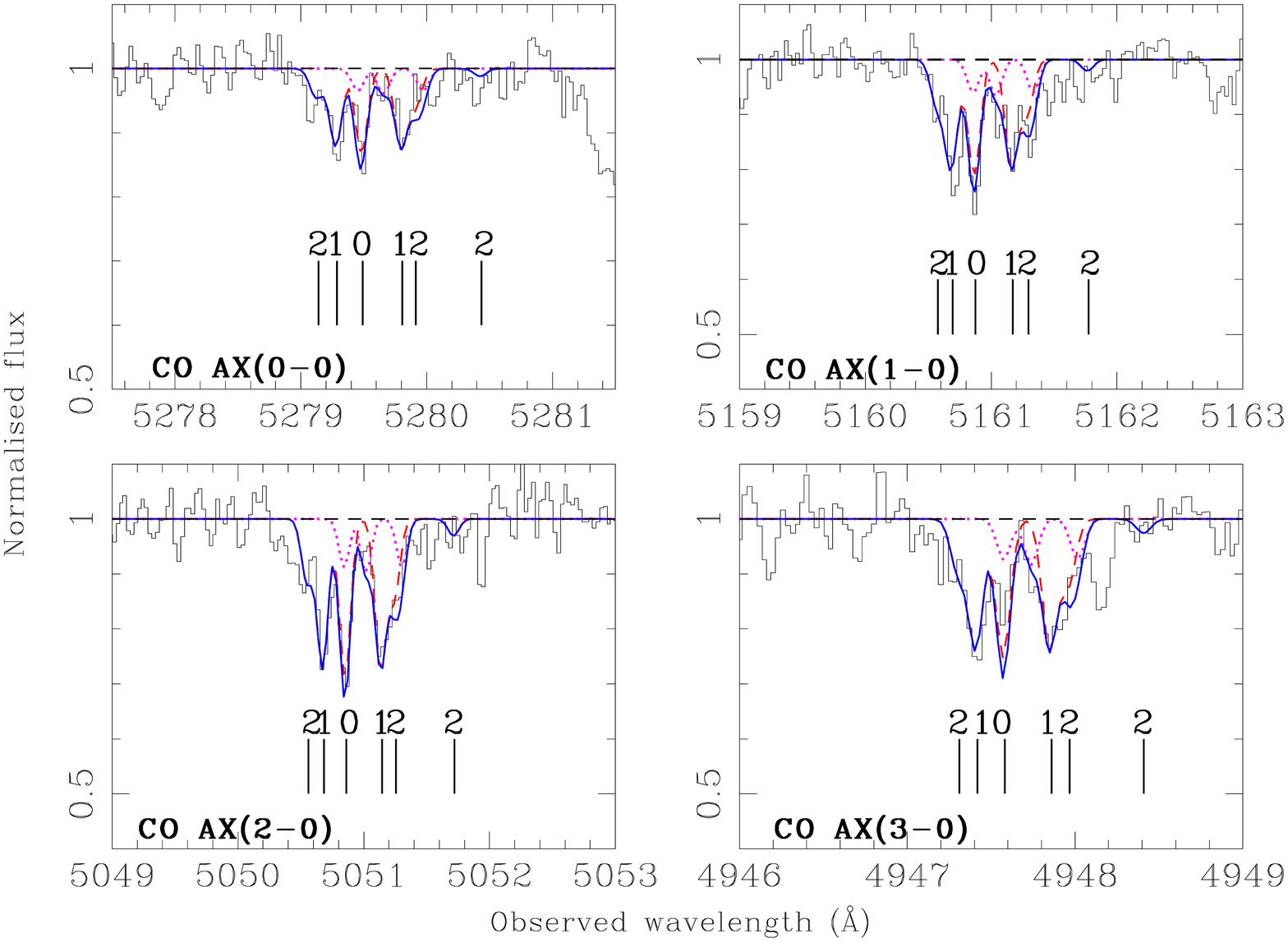}
   \end{center}
   \caption{Voigt profile fits (blue curves) to the 4.7~$\mu$m CO bands at $z_{\rm abs}$ = 2.4185 toward a quasar SDSS J143912.04+111740.5 (black histograms). The vertical lines with numbers are the locations of different CO transitions from different $J$ levels. The figure is taken from \citet{2008A&A...482L..39S}.}
   \label{fig:2_10_1}
\end{figure}

Recently, carbon monoxide (CO) molecular bands have also been detected in quasar spectra through the detection of A-X band absorption lines (Fig.~\ref{fig:2_10_1}), although the sample size ($\sim$10) is still small \citep{2008A&A...482L..39S, 2018A&A...612A..58N, 2019A&A...625L...9G}. For example, the first discovered CO-bearing DLA system at $z$ $\sim$ 2.42 has an extremely large molecular fraction ($f$ $\sim$ 0.27), a large metallicity with a dust depletion, and a column density ratio of N(CO)/N(H$_2$) = 3$\times$10$^{-6}$ \citep{2008A&A...482L..39S}. Thus, CO-bearing DLAs as well as H$_2$-bearing ones are crucial for studying the star/galaxy formation history at high redshift.

Since column densities of CO molecular clouds can be measured with high accuracy of $\sim$0.02~dex by applying Voigt profile fit, we can not only estimate their physical properties like metallicity and molecular fraction, but use them for measuring a temperature of the cosmic microwave background radiation (CMBR) as follows. Assuming that the relative level populations of the CO rotational levels are thermalized by the CMBR, \citet{2008A&A...482L..39S} measured the CMBR temperature at $z$ $\sim$ 2.42 as T$_{\rm CMBR}$ = 9.15$\pm$0.72~K, while 9.315$\pm$0.007~K is expected from the theory (i.e., T$_{\rm CMBR}$($z$) = T$_{\rm CMBR}$($z$=0)$\times$(1+$z$)). Among several methods of measuring T$_{\rm CMBR}$ at high redshift \citep{2022Natur.602...58R}, this is one of the most accurate measurements.

In the past studies including \citet{2008A&A...482L..39S}, extra-Galactic CO molecular clouds have been detected through the A-X bands that correspond to the electron transition in the ultraviolet region in the rest frame (optical regions in the observed frame). They are also accessible through the detection of absorption lines corresponding to rotational and vibrational (ro-vib) transitions at $\lambda$ $\sim$4.7~$\mu$m in the NIR region in the rest frame (MIR region in the observed frame). 
The high resolution MIR spectrograph with GREX-PLUS (covering 10--18$\mu$m) will for the first time enables us to start a systematic search of the 4.7~$\mu$m CO bands in ESDLAs at high redshift.

\subsection{Required observations and expected results}

\begin{figure}[ht]
   \begin{center}
   \includegraphics[width=150mm,angle=0]{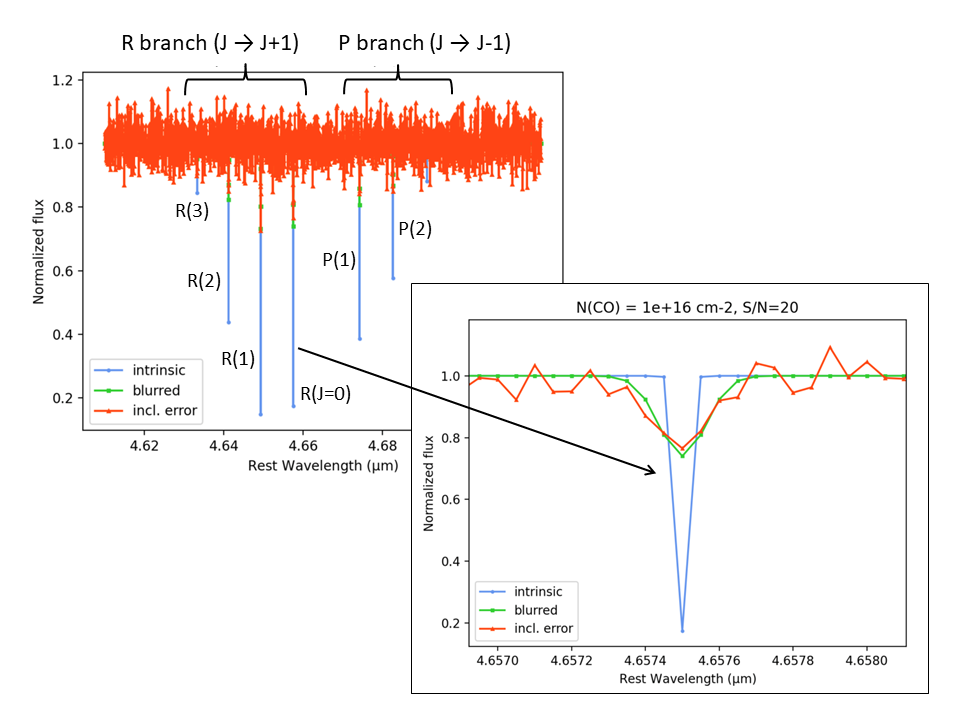}
   \end{center}
   \caption{Synthesized spectrum around the 4.7~$\mu$m CO ro-vib bands at $z$ = 2.2. We assume log($N_{\rm CO}$/cm$^{-2}$) = 16, T$_{\rm kin}$ = 100~K, and n(H$_2$) = 18 cm$^{-3}$, following \citet{2008A&A...482L..39S}, \citet{2019MNRAS.490.2668B}, and \citet{2019A&A...625L...9G}. Blue curve denotes the intrinsic spectrum with thermally broadened line width, while green and red curves are those after applying convolution with $R$ = 30,000 and adding noise to produce S/N = 20 pixel$^{-1}$ spectrum. Both R and P branches are clearly detected. We iterate model fits to the synthesized spectrum using four free parameters (T$_{\rm CMBR}$, T$_{\rm kin}$, n(H$_2$), and log$N_{\rm CO}$) after changing a seed for noise. The accuracy of T$_{\rm CMBR}$ measurement ($\delta$T) is very dependent on S/N ratio; $\delta$T $\sim$ 1~K or 0.1~K if S/N = 20 or 1000 pixel$^{-1}$.}
   \label{fig:2_10_2}
\end{figure}

In this proposal, we aim to detect the 4.7~$\mu$m CO ro-vib bands in ESDLAs, which will be the first systematic survey of the 4.7~$\mu$m CO bands at high redshift. We also attempt to measure T$_{\rm CMBR}$ based on the relative strength of several transitions of both R branch (J $\rightarrow$ J+1) and P branch (J $\rightarrow$ J-1), and verify the past measurements of T$_{\rm CMBR}$. To achieve these goals, both high spectral resolution ($R$ $>$ 30,000) and high signal-to-noise ratio (S/N $>$ 20 pixel$^{-1}$) are required. As a feasibility test, we perform a simulation as follows: 1) we assume an ESDLA system with CO column density of log($N_{\rm CO}$/cm$^{-2}$) = 16 at $z$ = 2.2 whose kinetic temperature of T$_{\rm kin}$ = 100~K, gas volume density of n(H$_2$) = 18 cm$^{-3}$ referring to the past studies \citep{2008A&A...482L..39S, 2019MNRAS.490.2668B, 2019A&A...625L...9G}, 2) we assume that CO rotational excitation is dominated by radiative pumping by the CMBR at $z$ = 2.2 (T$_{\rm CMBR}$ = 8.74~K) but partially contributed by collisional excitation, 3) we run the non-LTE radiative transfer code RADEX \citep{2007A&A...468..627V} for the modeled absorption system to calculate the column densities of CO molecules at ground and excited levels, 4) we synthesize a spectrum around the 4.7~$\mu$m CO ro-vib bands with thermally broadened line width, convolve it assuming a spectral resolution of $R$ = 30,000, and then add noise to produce a spectrum with S/N = 20 pixel$^{-1}$ (Fig.~\ref{fig:2_10_2}), and finally 5) we repeat to fit models to the synthesized spectrum using four free parameters (T$_{\rm CMBR}$, T$_{\rm kin}$, n(H$_2$), and log$N_{\rm CO}$). The accuracy of T$_{\rm CMBR}$ measurement ($\delta$T) is very dependent on S/N ratio. For example, $\delta$T $\sim$ 1~K (similar to those measured in the past studies) if S/N = 20 (i.e., the minimum S/N ratio we require) and $\delta$T $\sim$ 0.1~K if S/N = 1000.

If we assume WISE band3 magnitude (at $\sim$12$\mu$m) of W3 = 9.5 ($\sim$1~mag fainter than one of the brightest quasars at $z$ $>$ 2, J13260399+7023462 \citep{2020ApJ...899...76J}) as a target's magnitude, 10 hours exposure would provide a $R$ $\sim$ 30,000 spectrum with S/N $\sim$ 20 pixel$^{-1}$ based on the sensitivity curve of GREX-PLUS high-resolution mode for a point source. Although such bright quasars with W3 $\lesssim$ 9.5 are still rare, Gaia-assisted quasar survey using a novel combination of astrometry from Gaia-DR3 and photometry from optical, NIR, and MIR missions effectively continues to discover bright quasars even if they are heavily dust-reddened (e.g., \citealt{2018A&A...615L...8H}). GREX-PLUS also plans to conduct its own survey of dust-obscured AGNs. These surveys will provide a number of quasars that are bright enough for our project.

\subsection{Scientific goals}

We will conduct high-resolution spectroscopy ($R$ = 30,000) of several MIR bright quasars (W3 $\lesssim$ 9.5) to search for the 4.7~$\mu$m CO ro-vib bands in high redshift ESDLAs. This will be the first systematic search of the 4.7~$\mu$m CO band to our knowledge. If we successfully acquire MIR spectra with high enough S/N ratio, we measure T$_{\rm CMBR}$ based on the relative strength of several transitions of both R and P branches. We have already confirmed this is feasible if spectra with S/N $>$ 20 pixel$^{-1}$ is available. The total exposure time to acquire spectra with S/N $\sim$ 20 pixel$^{-1}$ is $\sim$10~hours for quasars with W3 $\sim$ 9.5. We will be able to place more stringent constraints on T$_{\rm CMBR}$ at high redshift with higher quality spectra, which will be a direct crucial test of the standard $\Lambda$CDM cosmology.

\begin{table}[ht]
    \begin{center}
    \caption{Required observational parameters.}
    \label{tab:2_10}
    \begin{tabular}{|l|p{9cm}|l|}
    \hline
     & Requirement & Remarks \\
    \hline
    Wavelength & $\sim$14$\mu$m (10--18$\mu$m) & $a$ \\
    \hline
    Spatial resolution & $<1$ arcsec & \\
    \hline
    Wavelength resolution & $\lambda/\Delta \lambda>30,000 $ & $b$ \\
    \hline
    Field of view & 8$^{\prime\prime} \times 35^{\prime\prime}$ & \\
    \hline
    Sensitivity & 4.2~mJy (1hour, $5\sigma$, point-source, low BG) & $c$\\
    \hline
    Observing field & N/A & \\
    \hline
    Observing cadence & N/A & \\
    \hline
    \end{tabular}
    \end{center}
    $^a$ Any wavelength is useful as far as redshifted 4.7~$\mu$m CO bands are covered.\\
    $^b$ $R$ = 30,000 is enough, but higher resolution is better since 4.7~$\mu$m CO lines are probably not resolved if we assume they are thermally broadened with T$_{\rm kin}$ = 100~K.\\
    $^c$ At least S/N = 20 pixel$^{-1}$ at $\lambda$ $\sim$ 14$\mu$m is required with a reasonable observing time of $\leq$ 10 hours.
\end{table}

\clearpage
\section{Extremely Metal-Poor Galaxies}
\label{sec:empg}

\noindent
\begin{flushright}
Kimihiko Nakajima$^{1}$
\\
$^{1}$ NAOJ 
\end{flushright}
\vspace{0.5cm}

\subsection{Scientific background and motivation}

Galaxy mass is thought to be a fundamental quantity which governs the evolution of galaxies.
Theoretically, galaxy formation models based on $\Lambda$CDM cosmology predict that
galaxies grow through subsequent mergings of lower-mass objects and that
galaxy properties are largely determined by their masses through mass-dependent processes
at work in their evolution
(e.g., \citealt{blumenthal1984,davis1985,bardeen1986}).
Therefore, observations of the mass-dependence of galaxy properties back in cosmic time
are crucial to understanding how galaxies evolve to acquire present-day properties.
Particularly, low-mass galaxies at high redshifts are interesting since they are likely to be building blocks of massive galaxies seen in later epochs.

Another key property of galaxies is gas-phase metallicity.
This can be used to investigate complex physical processes that regulate galaxy evolution,
such as star-formation/explosion history, outflow of metal-enriched gas, and
infall of metal-poor intergalactic medium (IGM) gas.
Despite its complexity, a clear relationship between stellar mass and metallicity
has been recognized for local galaxies over 4 orders of magnitude in stellar mass
(e.g., \citealt{tremonti2004,AM2013}), where a galaxy with larger mass have a
higher gas-phase metallicity.
A similar mass-metallicity relation has been observed for higher redshift galaxies up to $z\sim 3.5$
with a tendency that galaxies present lower metallicities at higher redshifts for a given stellar mass
(\citealt{MM2019} for a review).
More recently, the early JWST spectroscopic data provide the metallicity measurements for galaxies at $z=4-10$, suggesting the presence of mass-metallicity relation with a small evolution from $z=2-3$ to $z=4-10$, albeit with a potential deficit of metallicity at $z>8$ (e.g., \citealt{nakajima2023}).
Such a clear relationship between mass and metallicity, and its secondary dependence on
star formation rate (e.g., \citealt{mannucci2010}) suggest that galaxies have
metallicity equilibrium conditions for the balance between star formation, gas outflows and inflows
(e.g., \citealt{lilly2013}).
Several state-of-the-art cosmological simulations are able to reproduce the metallicity observations
in the local universe and at redshifts up to $z\sim 2-3$,
but have dissimilar predictions for much higher redshifts in the early universe,
such as the slope of the mass-metallicity relation, metallicities typically expected in the low-mass regime, 
and their evolution across cosmic time
(e.g., \citealt{ma2016,derossi2017,torrey2019,langan2020,ucci2021,wilkins2022,nakazato2023}).
Indeed, the recent JWST studies at $z=4-10$ (e.g., \citealt{nakajima2023}) suggest a possible inconsistency of the mass-metallicity relation with the existing simulations at the low-mass end ($\lesssim 10^8\,M_{\odot}$), although the sample size of the low-mass galaxies remains small. 
The potential break of the mass-metallicity-star formation rate relation at $z>8$ also needs to be examined with a larger sample. 
Determining the metallicity relationships at high redshifts including the low-mass regime is crucial to constrain the models
and hence the physics affecting galaxy evolution across cosmic time.

In the coming decade, there will be powerful infrared observing instruments that allow us to
explore galaxies in the early universe, such as JWST, Euclid, Roman Space Telescope (Roman in short),
and GREX-PLUS.
Using the near- and mid-infrared multi-band imaging data, one can efficiently select high redshift galaxy
candidates by detecting the Lyman break at the wavelength of $0.1216\times (1+z)\,\mu$m.
Galaxies selected in this way, called LBGs, will be followed up
spectroscopically with JWST, ALMA, and the next-generation extremely large telescopes
to be used for the early galaxy evolution studies including the chemical enrichment.
However, LBGs are UV-selected galaxies and provide an UV-bright sample with modest stellar masses,
i.e. not ideal for a systematic sampling of low-mass galaxies.
In order to fully address the early galaxy evolution, it is essential to explore galaxies in the low-mass,
low-metallicity regime as a representative population of galaxies in the early universe.

\subsection{Required observations and expected results}
\begin{figure*}[htb]
  \centering
    \includegraphics[bb=0 0 879 284, width=0.99\textwidth]{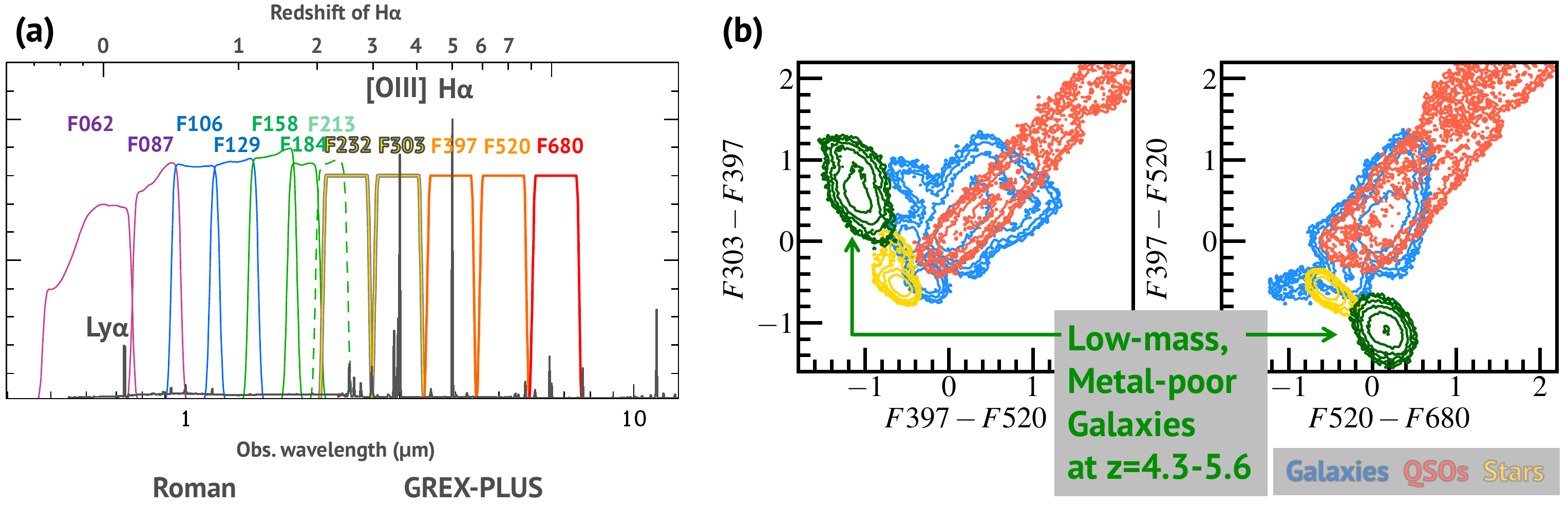}
    \caption{
      \textbf{(a:)} A modeled spectrum of low-mass, metal-poor galaxy
      at $z=5$ (gray) along with the available filters of 
      Roman Space Telescope (up to $2\,\mu$m) and 
      GREX-PLUS (from $2$ to $8\,\mu$m based on the current plan).
      The infrared bands allow us to perform systematic searches of low-mass galaxies
      over $z=2-6$
      by capturing the characteristic intense emission lines such as H$\alpha$
      imprinted on the broadband photometric colors as shown in Panel (b).
      Combining with Lyman-break / Ly$\alpha$ provided by Roman, 
      such a low-mass galaxy search can be extended to $z=6-8$.
      \textbf{(b:)} 
      Simulated distributions of low-mass, metal-poor galaxies at $z=4.3-5.6$ (green) 
      on the color-color diagrams using the GREX-PLUS bands,
      ensuring their efficient selection and discrimination from the other populations 
      of evolved / different redshift galaxies (blue), QSOs (orange), and galactic stars (yellow)
      solely using the photometric data.
      \label{fig:empg_spectrum_z5}
    }
\end{figure*}

We propose GREX-PLUS, which covers from 2 to 8\,$\mu$m with 5 different photometric bands in a seamless way, 
can uniquely provide a systematic sample of low-mass, metal-poor galaxies
over the redshift range of $z=2-6$ (and beyond with Roman) 
by exploiting the rest-frame optical intense emission lines.
Figure \ref{fig:empg_spectrum_z5} presents a model spectrum of low-mass, metal-poor galaxies at $z=5$
associated with intense emission lines such as
the Hydrogen Balmer lines (H$\alpha$, H$\beta$, etc) and the metal lines (e.g., [OIII]$\lambda\lambda5007,4959$)
in the rest-frame optical wavelength regime.
The presence of such intense emission lines would boost the broadband photometry of GREX-PLUS,
presenting characteristic photometric colors.
Using the low-mass, metal-poor galaxy templates and pseudo observations,
Panel (b) illustrates we can isolate candidates of such low-mass, metal-poor galaxies at high redshifts
from the other populations (evolved / different redshift galaxies, QSOs, and Galactic stars)
efficiently using the GREX-PLUS photometric bands.
Note that two bands longer than H$\alpha$ are ideally needed to remove strong [OIII] emitting objects at slightly higher-redshifts by confirming no strong emission lines in the longer bands.
This novel technique was originally adopted by \citet{kojima2020} for local ($z<0.03$) extremely metal-poor galaxy search with the Subaru/Hyper Suprime-Cam's wide and deep data. Developing a machine learning classifier, \citet{kojima2020} successfully construct a metal-poor galaxy sample in the local universe. 
\citet{nishigaki2023} extend the technique towards higher redshift at $z=4-5$ using the early JWST photometry data, and demonstrate its novel application at high redshift, albeit with the limited survey volume.
We will extend the technique with GREX-PLUS (and with Roman) to perform a more systematic sampling of low-mass, metal-poor galaxies at high redshift.

For a reliable classification, two of the most intense emission lines, H$\alpha$ and [OIII]$\lambda 5007$,
should fall in two different photometric bands next to each other,
and not in a single band at any redshift.
The current conceptual design, with the 5 photometric bands as tentatively shown in Figure \ref{fig:empg_spectrum_z5},
perfectly meets this criterion.
The current band set (i.e., GREX-PLUS solely) allows us to search for low-mass, metal-poor galaxies in the 3 redshift ranges;
from $z=2.2$ to $2.8$, from $z=3.1$ to $4.1$, and from $z=4.3$ to $5.6$ based on the above method.
If combined with the $\sim 1\,\mu$m data of Roman, which provides an additional redshift constraints by capturing the rest-frame UV spectroscopic signatures of the Lyman break and/or Ly$\alpha$, the method can be further extended towards higher redshift up to $z=8$, as only one band longer than H$\alpha$ becomes sufficient to capture the stellar continuum without the redshift uncertainty. 
This science case therefore has strong synergy with Roman, but never happens at $z>2$ with Roman alone (i.e., without GREX-PLUS).

Table \ref{tab:empg_expectations} lists the rough numbers of galaxies expected to be found
in each redshift range depending on different survey depths and areas.
For the estimation, we assume (i) the number density of metal-poor galaxies explored at $z=0$ \citep{kojima2020}, (ii) the evolution of mass-metallicity relation with redshift \citep{torrey2019,sanders2021}, and (iii) the evolution of stellar mass function with redshift \citep{baldry2012,song2016,davidzon2017,stefanon2021}.
We only count galaxies whose stellar continuum at wavelengths longer than H$\alpha$
can be detected with the corresponding photometric band of GREX-PLUS.
According to the estimation, the survey design covering the widest area with a modest depth will result in the largest sample of low-mass, metal-poor galaxies at all redshift ranges, about 4k galaxies each at $z\sim 2.5$ and $\sim 3.5$, 1k galaxies at $z\sim 5$, and $\sim 100$ galaxies at $z\sim 7$ whose numbers are minimum necessary even at the highest redshift to systematically examine the detailed properties. 
We therefore request to keep the current Wide survey design for this topic, ideally with deeper depths at $\lambda > 5\,\mu$m to probe much lower-mass systems at $z>5$.
Table \ref{tab:empg_requirements} summarizes the required observational parameters for this science case.

\subsection{Scientific goals}
Building the precious sample of low-mass galaxies at $z=2-6$ provided with GREX-PLUS
(and up to $z=8$ with GREX-PLUS + Roman as explained above),
and comparing with the LBG sample, 
we can pursue key scientific goals including the following three:
\begin{itemize}
  \vspace{-1.mm}
  \setlength{\itemsep}{0cm}
  \setlength{\parskip}{0.25em}
\item[$\bullet$] Determining the mass-metallicity relation covering the low-mass regime to discuss the relevant physical processes.
\item[$\bullet$] Characterizing the other ISM properties and the nature of the ionizing spectrum in early galaxies to discuss the stellar population and the ionizing properties.
\item[$\bullet$] Quantifying the dark matter halo properties of low-mass galaxies through clustering analysis to discuss the descendants.
\end{itemize}
For the first two objectives, we need spectroscopic follow-up observations ideally in the GREX-PLUS wavelength regime to observe the rest-frame optical spectra. JWST is supposed to play an essential role in the follow-ups. If unavailable, we can use the photometric color excesses to infer the strengths of [OIII]+H$\beta$ and H$\alpha$ to examine the nebular properties (e.g., \citealt{stark2013,smit2014,roberts-borsani2016,bouwens2016,nishigaki2023}).
The strong optical emission lines are well-calibrated to estimate gas-phase metallicities (e.g., \citealt{nakajima2022} for the latest updates).
Moreover, we can use the extremely large telescopes such as TMT to examine the nebular properties with the rest-frame UV spectra (e.g., \citealt{nakajima2018}).

As listed in the second item, we can address another important question of the nature of the ionizing spectrum
with the detailed spectroscopic follow-up observations.
Galaxies hosting the first generation of stars (PopIII stars) are expected to be found between
the end of the dark age ($z=20-30$) and the epoch of reionization ($z\sim 7$), and thought
to play an important role in reionizing the Universe and in the subsequent structure formation.
Low-mass, metal-poor galaxies explored with the above method may contain such a primordial structure.
We can use the spectroscopic diagnostics as proposed by e.g., \citet{schaerer2003,inoue2011,NM2022}
to explore galaxies hosting PopIII stars and discriminate from another important class of primeval objects: 
direct collapse black holes (DCBHs), using the rest-frame optical and UV spectra.

Finally, we can directly discuss the link between such low-mass, metal-poor galaxies in the early universe and the evolved galaxies found in later epochs via the dark matter halo properties. Such investigations will be of particular importance to discuss whether they are likely building blocks of galaxies in later epochs including our Milky Way.

\begin{table}[htb]
  \begin{center}
    \caption{Expected numbers of low-mass, metal-poor galaxies with the current survey plans.}
    \label{tab:empg_expectations}
    \renewcommand{\arraystretch}{1.3}
    \begin{tabular}{|l|c|c|c|c|c|c|c|c|c|c|c|c|}
      \hline
      Redshift & \multicolumn{3}{c|}{2.2--2.8} & \multicolumn{3}{c|}{3.1--4.1} & \multicolumn{3}{c|}{4.3--5.6} & \multicolumn{3}{c|}{6.0--7.7$^{(\dag)}$} \\
      \hline
      Survey Type & D & M & W & D & M & W & D & M & W & D & M & W \\
      Depth (AB) & 27.0 & 25.7 & 24.3 & 27.1 & 25.8 & 24.4 & 25.2 & 23.9 & 22.5 & 24.0 & 22.7 & 21.3 \\
      Area (deg$^{2}$) & 10 & 100 & 1000 & 10 & 100 & 1000 & 10 & 100 & 1000 & 10 & 100 & 1000 \\
      \hline
      $N^o$ of galaxies & 100 & 800 & 4k & 100 & 800 & 4k & 30 & 200 & 1k & 10 & 80 & 100 \\
      \hline
    \end{tabular}
  \end{center}
  $\dag$ The highest-redshift search requires an additional redshift constraint such as Lyman break and Ly$\alpha$ provided by Roman.
\end{table}

\begin{table}[htb]
  \begin{center}
    \caption{Required observational parameters.}
    \label{tab:empg_requirements}
    \begin{tabular}{|l|p{9cm}|l|}
      \hline
      & Requirement & Remarks \\
      \hline
      Wavelength & 2--8 $\mu$m & \multirow{2}{*}{$a$} \\
      \cline{1-2}
      Spatial resolution & $<1$ arcsec & \\
      \hline
      Wavelength resolution & $\lambda/\Delta \lambda=3$--$4$ & $b$ \\
      \hline
      Field of view & 1000 degree$^2$, F232=24.2, F303=24.3, F397=24.4, & \multirow{2}{*}{$c$}\\
      \cline{1-1}
      Sensitivity & F520=22.5, F680=21.3 ABmag ($5\sigma$, point-source) & \\
      \hline
      Observing field & Fields where deep imaging data at $\lambda<2$ $\mu$m are available. & $d$ \\
      \hline
      Observing cadence & N/A & \\
      \hline
    \end{tabular}
  \end{center}
  $^a$ The wavelength coverage at $\lambda>5\,\mu$m is especially unique and crucial for high-redshift galaxy searches. \\
  $^b$ H$\alpha$ and [OIII]$\lambda 5007$ should fall in two different photometric bands next to each other, and not in a single band at any redshift ($\lambda=(5007+6563)/2\times(1+z)$\,\AA, $\Delta\lambda=(6563-5007)\times(1+z)$\,\AA).\\
  $^c$ The Wide survey plan in the current conceptual design will result in the largest sample of low-mass, metal-poor galaxies at high-redshifts. With deeper observations at $\lambda > 5\,\mu$m, we would be able to probe much lower-mass systems at $z>5$.\\
  $^d$ Deep fields observed with Roman would be ideal.
\end{table}

\clearpage
\section{Magellanic Clouds}
\label{sec:magellanicclouds}

\noindent
\begin{flushright}
Takashi Shimonishi$^{1}$
\\
$^{1}$ Niigata University
\end{flushright}
\vspace{0.5cm}

\subsection{Scientific background and motivation}
The Large and Small Magellanic Clouds (LMC and SMC) are the nearest star-forming low-metallicity dwarf galaxies. 
Their proximity \citep[about 50 kpc for the LMC and 60 kpc for the SMC;][]{Pie13, Gra14} enables us to spatially resolve the distribution of individual stars with a moderate angular resolution (1 arcsec $\sim$0.24 pc at the LMC). 
A nearly face-on viewing angle of the LMC ($\sim$27 degree) and a large separation from the Galactic-disk direction allows us to obtain a bird's eye view on the distributions of stars and interstellar medium (ISM) in a galaxy. 
Their low metallicity environments \citep[$\sim$1/2-1/3 for the LMC and $\sim$1/5-1/10 for the SMC; e.g.,][]{Rus92} play an important role as a laboratory to investigate various astrophysical and astrochemical phenomena in the decreased-metallicity environment, which is similar to the past metal-poor universe. 

Owing to these uniqueness, various types of survey observations have been carried out towards the Magellanic Clouds. 
Infrared surveys are especially useful for the detection and classification of embedded or low-temperature objects, such as young stellar objects (YSOs) and mass-losing late type stars. 
Studies of extragalactic YSOs made substantial progress with the advent of high-sensitivity space-borne infrared telescopes such as Spitzer, AKARI, and Herschel. 

Nearly $\sim$1000 or more YSO candidates are newly identified in the LMC and SMC with the aid of whole-galaxy infrared imaging surveys \citep[e.g.,][]{Whin08,GC09,Kat12,Sew13} and follow-up spectroscopy \citep[e.g.,][]{ST,ST13,Sea09,Woo11,Oli13}. 
Infrared spectroscopic data of LMC/SMC YSOs are used to investigate the chemistry of ices and dust in the protostellar envelope \citep[e.g.,][]{ST,ST10,ST16,Oli09,Oli11,Sea11}. 
The infrared data are also used to categorize the evolutionary stage of massive YSOs \citep[e.g.,][]{Sea09}. 
Those studies have provided unique targets for follow-up observations in the radio regime with ALMA, which can spatially resolve molecular cloud cores at the distance of the Magellanic Clouds. 
Based on the synergy between infrared telescopes and ALMA, we have now started to study various "core-scale" phenomena related to star formation, such as protostellar outflows \citep{Fuk15,ST16b,Tok22} and emergence of a hot molecular core \citep{ST16b,ST20,ST23,Sew18,Sew22a}. 

Clearly, studies of the star formation and ISM with a star-forming core-scale are extending to extragalactic and low-metallicity environments. 
However, although a large number of extragalactic YSOs are identified in the LMC/SMC in the past 15 years, current observations are limited to luminous (mostly high-mass) YSOs due to the sensitivity limit of infrared surveys. 

\begin{figure}[thp!]
\begin{center}
\includegraphics[width=11.0cm]{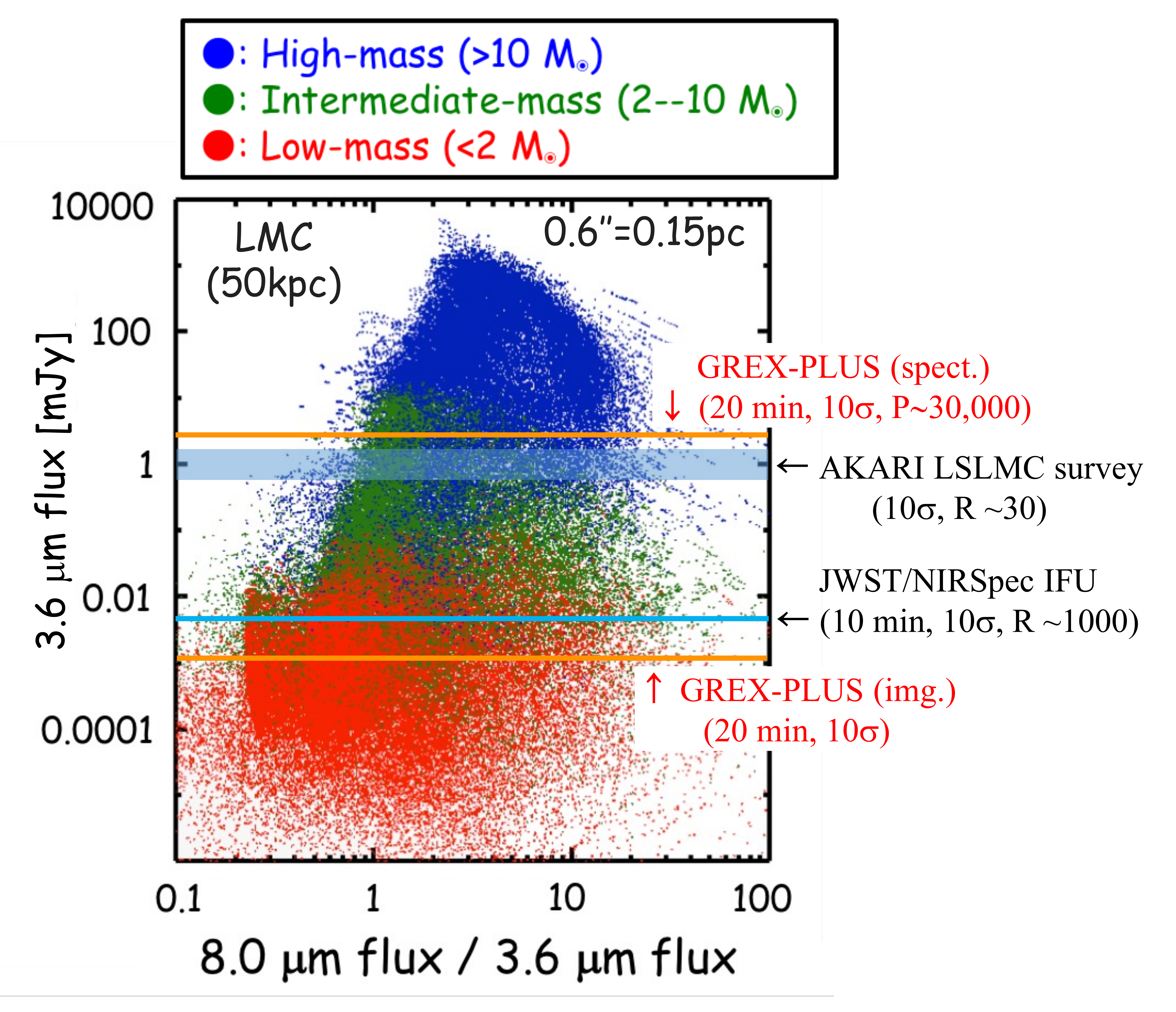}
\caption
{Theoretically-predicted infrared fluxes and colors of YSOs at the distance of the LMC (50 kpc), based on \citet{Rob07}. 
Plot symbols are color-coded depending on YSO's mass; high-mass ($>$10 M$_{\odot}$, blue), intermediate-mass (2 -- 10 M$_{\odot}$, green), and low-mass ($<$2 M$_{\odot}$, red). 
The solid orange lines indicate the expected spectroscopic (upper) and imaging (lower) sensitivities of GREX-PLUS. 
The spectroscopic sensitivities of the AKARI LSLMC survey \citep{ST13} and JWST/NIRSpec/IFU (10 minutes on source) are also shown. 
}
\label{tab:f1_ST}
\end{center}
\end{figure}

\begin{figure}[thp!]
\begin{center}
\includegraphics[width=13.0cm]{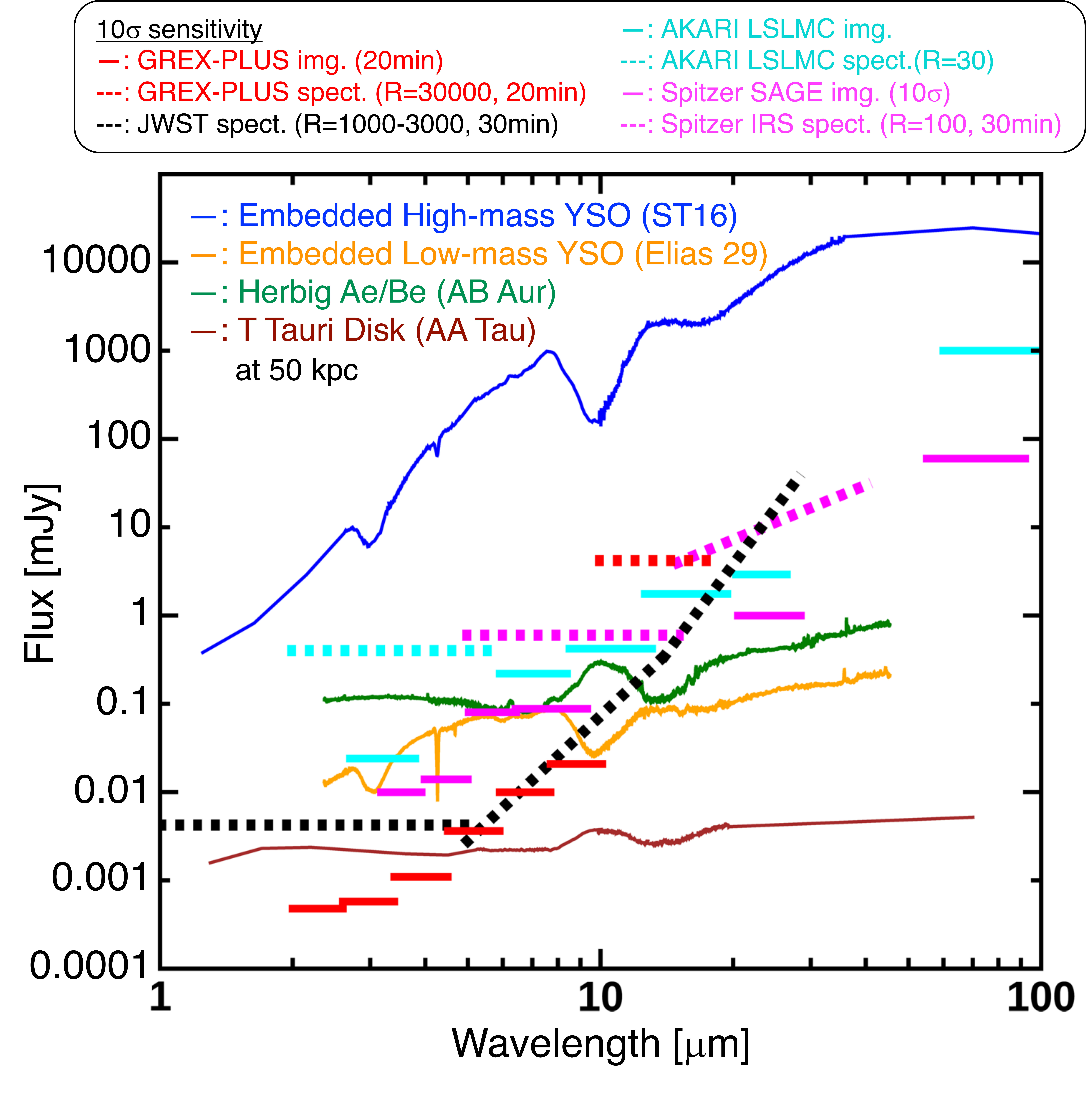}
\caption
{Spectral energy distributions of various objects related to star- and planet-formation scaled at the distance of the LMC; (blue) embedded high-mass LMC YSO (ST16), (yellow) embedded low-mass Galactic YSO (Elias 29), (green) Herbig Ae/Be star (AB Aur) as an intermediate-mass protoplanetary disk source, (brown) T tauri star (AA Tau) as a low-mass protoplanetary disk source. 
The wavelength coverages and sensitivities of selected infrared telescopes are shown by the horizontal lines (solid: imaging, dotted: spectroscopy), where those of GREX-PLUS are shown in red. 
}
\label{tab:f2_ST}
\end{center}
\end{figure}

\subsection{Required observations and expected results}
GREX-PLUS is expected to extend the mass limit of extragalactic YSO surveys down to the low-mass regime with its unprecedented survey efficiency and high-sensitivity in near-/mid-infrared wavelengths. 
Figure \ref{tab:f1_ST} shows theoretically-predicted infrared fluxes and colors of YSOs at the distance of the LMC \citep[based on the SED model of][]{Rob07}. 
It is shown that infrared imaging survey sensitivity of GREX-PLUS is capable of detecting low-mass YSOs in the Magellanic Clouds. 
Figure \ref{tab:f2_ST} shows wavelength coverages and sensitivities of GREX-PLUS and the past infrared space telescopes in comparison with the expected spectral energy distributions of star-forming objects at the distance of the LMC. 
A mid-infrared wavelength coverage of GREX-PLUS is important for characterizing the nature of star-forming objects, because embedded YSOs have their intensity peak in the mid- to far-infrared region, and detecting such a significant infrared excess is key to identify YSOs. 
Those figures indicate that GREX-PLUS will push down the mass limit of extragalactic YSO surveys to the low-mass regime. 
Such a survey should allow us to understand galaxy-scale star-formation activities in a wide range of protostellar masses. 

For such a study, a multiband (2--10 $\mu$m) survey towards a whole region of the LMC ($\sim$60 deg$^2$) and the SMC ($\sim$30 deg$^2$) is preferred. 
Imaging survey towards the Magellanic Bridge is also interesting. 
Mid-infrared sensitivity is important for the detection of embedded objects. 
For example, with the photometric sensitivity of 0.001 mJy at 3--4 $\mu$m and 0.01 mJy at 8--10 micron, an embedded low-mass Class I protostar such as Elias 29 will be detectable at the LMC.

High-dispersion mid-infrared spectroscopic capability of GREX-PLUS will provide us with a unique opportunity to probe chemical compositions of warm molecular gas in the vicinity of protostars. 
It is known that embedded YSOs exhibit various infrared absorption bands due to the associated molecular gas and ice. 
High-dispersion mid-infrared spectroscopy is a powerful tool to resolve ro-vibrational transitions of high-temperature gaseous molecules \citep[e.g.,][]{Boo98,vD04,Lah06,Dun18,Ind20}. 
A variety of astrochemically important molecular species, such as H$_2$O, CO$_2$, C$_2$H$_2$, HCN, CH$_4$, and NH$_3$, are detectable in mid-infrared. 
Especially, for "radio inactive" molecules such as CO$_2$, CH$_4$, C$_2$H$_2$, the infrared spectroscopy is the only way to directly detect them. 
In addition, for gaseous H$_2$O, ground-based radio observations are largely affected by Earth's atmosphere and difficult. 
Those molecules are expected to be abundant in a high-temperature region of the protostellar envelope. 
Investigating their abundances and chemical diversities based on a statistical number of YSO samples is essentially important for understanding the emergence of the chemical complexity in the low-metallicity ISM. 

At the distance of the Magellanic Clouds, the spectroscopic sensitivity of GREX-PLUS allows us to probe almost high-mass YSOs by high-dispersion spectroscopy. 
About 300 infrared sources are confirmed to be high-mass YSOs in the LMC/SMC by infrared spectroscopy with Spitzer and AKARI. 
Systematic high-dispersion spectroscopic survey towards those potential targets are interesting. 
Such observations will provide us with a statistical data on chemistry of forming stars located at a uniform distance, 
and will unveil how the chemical compositions of the protostellar envelope varies depending on stellar luminosities, evolutionary stages, and local environmental characteristics such as metallicity, radiation field strength, degree of clustering, or location within a galaxy.

\subsection{Scientific goals}
In summary, GREX-PLUS observations of the Magellanic Clouds will provide us with unprecedented data set for on-going star formation and chemical evolution of the ISM in our neighborhood low-metallicity galaxies. 
A whole-galaxy infrared imaging survey will unveil low-mass to high-mass star formation activities across the LMC/SMC, while high-dispersion spectroscopy towards known high-mass YSOs will unveil the chemical receipt of stars and planets at low metallicity and its relation with the local interstellar conditions. 
We also note that the outer Galaxy is another nearby important laboratory for low-metallicity star-formation and ISM studies \citep[e.g.,][]{Bra07,Yas08,Izu17,Izu22,ST21}, and infrared imaging/spectroscopic surveys for such a region will also be an interesting science case for GREX-PLUS. 

In this chapter, we have discussed GREX-PLUS's science cases related star-formation and astrochemistry in the Magellanic Clouds. 
We finally note that other potential science cases, such as those related mass-losing late type stars, diffuse interstellar bands, polycyclic aromatic hydrocarbons, or supernova remnants, would also be important for GREX-PLUS observations of the Magellanic Clouds and should be discussed in future.

\chapter{Galactic and Planetary Sciences}
\label{chap:galacticplanetary}

\section{Theoretical Perspective}
\label{sec:PS_theoreticalperspective}

\newcommand{\HK}[1]{{\color{red}#1}}

\noindent
\begin{flushright}
Hiroyuki Kurokawa$^{1}$
\\
$^{1}$ University of Tokyo
\end{flushright}
\vspace{0.5cm}

\subsection{Introduction}

\begin{figure}
    \centering
    \includegraphics{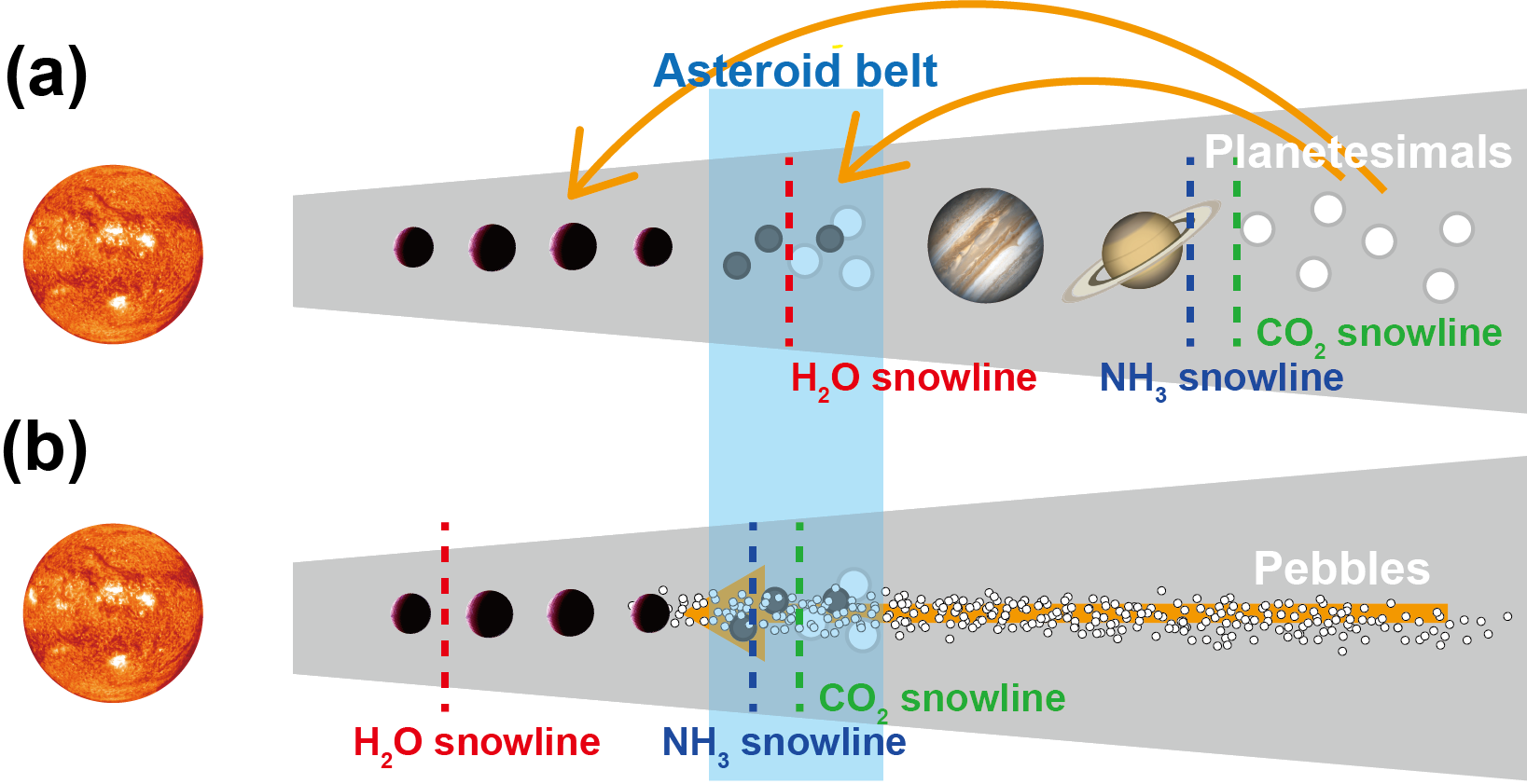}
    \caption{Schematic illustration of two different modes of volatile delivery to terrestrial planets. (a) Volatile delivery by planetesimals before and/or after the dissipation of protoplanetary disk gas. (b) Volatile delivery by pebbles following the migration of snow lines before the dissipation of protoplanetary disk gas.}
    \label{fig:Fig_3_1_1}
\end{figure}

GREX-PLUS will be able to perform key observations of protoplanetary disks, the solar system small bodies, and planets in the solar and extrasolar systems to understand the origins of planets, their habitability, and building blocks for life. Such scientific goals fit ISAS's space science and exploration roadmap toward exploring life in the universe. In Section \ref{sec:PS_theoreticalperspective}, we summarize the scientific background and theoretical perspective on how life-essential volatile elements were delivered to terrestrial planets to motivate the observations with GREX-PLUS instruments to unveil dynamic evolution of snow lines in protoplanetary disks, volatile-rich small bodies in the solar system, and giant planets in the solar and extrasolar systems (see the following subsections for details). 

\subsection{Protoplanetary disks}
\label{subsec:ppd}

Planets form in protoplanetary disks surrounding host stars, and thus constraining their properties is important for understanding the origin of planetary systems. How dust particles, which ultimately form planets, are transported in disks and aggregate into larger bodies is not fully understood, but planet formation theory suggests that the locations of snow lines, where molecules of interest (most importantly H$_2$O, but others as well) condense, in a protoplanetary disk control the sizes and compositions of resulting planets (Figure \ref{fig:Fig_3_1_1}). Sublimation of volatiles change the composition of solid materials \citep[e.g.,][]{Hayashi+1981} and thus planets to form \citep[e.g., rocky vs. icy planets,][]{2004ApJ...604..388I}. Moreover, change in gas and dust surface densities associated to snow lines influences the efficiency of dust aggregation and, ultimately, the architectures of planetary systems. Classical planet formation models assume that increased dust surface density beyond the H$_2$O snow line is preferable for giant planet formation \citep[e.g.,][]{Hayashi+1981,2000ApJ...537.1013I,2008ApJ...673..502K}. More recent models that take dust migration \citep[the inward drift due to aerodynamic drag induced by disk gas rotating with sub-Keplerian speed,][]{1976PThPh..56.1756A,1977MNRAS.180...57W} into account also suggest that sublimation at the snow line causes dust pileup and thus efficient aggregation \citep[e.g.,][]{2011ApJ...733L..41S,2016ApJ...821...82O}. 

The locations of snow lines in protoplanetary disks are poorly understood. A snow line in a disk is defined as the location where the partial pressure of the molecule of interest becomes equal to its saturation vapor pressure; therefore, the temperature matters. The temperature at the midplane of the disk is determined by i) the stellar irradiation to the disk surface and re-emission to the midplane and ii) release of gravitational potential energy of gas accreting onto the host star. Theoretical models suggest that the decline of the second component causes the temperature decrease and, consequently, inward migration of snow lines with time. Notably, in the late stage of protoplanetary disk evolution (a few Myrs), the H$_2$O snow line is predicted to be located inside the inner boundary of the habitable zone, namely, even inside Earth's orbit \citep{2007ApJ...654..606G,Oka+2011}. A recent study considering non-ideal MHD effects on disk gas accretion showed that accretion energy is released at the high altitudes of the disk and thus the temperature at the disk midplane is even lower from the earlier stages than previously thought \citep{Mori+2021}. Such configuration of the H$_2$O snow line with respect to the habitable zone can drastically changes our understanding of how Earth and other terrestrial planets formed. Once the H$_2$O snow line passed the orbits of protoplanets, migration and accretion of mm- to cm-sized icy dust particles (icy pebbles) can easily supply water comparable to Earth's ocean mass \citep[Figure \ref{fig:Fig_3_1_1}b,][]{Sato+2016,Morbidelli+2016}. Such efficient supply of water is, if happened, thought to even change the question on Earth's habitability from \textquotedblleft how did Earth acquire its water?" into \textquotedblleft how did Earth avoid to become icy giants?" Moreover, there are uncertainties in the theoretical predictions for effective snow line locations depending on formation of clathrates and salts \citep{2014ApJ...796L..28M,2020Sci...367.7462P}; therefore, observational constraints are needed.

The mid-infrared high resolution spectrometer onboard GREX-PLUS will be able to inform the position of the H$_2$O snow line in the midplane regions of protoplanetary disks by observing the Doppler shift of water vapor emission lines caused by Keplerian motion (see Section~\ref{sec:protoplanetarydisks}). Previous attempts to constrain the location of the H$_2$O snow line were more sensible to disk surfaces, where the snow line is further from the host star than in the midplane. The locations of other volatile species including NH$_3$, H$_2$S, CO$_2$ snow lines will, if detected, further test the theoretical predictions for snow line locations and constrain how different compositions of solid materials are distributed in disks and whether icy pebbles can contribute significantly to the supply of volatile elements essential for life.

\subsection{Small bodies in the solar system}

Asteroids and comets (hereafter, together called small bodies) in the solar system record the origin of Earth and its habitability. Small bodies have been thought to be remnants of planet formation stages and also to have delivered water and other life-essential elements to terrestrial planets. Classical planet formation theory assumed that planets form hierarchically from dust through km-sized planetesimals \citep[namely, small bodies,][]{1985prpl.conf.1100H}. The terrestrial planet forming region was assumed to be volatile free, and thus Earth was thought to have acquired these elements by accreting small bodies which originate from the outer solar system \citep[Figure \ref{fig:Fig_3_1_1}a, e.g.,][]{2000M&PS...35.1309M}. Recent theory, where migration of dust particles and snow lines are taken into account, points to another possibility: volatile delivery by icy pebbles (Figure \ref{fig:Fig_3_1_1}b, Section \ref{subsec:ppd}).

The two modes of volatile delivery (planetesimals vs. pebbles) would leave distinct record on small body composition and population and thus potentially be distinguished with asteroid observations. In the planetesimal migration scenario, formation and migration of giant planets cause gravitational scattering of small bodies, which leads to bodies originally formed at and beyond the giant planet forming region being transported to the inner solar system including the terrestrial planet forming region and the current asteroid belt. Therefore, some classes of volatile-rich asteroids in the asteroid belt should have the similarity in the composition with the Trans-Neptunian objects \citep[e.g.,][]{2021ApJ...916L...6H} for various sizes. On the other hand, inward migration and accretion of icy pebbles onto terrestrial planets would also cause accretion onto small bodies; for instance, ammonia on the dwarf planet Ceres (the largest body in the asteroid belt) has been interpreted as a potential consequence of pebble accretion \citep{2015Natur.528..241D,2019ESS.....431719N}. Because the accretion efficiency is highly dependent on the mass of the target \citep{2012A&A...544A..32L,2016A&A...586A..66V}, icy pebble accretion would cause larger asteroids being more volatile-rich \citep{2019ESS.....431719N}.

The near-infrared camera and the mid-infrared high resolution spectrometer onboard GREX-PLUS will be able to survey the presence/absence of important ices and minerals on small bodies down to $\sim$10 km in diameter, which is much smaller than the sizes probed with previous spectral observations in these wavelengths. If a body shows spectral features of H$_2$O ice or hydrated minerals, the body should have formed beyond the H$_2$O snow line \citep{2012Icar..219..641T,2015aste.book...65R,2019PASJ...71....1U}. Similarly, ammonia ice and ammoniated phyllosilicates suggest the formation beyond the NH$_3$ snowline; CO$_2$ ice and carbonate minerals suggest the formation beyond the CO$_2$ snow line \citep{2012ApJ...752...15O,2017AJ....153...72V,2020Sci...367.7462P,2022AGUA....300568K}. These key ices and minerals to understand the dominant mode of volatile delivery show diagnostic spectral features in near- to mid-infrared wavelengths which will be covered by the GREX-PLUS instruments (see Section \ref{sec:icysmallbodies}). 

\subsection{Giant planets in the solar and extrasolar systems}

Formation and migration of giant planets in planetary systems determine how volatile elements are delivered to terrestrial planets (Figure \ref{fig:Fig_3_1_1}). A giant planet forming in a protoplanetary disk can cave a gap in the gas disk by its gravity \citep[e.g.,][]{1986ApJ...307..395L}. Because such a gas gap induce a positive pressure gradient at its edge, gas starts to rotate with super-Keplerian speed. Consequently, the direction of pebble migration changes from inward to outward, halting icy pebble accretion to terrestrial planets in inner orbits \citep{Morbidelli+2016}. Both growth and migration of giant planets cause gravitational scattering of small bodies and subsequent accretion of volatile-rich asteroids onto terrestrial planets \citep{2011Natur.475..206W,2017Icar..297..134R}.

Chemical compositions of giant planets record their formation locations in protoplanetary disks and, consequently, migration history. Observational constraints and theoretical modeling points to the fact that giant planets form by the \textquotedblleft nucleated instability" -- a solid core first form, and then the core starts to accrete gas rapidly once it becomes massive enough \citep[e.g.,][]{2000ApJ...537.1013I}. Compositions of gas and solid phases change across volatile snow lines, leaving diagnostic signatures in atmospheric compositions of giant planets (see Sections \ref{sec:exoplanetatmosphere} and \ref{sec:solarsystemplanet}). Atmospheric H, C, N, and O ratios of short-period extrasolar planets have been considered to reflect their formation locations relative to the relevant snow lines \citep[H$_2$O, NH$_3$, CO$_2$, CH$_4$, CO, and N$_2$; e.g.,][]{2011ApJ...743L..16O,2016ApJ...833..203P,Notsu+2020}. The diagnosis of long-period giant planets including the solar system ones is less straightforward than that of short-period ones because condensation in the depth of the atmosphere changes the chemical composition in gas phase and limits the detectability of the condensable molecules. Nevertheless, uniform enrichment of C, N, O, S, P, and noble gases with respect to their solar abundances of Jupiter's atmosphere has been interpreted as a signature of its formation beyond the N$_ 2$ snow lines \citep{2019AJ....158..194O}. Solar-like $^{14}$N/$^{15}$N ratios of Jupiter's and Saturn's atmospheres also supports their formation beyond the N$_2$ snow line \citep{2014Icar..238..170F,2014ApJ...796L..28M}. Possible super-solar S/N ratios of Uranus' and Neptune's atmospheres have been discussed to inform their formation between the H$_2$S and N$_2$ snow lines \citep{2020RSPTA.37800107M}.

The mid-infrared high resolution spectrometer onboard GREX-PLUS will be able to acquire atmospheric spectra of solar and extrasolar giant planets to constrain molecular abundances of interest (See Sections \ref{sec:exoplanetatmosphere} and \ref{sec:solarsystemplanet}). The informed molecular abundances including H$_2$O and NH$_3$ combined with chemical network modeling would constrain bulk elemental abundances and consequently formation locations of short-period extrasolar planets. Detection of NH$_3$ and H$_2$S in the atmospheres of the solar system ice giants would be extra success, but GREX-PLUS will provide better estimate of atmospheric temperature structures by observing H$_2$ lines, which are critical information for constraining molecular abundances with observations in other wavelengths \citep[e.g., near-infrared,][]{2018NatAs...2..420I,2019Icar..321..550I}. Reconstructed atmospheric elemental compositions would enable us to constrain formation and migration history of giant planets and, consequently, volatile delivery to terrestrial planets.


\clearpage
\section{Protoplanetary Disks}
\label{sec:protoplanetarydisks}

\noindent
\begin{flushright}
Shota Notsu$^{1,2}$, 
Hideko Nomura$^{3}$
\\
$^{1}$ RIKEN
$^{2}$ University of Tokyo
$^{3}$ NAOJ
\end{flushright}
\vspace{0.5cm}

\subsection{Scientific background and motivation}
Protoplanetary disks are rotating accretion disks surrounding young newly formed stars (e.g., T Tauri stars, Herbig Ae/Be stars). 
They are composed of dust grains and gas, and contain all the material that will form planetary systems orbiting main-sequence stars. They are active environments for  creation of simple and complex molecules, including organic matter and H$_{2}$O. 

In the hot inner regions of disks, H$_{2}$O ice evaporates from dust-grain surfaces. In contrast, it is frozen out on dust-grain surfaces in the outer cold parts of disks. The border of these two regions is the H$_{2}$O snowline \citep{Hayashi+1981}. Outside the H$_{2}$O snowline, the solid material is enhanced with the supply of H$_{2}$O ice. In addition, dust grains covered with water ice mantles can stick even at higher collisional  velocities and efficient coagulation are promoted. Thus, the formation of gaseous planetary cores is promoted in such regions, and we can regard the H$_{2}$O snowline in disk midplane divides the regions of gas-giant planet and rocky planet formation. Icy planetesimals, comets, and/or icy pebbles coming from outside the H$_{2}$O snowline may bring water to rocky planets including the Earth (e.g., \citealt{Morbidelli+2016, Sato+2016}).
 Recent theoretical studies (e.g., \citealt{Oka+2011, Mori+2021}) suggested that the location of the H$_{2}$O snowline will change, if we change the physical conditions such as the luminosity of the central star, the mass accretion rate, and the dust-grain size distribution in the disk.
 Thus, observationally locating the positions of the H$_{2}$O snowline in protoplanetary disks is important, since it will give constraints on the current formation theories of planetesimals and planets, and will help to clarify the origin of water on rocky planets including the Earth.

Recently, high-spatial resolution observations with the Subaru Telescope and ALMA have revealed the detailed substructures in the dust emission from the disks, and signs of planet formation have found (e.g., \citealt{Andrews+2018}). Molecular emission line observations using ALMA have also spatially resolved the positions of CO snowlines (e.g., \citealt{Qi+2013, Qi+2019}). CO evaporates from the dust-grain surfaces into the gas phase with lower desorption temperature than that of water, and thus the CO snowline is located at a larger radius, making it relatively easy to observe. Molecular emission lines of water from protoplanetary disks have been detected by the Spitzer Space Telescope and the Herschel Space Observatory in the mid- and far-infrared wavelengths, respectively (e.g., \citealt{Carr+2008, Hogerheijde+2011, Blevins+2016, vanDishoeck+2021}). However, these emission lines are mainly emitted from water molecules at the hot disk surface and at the cold outer disk outside the water snowline, and the position of the water snowline in the disk midplane has not been directly located.

The formation processes of large organic molecules, which could lead to the origin of life, is another interesting topic. In star-forming regions, complex organic molecules have been observed spectroscopically mainly in the (sub-)millimeter wavelengths (e.g., \citealt{Booth+2021, Yang+2021}), but the maximum number of atoms in each molecule is about 10 at most. On the other hand, various large organic compounds, including amino acids, have been detected in comets and meteorites in the solar system. It is still unclear how the large organic compounds found in the inner solar system objects are formed from molecules found in star-forming regions. Complex organic molecules are thought to be formed on dust-grain surfaces. Grain-surface reactions behave differently in the low-temperature and warm regions (e.g., \citealt{Walsh+2014, Notsu+2022}). At low temperatures, hydrogen diffuses mainly on the dust-grain surface and reacts with other grain-surface molecules to form hydrogenated molecules. In the warmer regions, some of the hydrogenated molecules are destroyed by UV radiation to form radicals, which react with each other to form more complex organic molecules. Investigating the distributions of organic molecules in the warm regions will lead to clarify the formation processes of prebiotic molecular species in the planet-forming regions.
The recent ALMA observations with high sensitivity have made it possible to detect complex organic molecules that were previously undetectable in protoplanetary disks (e.g., \citealt{Oberg+2015, Walsh+2016, Booth+2021, Brunken+2022}).
They trace the emission of organic molecular lines from the outer region in the disk (around a few tens of au), where the cold dust-grain surface reactions are active, as well as the emission from the inner hot region of the disks around Herbig Ae/Be stars.

Here, we propose to detect water snowlines in various protoplanetary disks by survey observations of water emission lines using GREX-PLUS in order to elucidate the evolution of water snowline positions.
In addition, we also propose to observe complex organic molecular lines emitted from the inner region of the disks in order to constrain their formation processes in planet forming regions.

\subsection{Required observations and expected results}
Based on detailed physical and chemical structure models of protoplanetary disks, we investigated the candidate water emission lines that trace water snowline positions in the disks \citep{Notsu+2016, Notsu+2017, Notsu+2018}. We found that emission lines with small Einstein A coefficients ($\sim10^{-6}-10^{-3}$) and relatively high excitation energies ($\sim1000$ K) are suitable for identifying the location of the water snowline near the equatorial plane of the disk. Such water emission lines are widely distributed from the mid-infrared to sub-millimeter wavelengths, in the GREX-PLUS wavelength coverage, the water emission line at 17.75 $\mu$m can be used to trace the position of the water snowline. The water snowline in a disk with a central star of solar mass is difficult to observe with spatially resolved imaging even with ALMA. However, because of the Keplerian rotation of the disk, the emiting region can be identified by analyzing the Doppler-shifted emission profiles, even if spatially unresolved. 

\begin{figure}[h!]
\centering
\includegraphics[width=0.7\linewidth]{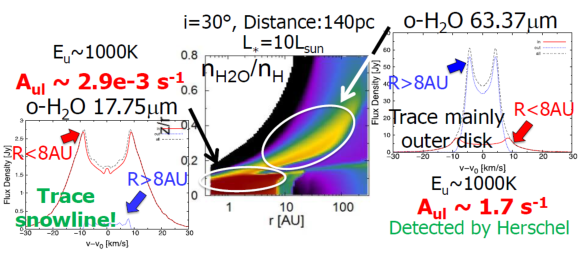}
\includegraphics[width=0.25\linewidth]{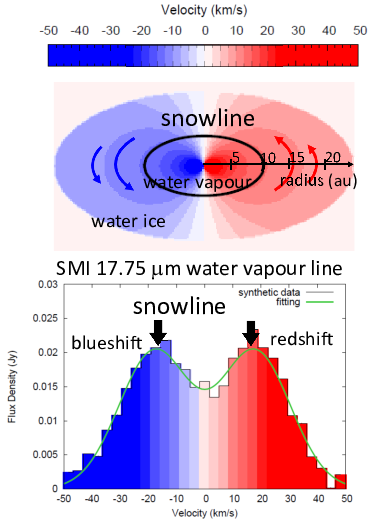}
\caption{({\it Left}) Model calculations of the distribution of gas-phase water in a protoplanetary disk and their emission line spectra \citep{Notsu+2017}. The 63.37 $\mu$m emission lines detected by Herschel mainly trace the hot surface layer of the outer disk, whereas the 17.75 $\mu$m emission lines, observable by GREX-PLUS, trace the position of the water snowline. ({\it Right}) By analysing emission profiles that trace the Doppler shift due to the Keplerian rotation, the emission region, i.e. the water snowline, can be located using high spectral resolution of R=29,000 even if it cannot be spatially resolved \citep{Kamp+2021}.}
\end{figure}

Simulated observations based on model calculations show that the location of the water snowline can be identified from the line profiles with a wavelength resolution of R=29,000. We also investigated the sensitivity required for detection using disk models with various physical quantities such as dust size, disk mass, and central star luminosity \citep{Kamp+2021}. Using a model with typical physical quantities, water line emission from a disk with a central star of about solar mass in a star-forming region near the solar system (d = 150 pc) can be detected by GREX-PLUS with an integration time of 50 min. The water line emission from a disk with a central star of 2-3 times the solar mass can be detected by GREX-PLUS with an integration time of about 30 minutes, even in star-forming regions in giant molecular clouds at a distance of d=420 pc from the solar system. We expect to detect target water emission lines with JWST's medium-dispersion spectroscopic observations, and to identify water snowline positions by analyzing the emission line profiles obtained with GREX-PLUS high-dispersion spectroscopic observations.

The GREX-PLUS's wavelength bands includes transition lines of complex organic molecules such as CH$_{3}$CN and CH$_{3}$OH as well as small organic molecules such as HCN and C$_{2}$H$_{2}$. Model calculations of CH$_{3}$CN molecular emission lines from a disk with a central star of solar mass show that at the distance of TW Hya (d=60 pc), emission lines can be detected with an integration time of a few hours. The GREX-PLUS observations contribute to clarify the composition of organic molecules formed by the grain-surface reactions in the warm region, which was difficult to detect by ALMA. Emission lines of small organic molecules such as HCN and C$_{2}$H$_{2}$ have already been detected with the Spitzer Space Telescope (e.g., \citealt{Carr+2008, Pontoppidan+2010}). The analyses of the emission line profiles obtained by GREX-PLUS's high dispersion spectroscopy is expected to provide information on the spatial distribution of the C/O and N/O elemental composition ratios in the disk. Since water is a major carrier of oxygen elements both in gas and icy-phase, the C/O elemental composition ratio in the gas phase (and also solid phase) is expected to change significantly inside and outside the water snowline. On the other hand, the C/O elemental composition ratios in the gas and solid phases are also affected by chemical reactions. Since the elemental compositions of gas-giant planets are considered to reflect the elemental compositions of protoplanetary disks, statistically investigating the spatial distribution of the elemental composition ratios in disks and comparing with those of short-period gas-giant planets will provide constraints on where in the disks such gas-giant planets formed (e.g., \citealt{Notsu+2020, Notsu+2022}).

\subsection{Scientific goals}
The line survey observations will cover about 100 protoplanetary disks in the relatively nearby solar-mass and intermediate-mass star forming regions such as Taurus, $\rho$ Oph, Chameleon, Upper Sco, Orion, Perseus, Serpens, Aquila, etc., for total observation time of about 300 hours. We will study the time evolution of the position of water snowlines and clarify the formation processes of planetary systems and how to supply water and organic molecules to rocky planets. In addition, we will detect small organic molecules such as HCN and C$_{2}$H$_{2}$ and complex organic molecules (CH$_{3}$CN and CH$_{3}$OH) in the nearby disks. This will reveal the distribution of elemental compositions such as C/O in planet forming regions and the complex formation process of organic molecules on warm grain-surfaces, and discuss the formation process of short-period gaseous planets and molecular species originating life. Thanks to the wide wavelength coverage (10--18 $\mu$m) which include water and various target molecular lines, we expect to achieve these science goals from the same line survey observations.

\begin{table}[h]
    \label{tab:protoplanetarydisks}
    \begin{center}
    \caption{Required observational parameters.}
    \begin{tabular}{|l|p{5cm}|l|}
    \hline
     & Requirement & Remarks \\
    \hline
    Wavelength & 10--18 $\mu$m & The target water line is at 17.75 $\mu$m \\
    \cline{1-2}
    \hline
    Wavelength resolution & $\lambda/\Delta \lambda>29,000$ & To resolve the double peaks of line profiles \\
    \hline
    Sensitivity & $<$ 5 mJy (5$\sigma$, 1h) & \\
    \hline
    Observing field & Nearby star-forming regions with $d<500$ pc &  \\
    \hline
    \end{tabular}
    \end{center}
\end{table}

\clearpage
\section{Interstellar Molecules}
\label{sec:interstellarmolecules}

\noindent
\begin{flushright}
Yasuhiro Hirahara$^{1}$
\\
$^{1}$ Nagoya University
\end{flushright}
\vspace{0.5cm}

\subsection{Scientific background and motivation}

high resolution spectroscopic observations provide useful information about the physical and chemical conditions in the interstellar medium (ISM), such as interstellar molecular clouds and circumstellar envelopes. So far, more than 270 interstellar molecular species have been identified, showing the complex chemical processes in the variety of ISM. For the identifications of the interstellar molecules, contributions of the pure rotational molecular transition in the microwave region were the most prominent both in the spectroscopic observations and the laboratory experiments, especially for the short-lived species, such as ionic and radical species. In the mm- and submm- wavelength region, so called the radio region, the heterodyne techniques make it quite easily to achieve high spectral resolution, which is the most important for the precise assignment of the spectra. Rotational transitions can only be observed for the gas-phase molecules with permanent electric dipole moment. A few molecules are identified in ISM through their electronic transitions mostly in the optical and the ultraviolet region. In general, molecules except radical species have dissociative excited states, which makes it rare and difficult to observe their electronic transitions.

Since all molecules have vibrational transitions mostly observed in the infrared (IR) wavelength regions, the IR spectroscopic observations has a significant advantage. Many infrared spectrum database are published and updated especially for the chemical analyses of the complex organic molecules. In particular, the mid-infrared region is called the ``fingerprint region'' because transitions accompanying vibrations of various functional groups of organic materials appear in this region, and hence it is a useful band used in spectroscopic analysis to determine the structure of the carbon skeleton. Besides the organic compounds, Si-O skeletal vibration modes in various silicates also appear in this region, providing a wealth of information on the structure of mineral or amorphous solid silicates. When focusing on the observations of ISM, the infrared vibration spectroscopy is the unique method for their identification for the non-polar gas-phase molecules, such as H${_2}$, CH${_4}$, CO${_2}$ and SiH${_4}$, being complementary to the radio observation.

Spectroscopic observations, however, in the infrared region to date have detected only about 30 species of molecules, many of which were identified earlier in the rotational spectra in the radio frequency region. Only 25 chemical species, shown in Table 3.2, have been detected first in infrared observations.  The main reason for the paucity of past studies lies in the difficulty of high-sensitivity and high resolution spectroscopic observations at wavelengths longer than 8-13 $\mu$m (N-band) due to deep absorption by the Earth's atmosphere. In fact, as shown in Table 3.2, about half of the interstellar molecules have been discovered by space telescopes equipped with medium or low wavelength resolution spectrometers in the infrared region. Therefore, GREX-PLUS with unprecedentedly high wavelength resolution spectrometer will contribute to the progress of interstellar chemistry.

\begin{table}[ht]
    \label{tab:listofmolecules}
    \begin{center}
    \caption{Interstellar molecules detected by infrared spectroscopic observation
    (underline: first detection by IR space telescopes).}
    \begin{tabular}{ll}
    \hline \hline
    Type & Species \\
    \hline
    Simple hydrides, inorganic species & H$_2$, CH$_4$, CO$_2$, SiH$_4$ \\
    Aromatic molecules & \underline{C$_6$H$_6$ (benzene) [ISO], C$_{60}$, C$_{70}$ [Spitzer]} \\
    Linear carbon chains & C$_3$, C$_5$, C$_2$H$_4$, C$_2$H$_2$, C$_4$, C$_6$, \underline{HC$_4$H, HC$_6$H [ISO]} \\
    Molecular ions & H$_3$$^+$, HeH$^+$, \\
    & \underline{H$_2$O$^+$, H$_3$O$^+$, OH$^+$, SH$^+$, ArH$^+$, HCl$^+$, H$_2$Cl$^+$ [Herschel]} \\
    Radicals & CH$_3(^2A_2")$ \\
    \hline
    \end{tabular}
    References: [Spitzer]:\citet{2010Sci...329.1180C}, [ISO]:\citet{2001ApJ...546L.123C},
    
    [HERSHEL]:\citet{Indriolo_2015}, \citet{lis:hal-00633439}, \citet{2010A&A...521L..35B}
    \end{center}
\end{table}

The infrared vibration transitions of the target interstellar molecular species can be observed by resolving their rotational structures, and the wavelength resolution necessary to determine the molecular species assignment was investigated. As an example, we show the result for benzene, which was first detected by the infrared astronomical satellite ISO (Infrared Space Observatory) in Figure 3.3.

\begin{figure}
    \centering
    \includegraphics[width=14cm]{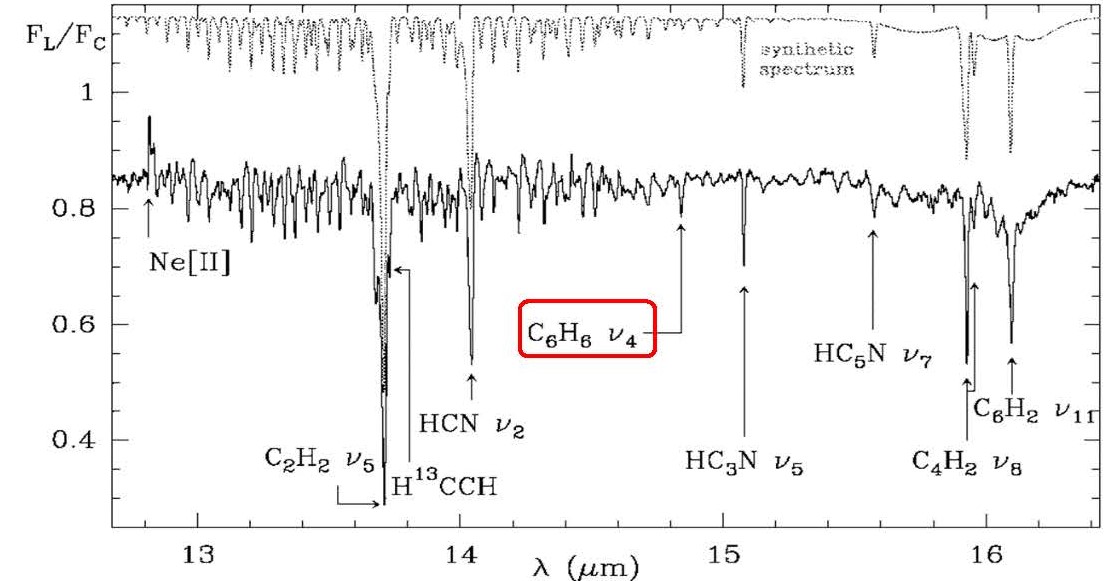}
    \caption{The infrared absorption spectrum of benzene (C$_6$H$_6$) and polyynes (C$_4$H$_2$, C$_6$H$_2$) detected for the first time, by the SWS (Short Wavelength Spectrometer) of ISO (Infrared Space Observatory) with $R\sim2,000$ toward the proto-planetary nebula CRL618 \citep{2001ApJ...546L.123C}.}
    \end{figure}

\underline{Science case I: Interstellar aromatic chemistry}

By the spectroscopic observation with the ISO short wavelength spectrometer (SWS, $R\sim2,000$), only the Q--branch ($\Delta J=0$) of the $\nu$$_4$ vibration rotational transition of C$_6$H$_6$, an oblate symmetric top molecule, is prominent, and the molecular species may not be reliably assigned. We calculated spectral shape of C$_6$H$_6$ for various wavelength resolutions by adopting the molecular constants for benzene \citep{DangNhu1989SpectralII}. As shown in Figure 3.4($\it{right}$), it is found that  all P-- and R-- branches of benzene $\nu$$_4$ band can be fully resolved for the resolution $R\geq28,000$. The high resolution observation for benzene will also contribute to increase the intensity of the spectra by a factor of $\sim5$, allowing precise estimation of the rotational excitation temperature of the molecular species. Therefore, mid-IR high resolution spectroscopic surveys of benzene and related small aromatic molecules as described below will contribute to the understanding of the material cycle and evolution in the universe. In the early model calculation for the synthesis mechanism \citep{woods2002synthesis}, benzene in CRL-618 can be produced through the ``bottom-up'' scheme with successive ion-molecule reactions of acetylene (C$_2$H$_2$) and protonated acetylene (C$_2$H$_3$$^+$), yet undetected ionic species which may be produced effectively in proto-planetary nebulae by the proton exchange reactions between (C$_2$H$_2$) and HCO$^+$. In their model, benzonitrile (C$_{6}$H$_{5}$CN) is predected to be produced with high abundance similar to benzene. Interestingly, the first detection of benzonitrile in interstellar media has been made in the low temperature dark cloud TMC-1 ($T$$_k$$\sim10$K) by the single-dish radiotelescopes of GBT 100m and NRO 45m \citep{2018Sci...359..202M}.

\begin{figure}
    \centering
    \includegraphics[width=14cm]{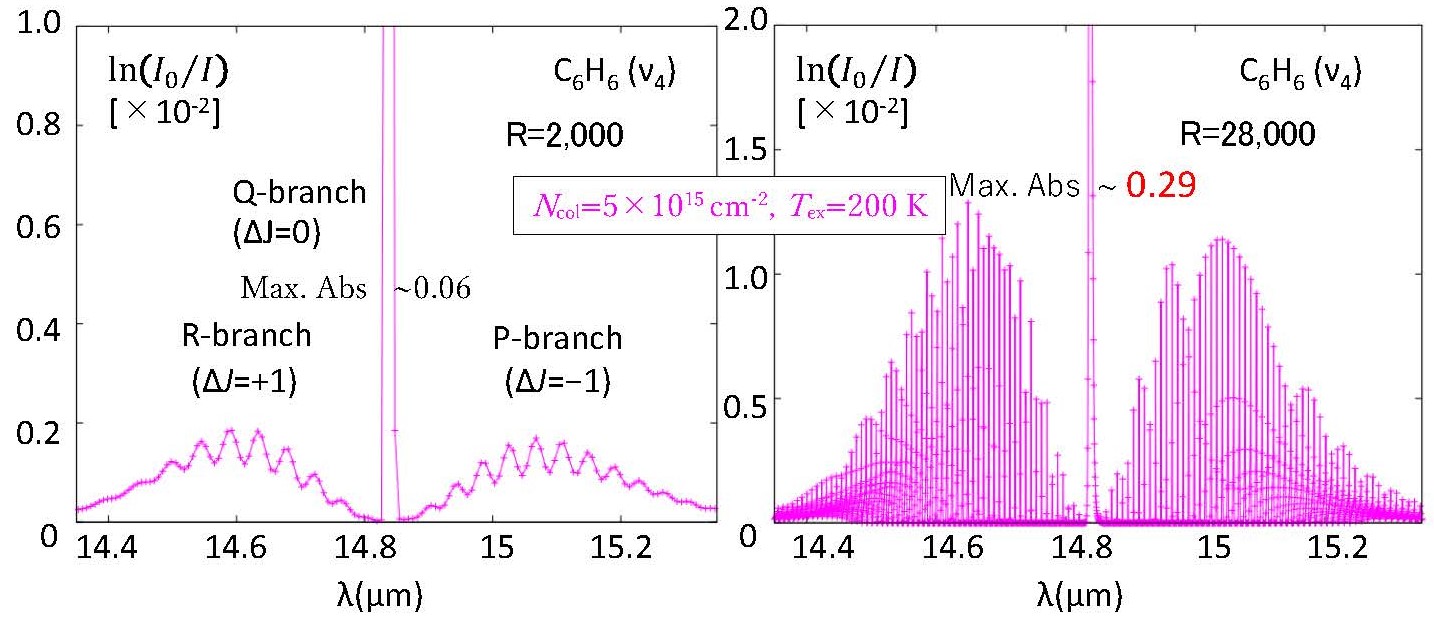}
    
    \caption{Simulated IR high-resolution spectrum of benzene (C$_6$H$_6$) $\nu$$_4$ vibration band in the case of the column density $\it{N}$$_{col}$=5$\times$10$^{15}$cm$^{-2}$ and $\it{T}$$_{ex}$=200K, with the wavelength resolution $\it{R}$=2,000 ($\it{left}$) and 28,000 ($\it{right}$) \citep{2017GradThesis.Sci.Nagoya-U.}.}
    \end{figure}

Followed by the successful detection of benzonitrile,  the first detection of the two isomeric aromatic compounds 1-- and 2--cyanonaphthalene (C$_{10}$H$_{7}$CN) in TMC-1 was reported by submillimeter-wave spectroscopy with the ALMA telescope \citep{2021Sci...371.1265M}. Soon after the discovery, the new aromatic molecules, as illustrated in Figure 3.5, ethynyl cyclopropenylidene (c-C$_{3}$HCCH), cyclopentadiene (c-C$_{5}$H$_{6}$), indene (c-C$_{9}$H$_{8}$), and ortho-Benzyne (o-C$_{6}$H$_{4}$) have been found in TMC-1 by using the Yebes 40 m single-dish radiotelescope (\citealp{2021A&A...649L..15C}, \citealp{2021A&A...652L...9C}).  Since the abundance of indene in TMC-1 is as high as 1.6 ${\times}$ 10$^{-9}$, the formation mechanism or the origin of the aromatic hydrocarbon molecules should be reconsidered.

\begin{figure}
    \centering
    \includegraphics[width=14cm]{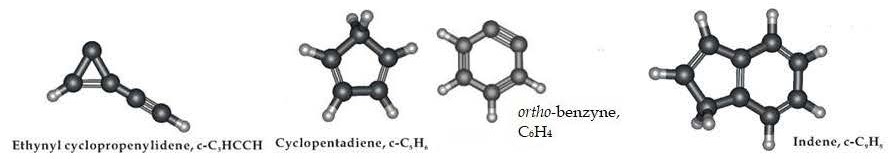}
    \caption{Molecular structures of aromatic hydrocarbons recently detected in the cold dark cloud TMC-1 (\citealp{2021A&A...649L..15C}, \citealp{2021A&A...652L...9C}).}
\end{figure}

On the other hand, the ``ultimate aromatic molecules'', fullerenes (C$_{60}$ and C$_{70}$), were detected by infrared spectroscopic observation with $R\sim500$ by the Spitzer Space Telescope toward the hydrogen-deficient young planetary nebula atmosphere Tc1, as shown in Figure 3.6 \citep{2010Sci...329.1180C}. Meanwhile, spectra of fullerenes have been detected toward the molecular cloud associated with the reflection nebulae NGC2023 \citep{2010ApJ...722L..54S} and NGC7023 \citep{2012PNAS..109..401B}. At present, spectra of fullerenes have been identified in many planetary and proto-planetary nebulae not only in our galaxy but also in the Small Magellanic Cloud, mainly based on the detailed analysis of the Spitzer archive data \citep{2014MNRAS.437.2577O}.

By applying a thermal emission model, \citet{2010Sci...329.1180C} obtained low and different excitation temperature, 330 K and 180 K for C$_{60}$ and C$_{70}$ in Tc1, respectively, suggesting that emission of fullerenes does not originate from free molecules in the gas phase, but from molecular carriers attached to solid material. Even if the attachment to the host solid is not strong enough for hindering rotational motion of guest fullerene molecules, it may be only possible to decompose the sharp Q-- branch features for the 17 and 19 $\mu$m bands of C$_{60}$ and 15 and 16 $\mu$m bands of C$_{70}$ by high wavelength resolution observation, as in the case of benzene spectrum by ISO. In theory, C$_{60}$ is a symmetric top rotor with a small rotational constant of B $\approx$ 0.0028 cm$^{-1}$, with nearly spherical shape with icosahedral ($\it{I_h}$) symmetry. Since all 60 carbon nuclei have zero spin for $^{12}$C$_{60}$, boson-exchange symmetry restrictions require that many of the rotational quantum levels disappear. These predicted ``missing'' levels result in a characteristic pattern of spectral line spacings may be a key to observing and assigning a rotationally resolved C$_{60}$ spectrum. In spite of the small B values, the spacings between the rotational levels are thin out for the P- and Q- branches of $^{12}$C$_{60}$. However, such  simplification in rotational structure only occurs for $^{12}$C$_{60}$. The relative concentrations of $^{12}$C$_{60}$, $^{13}$C$^{12}$C$_{59}$, and $^{13}$C$_{2}^{12}$C$_{58}$ are $\approx$ 0.51, 0.34, and 0.11, respectively, even when the isotopic abundance ratio [$^{13}$C]/[$^{12}$C] is as low as 0.011 (terrestrial value).  Since nearly half of the C$_{60}$  molecules will include at least one $^{13}$C atom, vibration spectrum of C$_{60}$ in the low temperature gas in the ISM, even if exist, will be seriously complicated.

Since the first discovery, the formation mechanism of fullerenes are the intriguing subject of discussion. According to the result of the chemical model calculation involving the UV irradiation from the PDR region \citep{2016A&A...588C...1B} , the ``top down'' scheme where larger carbon clusters shrink to reach C$_{60}$ may be more plausible than the traditional ``bottom-up'' approach. In the ``bottom-up'' scheme, the production of large aromatic molecules are built up from small molecules such as benzene and other aromatic molecules shown in Figure 3.5, and also as simpler hydrocarbon ions. Inspired by the successive detections of cyano-substituted aromatics in the cold molecular as described above, \citet{refId0} conducted the UV and VUV photodissociation experiment of  protonated benzonitrile (C$_{6}$H$_{5}$CNH$^{+}$). They found that the primary dissociation channel is the phenylium cation (C$_{6}$H$_{5}^{+}$) which is higyly reactive species to produce larger polycyclic aromatic hydrocarbons. The high wavelength resolution spectroscopic observation in 10--20 ${\mu}$m region toward PDRs may clarify the formation mechanism, by the new detections of pure non-polar aromatic hydrocarbons in the gas phase, size of which are in between benzene and fullerenes.  $\it{ex.}$ naphtalene (C$_{10}$H$_{8}$) and its cation \citep{doi:10.1021/j100199a011}, and phenalenyl radical (C$_{13}$H$_{9}$$\cdot$) \citep{doi:10.1021/ja206322n}, an oblate symmetric top, ($\it{D}$$_{3h}$) showing simple rotational structure for the high resolution vibration spectrum.  

\begin{figure}
    \centering
    \includegraphics[width=13cm]{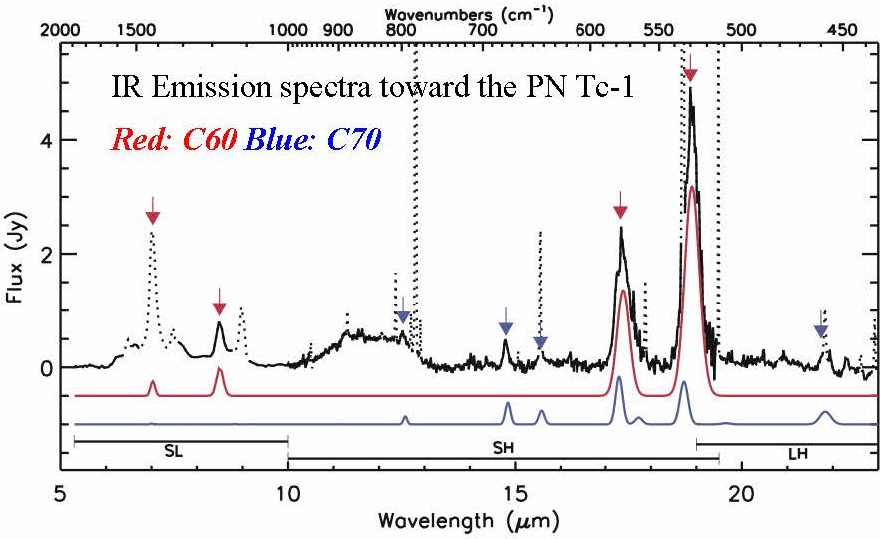}
    \caption{Infrared emission spectra of C$_{60}$ and C$_{70}$ toward the young planetary nebula Tc-1, continuum subtracted spectrum between 5 and 23 ${\mu}$m, by the Spitzer IRS \citep{2010Sci...329.1180C}.
The red and blue curves below the data are thermal emission models for all infrared active bands of C$_{60}$ and C$_{70}$ at temperatures of 330 K and 180 K, respectively.}
\end{figure}

\underline{Science case II: Search for the ``key carbo-ions'' for interstellar chemistry}

Figure 3.7 shows an exerpt of conventional ``build-up'' ion-molecule reaction network for the formation of organic molecules starting from simple molecules in an interstellar molecular cloud at low temperatures ($T\sim10K$) with hindered UV penetration to invoke fast ionization processes \citep{1992ChRv..92.1473,2013ChRv..113.8710A}. The hydrogen molecular ions (H$_2$$^+$) produced by high-energy cosmic rays quickly react with H$_2$ to form stable H$_3$$^+$, which then becomes a ``seed'' for the formation of more complex molecules via various ion-molecule reactions. Since the ``terminal ions'': CH$_3$$^+$, C$_2$H$_2$$^+$, C$_2$H$_3$$^+$, and ``the smallest aromatic hydrocarbon'', cyclic-C$_3$H$_3$$^+$ (cyclopropenyl cation) are very slow to react with ambient H$_2$, they react with atoms and molecules other than H$_2$ ($\it{ex}$. C, N, CO, H$_2$O, HCN) and electrons. In particular, ``carbo-ions'', as named by \citet{doi:10.1098/rsta.1988.0002}, at the branch points of the reaction network are predicted to be abundant, but all are still undetected yet in ISM. The performance of comprehensive exploratory observations in interstellar space under various environments is an important issue for the validation of reaction network models for interstellar chemistry. The terminated carbo-ions listed in Table 3.3 are chemically stable, highly symmetric in molecular structure, and thus have no or very small permanent dipole moment.

It is worth pointing out that the deep spectral searches for the carbo-ions by vibrational transition are of importance for the advance of laboratory astrophysics and astrochemistry.
As the earlier studies, the laboratory high-resolution absorption spectroscopy of CH$_3^+$($\nu_3$), C$_2$H$_2$$^+$($^2\Pi_u$)($\nu_3$) and C$_2$H$_3$$^+$($\nu_6$) have been reported uniquely by Prof. Takeshi Oka’s group, using difference laser spectroscopy around the 3.2 $\mu$m region (\citealp{doi:10.1063/1.454194} \citealp{doi:10.1063/1.451929}, \citealp{doi:10.1063/1.457612}). The resolution of the measured spectra are high enough to decompose rotational structure which is useful for the astronomical searches. Later, laboratory measurements for the vibration transitions lying ${\lambda}$\textgreater10 $\mu$m for the carbo-ions are conducted by \citet{2005PhRvL..94g3001A} for C$_2$H$_2$$^+$($^2\Pi_u$)($\nu$$_9$), and \citet{MARIMUTHU2020111377}, by applying the tunable mid-IR laser for photo dissociation and ion trap for the detection of fragment molecular ions. It is noted that the measured vibration spectra are highly sensitive but low wavelength resolution for clarifying rotational structures. Since the molecular rotational constants for carbo-ions listed in Table 3.3 is large, it is expected to obtain the best and first information on their intriguing molecular structures, by the GREX-PLUS high resolution observations with ${R\geq20,000}$.

\begin{figure}
    \centering
    \includegraphics[width=14cm]{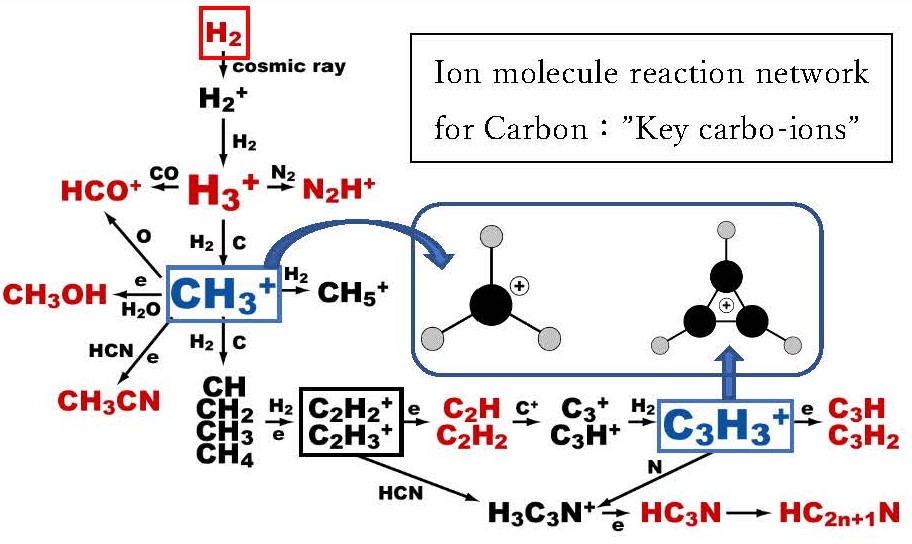}
    \caption{A chemical reaction network for the formation of organic molecules from simple molecules in an interstellar molecular cloud at low temperatures ($T\sim10K$)
    (\citealp{1992ChRv..92.1473},\citealp{2013ChRv..113.8710A}).
    }
\end{figure}

\begin{table}[ht]
    \label{tab:carboions}
    \begin{center}
    \caption{List of target key carbo-ions for spectroscopic search observation.}
    \begin{tabular}{lllll}
    \hline \hline
    Species & Transition & Wavelengh & Reference & Note\\
    \hline
    CH$_3$$^+$ & $\nu$$_3$, $\nu$$_4$ & 7.6$\mu$m & \citet{1985JCP...82...1...333} & --no experimental data--\\
    C$_2$H$_2$$^+$($^2\Pi_u$) & $\nu$$_9$ & 11.8$\mu$m & \citet{2005PhRvL..94g3001A} & Laser induced reaction + ion trap \\
    C$_2$H$_3$$^+$ & $\nu$$_7$ & 13.2$\mu$m & \citet{1986JChPh..85.3437L} & --no experimental data--\\
    C$_3$H$_3$$^+$ & $\nu$$_7$ & 11.0$\mu$m & \citet{MARIMUTHU2020111377} & IRPD spectoroscopy \\
    \hline
    \end{tabular}
    \end{center}
\end{table}

\subsection{Required observations}

\begin{table}[ht]
    \label{tab:ISMrequirements}
    \begin{center}
    \caption{Required observational parameters.}
    \begin{tabular}{|l|p{9cm}|l|}
    \hline
     & Requirement & Remarks \\
    \hline
    Wavelength & 11--18 $\mu$m & \\
    \hline
    Spatial resolution & $<3$ arcsec at ${\lambda=15\mu}$m &\\
    \hline
    Wavelength resolution & $\lambda/\Delta \lambda>30,000$ & \\
    \hline
    Field of view & N/A & \\
    \hline
    Sensitivity & Line sensitivity ${5\sigma -- 1 hour}$ & \\
    \hline
    Observing field & Our Galaxy & \\
    \hline
    Observing cadence & N/A & \\
    \hline
    \end{tabular}
    \end{center}
\end{table}

\clearpage
\section{Exoplanet Atmosphere}
\label{sec:exoplanetatmosphere}

\noindent
\begin{flushright}
Yuka Fujii$^{1}$, 
Yui Kawashima$^{2}$,
Taro Matsuo$^{3}$, 
Kazumasa Ohno$^{1,4}$
\\
$^{1}$ NAOJ,
$^{2}$ RIKEN, 
$^{3}$ Nagoya University
$^{4}$ University of California, Santa Cruz
\end{flushright}
\vspace{0.5cm}

\subsection{Scientific background and motivation}

Discoveries of exoplanets since 1990s have revolutionized our view of the universe by revealing the ubiquity of planetary systems and the wide diversity in their architectures. 
Understanding the trend and the origins of the diversity is crucial for the unified picture of planetary system formation that puts the solar system into a context. 

For this end, characterization of their atmospheric properties by detailed follow-up observations is expected to provide valuable clues. 
This is because the atmospheric composition of gaseous planets reflects their formation environment. For example, since the gaseous C/O ratio in the protoplanetary disk is expected to vary within the disk \citep[e.g.,][]{2011ApJ...743L..16O, Eistrup+16, Booth&Ilee19,Notsu+2020,Notsu+2022,Schneider&Bitsch21}, we can estimate where in the disk the planet captures its atmosphere by measuring the C/O ratio of the atmosphere.
So far, C/O ratios of the planetary atmospheres have been investigated mainly by constraining the abundance ratios of CO/$\mathrm{CH_4}$ to $\mathrm{H_2O}$ through low-resolution spectroscopy \citep[e.g.,][]{2011Natur.469...64M, 2016AJ....152..203L, 2020A&A...642A..28M}.
This is because those species are the main carbon- and oxygen-bearing molecules in the hydrogen-dominated atmospheres \citep[e.g.,][]{Lodders&Fegley02,Moses+13} and have prominent absorption features in near infrared wavelength accessible with the current observations.
While less abundant than CO, $\mathrm{CH_4}$, and $\mathrm{H_2O}$, there are also other good tracers of C/O ratio, such as $\mathrm{CO_2}$, HCN, and $\mathrm{C_2H_2}$, which are produced photochemically in the upper atmosphere \citep[e.g.,][]{2014RSPTA.37230073M, 2019ApJ...877..109K}.
These photochemical species can provide insights not only on C/O ratio but also the UV irradiation intensity from the host star, which plays vital roles in both atmospheric chemistry and planetary evolution accompanied with atmospheric escape.
However, their low absolute abundances hamper the detection by low-resolution spectroscopy.

In addition to C/O ratio, recent studies suggested that atmospheric N/O ratio also provides strong constrains on the planet formation environment \citep{Piso+16,Cridland+20,Ohno&Ueda21,Notsu+2022}.
This is because main nitrogen reservoirs in protoplanetary disk, namely NH$_3$ and N$_2$, have an order-of-magnitude difference between their abundances \citep{Oberg&Bergin21}, which produces a drastic radial variation of N/O ratio in the disk bordering across N$_2$ snowline.
Based on the atmospheric nitrogen abundance, several studies have suggested that solar system Jupiter might originally form at extremely cold environments of $<30~{\rm K}$ \citep{Owen+99,Oberg&Wordsworth19,Bosman+19,Ohno&Ueda21}.
In exoplanetary atmospheres, NH$_3$ and HCN provide a way to constrain the atmospheric nitrogen abundance \citep{Macdonald&Madhusudhan17,Ohno&Fortney22a,Ohno&Fortney22b}. 
However, low-resolution spectroscopy has not conclusively detected NH$_3$ and HCN in exoplanets.
This is possibly owing to the overlaps of their near infrared spectral features with those of O- and C-bearing species and prevalence of photochemical hazes in warm exoplanets that mute absorption features of gas molecules \citep[e.g.,][]{Gao+20,Dymont+21}.

To overcome the difficulty of the current observations of exoplanetary atmospheres, high resolution spectroscopy allows robust detection by distinguishing the feature of interest from the other molecules' features.  
Indeed, aforementioned minor species have recently started to be found in exoplanet atmospheres by high resolution spectroscopy with ground-based telescopes, which has advanced our understanding of the C/O ratios of exoplanetary atmospheres \citep{2021Natur.592..205G}.
The species of interest, especially C$_2$H$_2$, HCN, and NH$_3$, have stronger absorption features in the mid-infrared wavelength region, where significant telluric absorption hampers the observation from the ground.
Thus, utilizing the high resolution spectrograph (10--18~$\mu$m) mounted on GREX-PLUS, we can expect robust detection of the above species, 
which further advance our understanding of the atmospheric C/O ratio, N/O ratio, vertical mixing, and UV environment.

Due to the mid-infrared bandpass, high resolution spectroscopy of GREX-PLUS is also sensitive to the spectral features in thermal emission, for a wide temperature range of planets down to $\sim $500~K. 
The coolest ones are accessible only in the mid-infrared where their thermal emission is in the Jeans domain and the planet-to-star flux ratio becomes larger. 
high resolution spectroscopy of their thermal emission will be able to constrain the composition and the thermal structure of the atmospheres altogether with high fidelity for the first time. 
Furthermore, the high resolution spectroscopy allows us to resolve the Doppler shift of planetary thermal emission due to planetary orbital motion, which measures the radial velocity of planets. 
This suggests that planetary spectral features can be separated from that of the host star in a model-independent manner even for non-transiting (non-eclipsing) planets. 
Radial velocity measurements of non-transiting planets can  constrain that orbital inclination and hence their ``true'' masses. 
The fact that GREX-PLUS can probe non-transiting planets is  advantageous over JWST in probing relatively cool targets,  because the transit probabilities rapidly decreases as a function of the distance from the host star. 
The unique parameter space that mid-infrared high resolution spectroscopy of GREX-PLUS can probe will fill the gap between the two major categories of planets for which atmospheric observations are/will be actively performed: close-in transiting planets and more distant directly-imaged planets. 
While challenging, the radial velocity measurements of Jupiter-like planets also serve as a unique probe of the presence of (large) satellites or binary planets. 

The potential targets also include cooling gas giants in distant orbits that are still illuminating due to their intrinsic energy gained through gas accretion. Both the "hot-start" and "cold-start" models \citep[e.g.,][]{2007ApJ...655..541M} predict that these self-luminous gas giants with ages of less than 100 Myrs have the effective temperatures of more than 500K. The self-luminous gas giants will be excellent targets for GREX-PLUS in terms of improving the signal-to-noise ratio. Before the era of GREX-PLUS, several space telescope missions, such as JWST and Nancy Grace Roman Space Telescope (hereafter {\it Roman}) will search for self luminous gas giants around nearby moving groups at ages of 10 - 100 Myrs \citep{2020AJ....159..166U}. Note that the memberships of the nearby moving groups are well characterized by the high-precision proper-motion measurements of GAIA. 
For example, the Mid-InfraRed Imager (MIRI) of JWST provides several coronagraphic capabilities in the mid-infrared regime for the first time, which are sensitive to warm and temperate gas giants in the outer planetary system \citep{2021MNRAS.501.1999C}. 
In addition, {\it Roman} looks for the reflected light from young gas giants close to the central stars (less than 5 AU), thanks to the small inner working angle \citep{2018SPIE10698E..2IM}. 
Since both missions perform only high-contrast imaging (not high-contrast spectroscopy), GREX-PLUS will build good synergies with these missions.


\subsection{Required observations and expected results}

\subsubsection{Transmission spectra}

There could be two approaches to the observation of exoplanetary atmospheres using the high resolution spectograph of GREX-PLUS.
One is the transmission spectroscopy for transiting planets, which measures the absorption due to planetary atmosphere during the transit compared to the stellar spectrum outside the transit.
The cross-correlation method is often used to investigate the existence of chemical species in exoplanetary atmospheres \citep[e.g.,][]{2013MNRAS.436L..35B, 2017AJ....154..221N}.
This method calculates the cross-correlation between the observed spectrum and the template theoretical spectrum that predicts the wavelength position and magnitude of absorption lines for a specific chemical species. 
One can infer the presence or absence of each chemical species by looking at the goodness of the cross-correlation between the observed and template spectra.
In the transmission spectroscopy, the magnitude of the spectral features are scaled by atmospheric scale height. 
Thus, our prime targets are the planets with sizes ranging from that of Jupiter to Neptune that likely have hydrogen dominated atmospheres with large scale heights. 

\begin{figure}
    \centering
    \includegraphics[width=\textwidth]{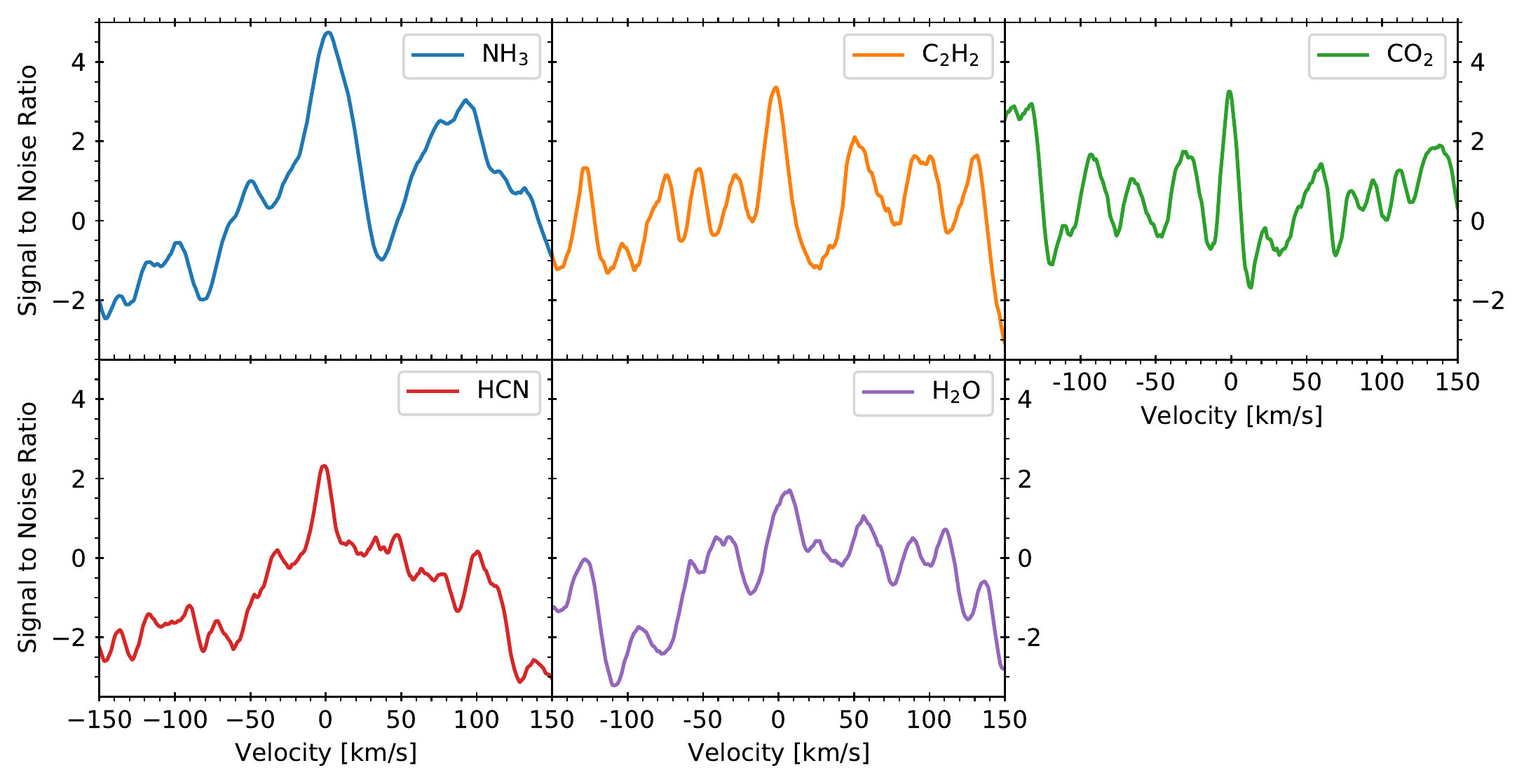}
    \caption{The results of the cross-correlation between the mock and template spectra for the case of 20 transit observations of a mini-Neptune planet GJ 1214b by GREX-PLUS.}
    \label{fig:ccf_transit}
\end{figure}

As an example,  we simulate the transit observation of a mini-Neptune planet GJ 1214b orbiting an M4.5 type host star at 14.65~pc by GREX-PLUS.
The transit duration is about one hour, and we consider the case where we observe 20 transits.
For the spectrum model, we have used the fiducial calculation case of \citet{2019ApJ...877..109K}. 
The abundance profiles of chemical species are calculated with the photochemical model of \citet{2018ApJ...853....7K}, assuming solar elemental abundance ratios and constant eddy diffusivity of $10^7$~$\mathrm{cm^2/s}$ throughout the atmosphere. 
As for the temperature profile, the analytical formula of \citet{2010A&A...520A..27G} is used. For more details, see \citet{2019ApJ...877..109K}.
In Figure~\ref{fig:ccf_transit}, we show the result of the cross-correlation between the mock and template spectra for species $\mathrm{NH_3}$, $\mathrm{C_2H_2}$, $\mathrm{CO_2}$, $\mathrm{HCN}$, and  $\mathrm{H_2O}$.
The detection significances are 4.7, 3.4, 3.3, 2.3, and 1.7$\sigma$ for $\mathrm{NH_3}$, $\mathrm{C_2H_2}$, $\mathrm{CO_2}$, $\mathrm{HCN}$, and  $\mathrm{H_2O}$, respectively.
(We note that for $\mathrm{H_2O}$, the existence of its absorption features in the wide range of the observational wavelength makes it somewhat challenging to detect by cross-correlation technique despite its large abundance.)
Thus, the high resolution transmission spectroscopy with the GREX-PLUS opens a new window to detect minor chemical species, such as NH$_3$, C$_2$H$_2$, and HCN, which have not been robustly discovered by the current low-resolution spectroscopy.

\subsubsection{Thermal emission spectra}

Another approach to exoplanet atmospheres is to detect thermal emission spectra of the planet. 
Even if the star and the planet are not spatially resolved, the features in planetary spectra may be identified in the composite spectra due to the distinct Doppler shift of the planetary spectra (for the application of this method to mid-infrared high resolution spectroscopy, see \cite{2021AJ....161..180F}). 
In what follows, we examine the viability of this method with GREX-PLUS mission using modeled thermal emission spectra and mock observations. 

\begin{figure}[tbh]
    \centering
    \includegraphics[width=0.5\textwidth]{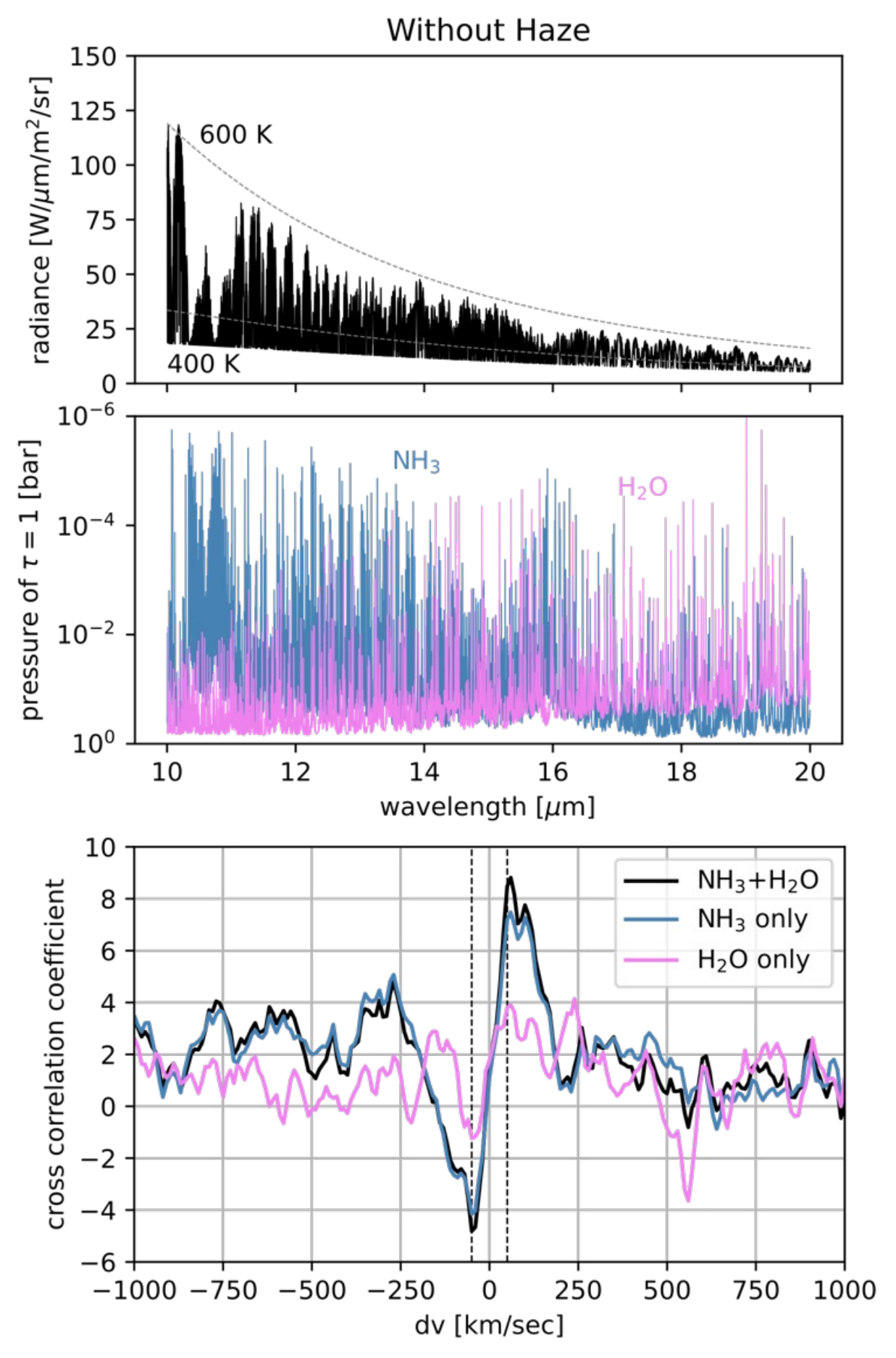}
    \caption{Top panel: model thermal emission spectrum of a Jupiter-like planet with equilibrium temperature of 500~K. Middle panel: pressure levels at $\tau =1$. Bottom panel: the result of cross-correlation analysis between the mock spectra and the model spectra, where mock data assumes a $\sim $500~K Jupiter-size planet around a Solar-type star at 20~parsec and that the 1~day observation is performed both at two quadrature phases. Two vertical dashed lines represents the radial velocities at two quadrature phases. }
    \label{fig:sp_corr_WJ500K}
\end{figure}

The top and middle panels of Figure \ref{fig:sp_corr_WJ500K} are a modeled thermal emission spectra of a warm Jupiter-sized planet (500~K, 1$R_J$, 1$M_J$) and the pressure levels where the opacity of each molecule from the top of atmosphere becomes unity. 
Here, we employed a 1-dimensional atmospheric profile assuming solar metallicity, where the temperature profile is taken from \citet{2014A&A...562A.133P} and the chemical profile is calculated using \citet{2018ApJ...853....7K}. 
Specifically, the mixing ratios of NH$_3$ and H$_2$O are approximately 10$^{-4}$ and 10$^{-3}$, respectively. 
Spectral features at $<$15$\rm{\mu}m $ are dominated by NH$_3$, while the H$_2$O features appear at longer wavelengths.

For mock observations, we place this planet around a Solar-type star at 20~parsecs from Earth. 
The target system is observed at two quadrature phases (corresponding to the maximum and minimum radial velocities) for 1~days each (any combination of phases would work as far as the two radial velocities are well separated). 
We have neglected the radial velocity of the central star, as it is much slower than that of the planet. 
Each spectrum is high-pass filtered and then the difference between the two are extracted. 
By cross-correlating this differential spectrum with the modeled planetary spectrum, we examine the detectability of the certain features of planetary spectra and measure the radial velocity of the planet. 

The cross-correlation between the mock data and theoretical models are shown in the bottom panel. 
The signal-to-noise ratio (SNR) of the peak of cross-correlation function can be $>5\sigma $ if the model is the correct one, while the model with NH$_3$ only can also lead to similar SNR. 
H$_2$O is harder to detect in this case. 

\begin{figure}[tbh]
    \centering
    \includegraphics[width=0.5\textwidth]{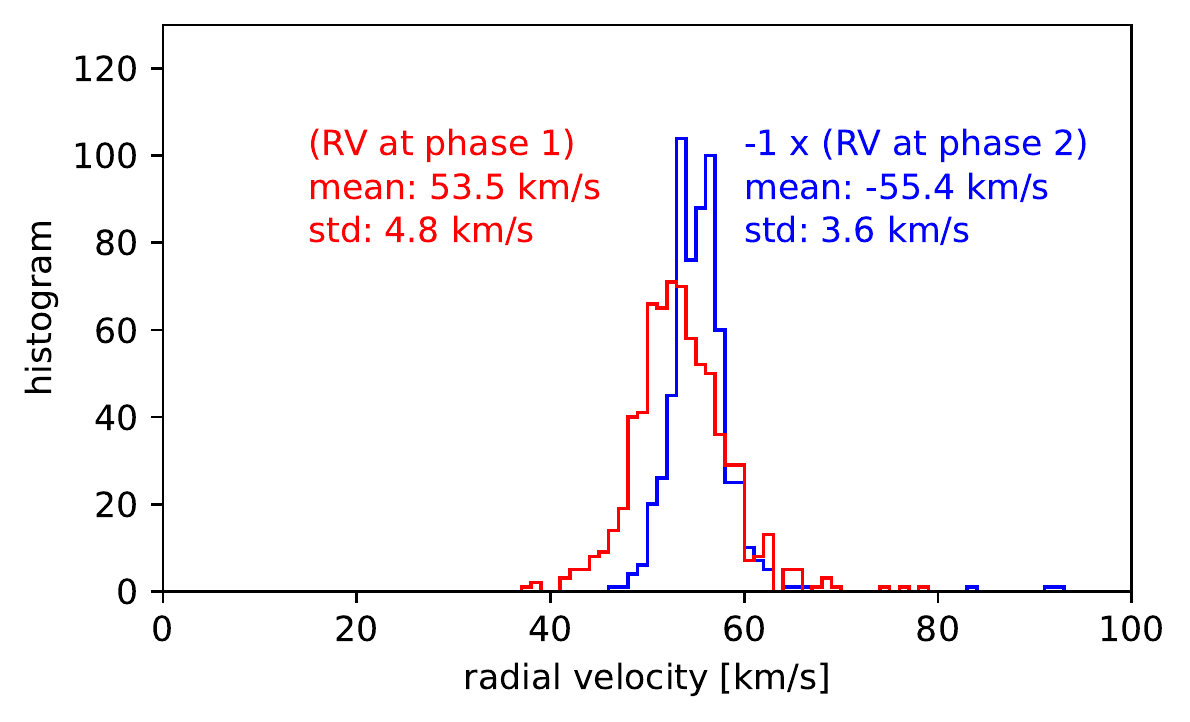}
    \caption{Probability distribution of planetary radial velocity at two quadrature phases based on a mock observation of a $\sim $500~K Jupiter-sized planet around a solar-type star at 20~pc, where each quadrature phase is observed for 1~day.}
    \label{fig:hist_rv_WJ500K}
\end{figure}

Figure \ref{fig:hist_rv_WJ500K} shows the distribution of the estimated radial velocity at these two phases (indicated by two colors), suggesting that the uncertainty in the radial velocity is a few km~s$^{-1}$. 
This can be compared with the typical orbital velocity (30-100~km s$^{-1}$), which implies that their orbital inclination can be well constrained. 

If an Earth-sized planet and a Neptune-sized planet are orbiting the planet at the orbit of Jovian moon Io, the resultant variations of radial velocity are approximately 50~m~s$^{-1}$ and 900~m~s$^{-1}$, respectively. 
The radial velocity variaition of binary planets as proposed by \cite{2014ApJ...790...92O} would be up to 17~km~s$^{-1}$. 
Monitoring the radial velocity at different epochs may allow us to constrain the presence of relatively large moons or binary planets.

While we did not assume a specific target in the study above, 
the analysis can be directly applied to, for example, 
Rho CrB b (17.5~pc, $M_p \sin i\sim 1.1 M_J$) and 
70 Vir b (17.9~pc, $M_p \sin i\sim 7.4 M_J$). 
Furthermore, cool Jupiter-like planets ($\gtrsim $300~K) around nearby M-type stars may also be investigated. 
For example, GJ~876~c, a Jovian planet around an M4 star at 4.7~pc, receiving incident energy flux similar to that of Earth, has similar star-to-planet contrast and similar photon flux to the case of a warm Jupiter around a solar-type star at 20~pc discussed above. 
Thus, atmospheric species on such planets may also be detected depending on the atmospheric properties. 

\begin{figure}
    \centering
    \includegraphics[width=0.5\textwidth]{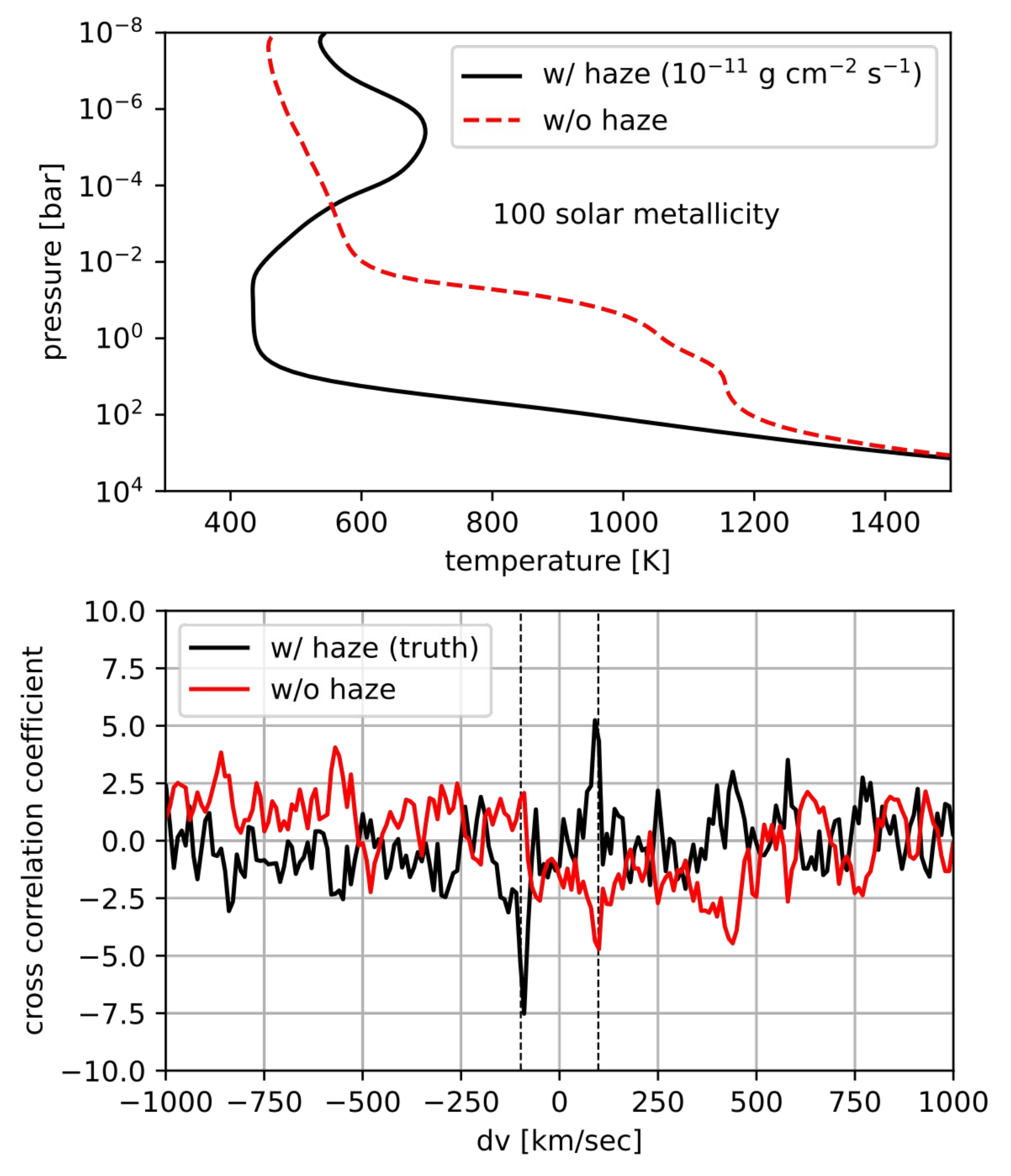}
    \caption{Top panel: modeled temperature profile of GJ1214b with and without haze layer based on Ohno \& Fortney in prep; the presence of haze layer of GJ1214b is implied from observed near-infrared transmission spectrum \citep{Kreidberg+14}. The haze layer would cause thermal inversion in the upper atmosphere. Bottom panel: cross-correlation coefficient between the mock data of GJ1214b and two modeled spectra with and without haze layer (black and red lines in the top panel). The mock data assumes the atmospheric profile with haze (black line in the top panel) and observations for 2.5 days at each quadrature phase. Two vertical dashed lines represents the radial velocities at two quadrature phases. While the model with haze layer has signals at the correct radial velocities with SNR $>$ 5, the model without haze layer has less significant signals. }
    \label{fig:GJ1214b_crr_haze}
\end{figure}

A more intensive use of telescope would be needed for thermal emission of Neptune-sized planets such as GJ~1214b. 
The detectability depends highly on the actual atmospheric structure.
To create the mock spectra, we computed the atmospheric pressure-temperature profile for 100$\times$ solar metallicity using a 1D radiative-convective equilibrium model \citep{McKay+89,Marley&Robinson15} and then postprocessed it by a publicly available photochemical model VULCAN \citep{Tsai+21}.
We also consider the possible presence of photochemical hazes by iteratively computing the the PT profile and vertical haze profile with an aerosol microphysical model (\citealt{Ohno&Okuzumi18,Ohno&Kawashima20}; Ohno \& Fortney in prep).
Our simulation suggests that $\gtrsim $5 days observations will be able to detect NH$_3$ on GJ~1214b. 
Note that thermal emission is sensitive to both the composition and the vertical profile of atmosphere. 
Cross-correlation analysis using model thermal emission spectra with and without haze layer, the former of which has thermal inversion in the upper atmosphere (e.g., \citealt{2015ApJ...815..110M}; \citealt{Lavvas&Aufaux21}; Ohno \& Fortney in prep), could provide an evidence of the haze layer (Figure \ref{fig:GJ1214b_crr_haze}).

\subsection{Scientific goals}

\begin{itemize}
    \item Detecting minor species in exoplanet atmospheres as a probe of atmospheric elemental abundances, vertical mixing, and/or UV environment. 
    \item Constraining the atmospheric thermal structure of the planet to infer the presence or absence of aerosols. 
    \item Measuring the radial velocity of the planet. 
\end{itemize}

\begin{table}
    \label{tab:exoplanet}
    \begin{center}
    \caption{Required observational parameters.}
    \begin{tabular}{|l|p{9cm}|l|}
    \hline
     & Requirement & Remarks \\
    \hline
    Wavelength & 10--18 $\mu$m & \multirow{2}{*}{} \\
    \cline{1-2}
    Spatial resolution & N/A & \\
    \hline
    Wavelength resolution & $\lambda/\Delta \lambda>30000. $ & \\
    \hline
    Field of view & N/A & \\
    \hline
    Sensitivity & ?? & \\
    \hline
    Observing field & N/A & \\
    \hline
    Observing cadence & N/A & \\
    \hline
    \end{tabular}
    \end{center}
\end{table}

\clearpage
\section{Solar System Planetary Atmosphere}
\label{sec:solarsystemplanet}

\noindent
\begin{flushright}
Hideo Sagawa$^{1}$
\\
$^{1}$ Kyoto Sangyo University
\end{flushright}
\vspace{0.5cm}

\subsection{Scientific background and motivation}
Observations of planetary atmospheres are important not only for studying the meteorology and climate, but also for understanding the formation and evolution of planets through gathering the knowledge on the atmospheric compositions. 
In addition, detail characterization of planetary atmospheres, such as the redox state and its chemical stability, the presence or absence of clouds and hazes, and atmospheric escape, etc., can provide essential information whether the planet can harbor life, i.e., habitability.

For a long while, people attempted to describe the atmospheres of other planets based on the knowledge that collected from the Earth's atmosphere. However, it is now obvious that the Earth's atmospheric conditions are not universal: our planet is not the standard reference for planetary atmospheres. Each planet has a different distance to the Sun, which means a different amount of solar radiation is injected into the atmosphere. The different planetary mass and rotation speed make a different gravitational acceleration and Coriolis force, which are two essential physical quantities in the atmospheric dynamics. Total amount of the atmosphere and its composition, including minor trace gases, are also different from each planet, and these parameters play key roles in the radiative transfer i.e., energy distribution in the atmosphere. There are many other things worth noting (including `non'-atmospheric phenomena such as the presence or absence of an intrinsic magnetic field, volcanic activity, and so on), and all these things connect to form the diversity of planetary atmospheres. More we know about such diversity, more significance the importance of expanding the comparative studies of each planetary atmosphere becomes.

Infrared spectroscopy has been served as the most effective tool in observing planetary atmospheres. Thermal infrared radiation from the planet is observed as a combination of continuum emission from the atmosphere and several absorption lines (emission lines, for some cases) induced by several specific molecular species in the atmosphere. Thus, we can obtain information on both the temperature structure and atmospheric composition by analyzing such infrared spectra.

There is no doubt that the temperature structure is one of the most fundamental physical quantities in planetary atmospheres, and the changes in the thermal structure with latitude and time (season) are the driving factors of various physical phenomena in planetary atmospheres such as atmospheric circulations. 
The gaseous and icy giant planets beyond Jupiter have internal heat sources, while it is not yet clear to what altitude the internal heat source affects the atmospheric temperature structure. For example, it seems that the internal heat less affects the weather layer of Uranus's atmosphere, while it is not the case for Neptune. So, the knowledge on the thermal structure of the atmosphere of gaseous and icy giant planets is also important to constrain their interior formation. 

The temperature structures of these outer solar system planets have been studied by a couple of studies, starting with the flyby observation by the Voyager spacecraft \citep[e.g.,][]{Lindal81, Lindal85, Conrath98}, followed by observations by infrared space telescopes and orbiting spacecraft \citep[e.g.,][]{Lellouch01, Orton14, Fletcher16}, but there are no observations that cover the vertical structure of temperature from low to high altitudes. 
In addition, the temporal coverage of these existing observations is rather sparse and limited, and they are incomplete for understanding seasonal changes of the atmospheres of the icy giant planets.

\begin{figure}[!t]
\centering
\includegraphics[width=0.88\linewidth]{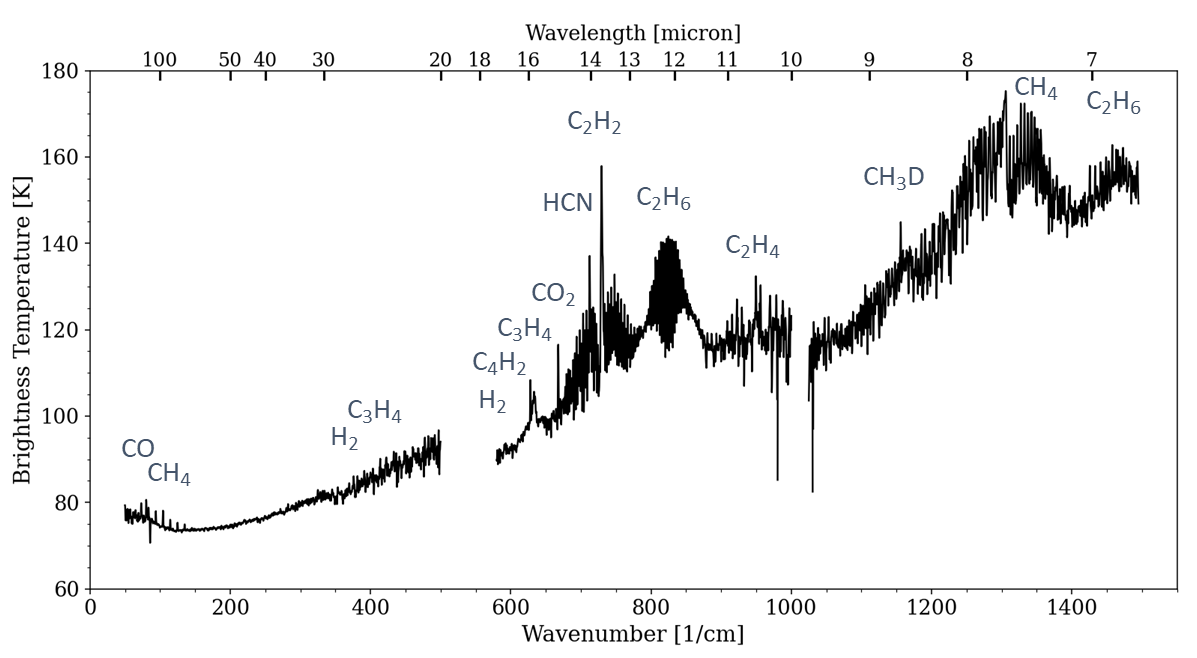}
\caption{An example of infrared spectra of Titan atmosphere observed by Cassini/CIRS. 
Spectral resolution of CIRS is $\sim$0.5~cm$^{-1}$.} 
\label{fig:SolSysAtmos_Titan}
\end{figure}

Observations of atmospheric compositions including isotopologues are essential for understanding the atmospheric chemistry. In particular, Titan’s atmosphere is of most interest among other solar system bodies, as its N$_2$-dominated and CH$_4$-rich reducing atmosphere holds one of the most complex atmospheric chemistry \citep{Horst17}. Past observations of Titan’s atmosphere by the CIRS (Composite Infrared Spectrometer) onboard the Cassini spacecraft have reported the presence of numerous organic molecules, including nitriles (HCN, HC$_3$N, CH$_3$CN, etc.) and hydrocarbons (C$_4$H$_2$, C$_2$H$_2$, C$_2$H$_6$, etc.), in the stratosphere (Figure~\ref{fig:SolSysAtmos_Titan}). These species showed strong spatial and temporal variabilities in their abundances which are interpreted as the effect of Titan’s atmospheric circulation and photochemistry \citep{Coustenis07, Coustenis10}. 
The observations by Cassini/CIRS only covered less than half of one Titan year (approximately 30 Earth years). New monitoring observations in future are desired to capture full characteristics of the seasonal variation in Titan’s atmosphere.

\subsection{Required observations and expected results}

\paragraph{(1) Derivation of the vertical profile of atmospheric temperature:} 
Infrared emission of the gaseous and icy giant planets contains spectrally broad absorption features caused by collision-induced absorption (CIA) of molecular hydrogen (H$_2$) and Helium, which are the two major constituents of the giant planets’ atmosphere. 
A few narrow spectral features due to the rotational transition of H$_2$ quadrupole are also observed. The intensity (opacity) of CIA and quadrupole line emissions depend on the abundance of H$_2$ molecules and the temperature distribution along the line-of-sight of observations. Since the amount (volume mixing ratio) of H$_2$ in the atmosphere is already known, the temperature can be retrieved from observations of these infrared spectra. It should be noted that the intensity of the quadrupole line emission also depends on ortho-para ratio of nuclear spin isomers of H$_2$, therefore an assumption on the ortho-para ratio of H$_2$ is required for this analysis (the ratio of 3:1, which is a value for the room temperature and pressure, is often adopted). 

\begin{figure}[!t]
\centering
\includegraphics[width=0.92\linewidth]{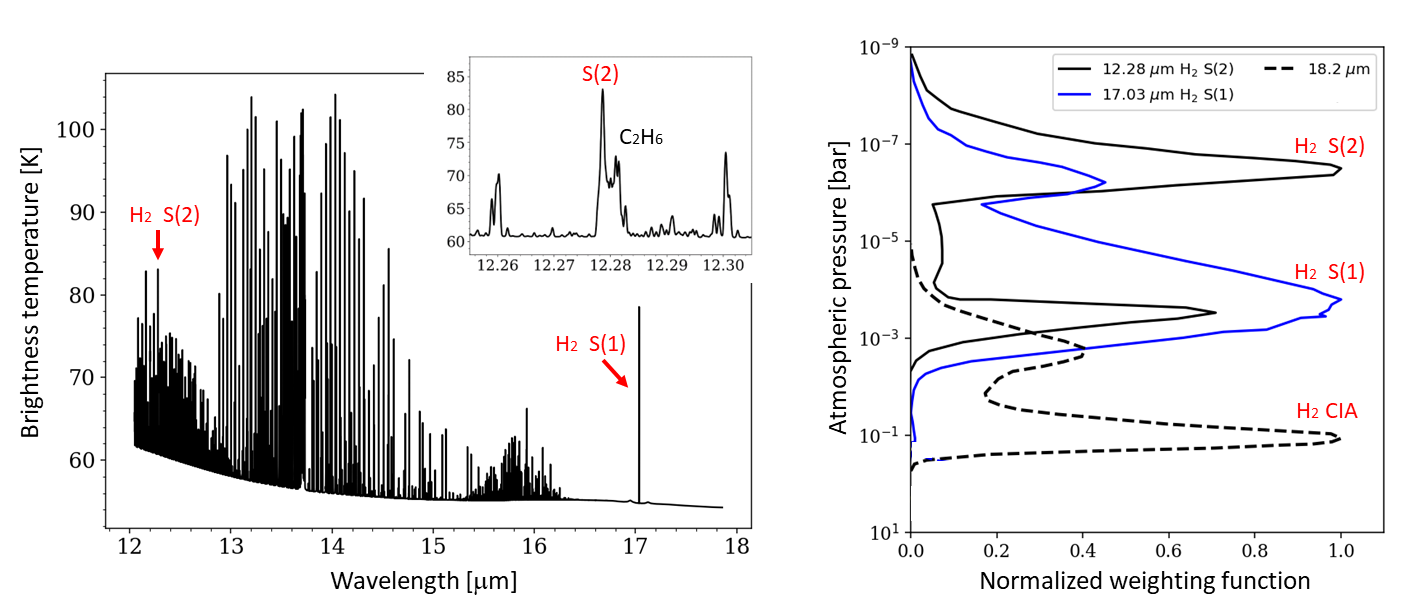}
\caption{Left: Simulated infrared spectrum of Uranus's atmosphere. A close-up to H$_2$ S(2) is shown in the upper right panel. A spectral resolving power of $\sim$30,000 is assumed. Right: Temperature weighting functions per some specific wavelengths.} 
\label{fig:SolSysAtmos_Uranus}
\end{figure}

As an example, Figure~\ref{fig:SolSysAtmos_Uranus} shows a simulated infrared spectrum of Uranus. A spectral resolving power of $\sim$30,000 is assumed. The H$_2$ quadrupole lines of S(1) and S(2) are observable at 17.04 and 12.28~$\mu$m, respectively. The absorption lines of C$_4$H$_2$, C$_2$H$_2$, and C$_2$H$_6$ are seen (not labelled in the plot, though) at the wavelengths shorter than 16~$\mu$m. 
The sensitivity to the temperature profile of Uranus’s atmosphere is estimated by calculating the weighting function (i.e., functional derivative of spectral radiance per wavelength with respect to the atmospheric temperature at each altitude).  A larger value of weighting functions indicates the observed spectrum has more sensitivity to the temperature at that level. 
The continuum emission can be used to constrain the temperature at the upper troposphere ($\sim$0.1 bar level, as seen in the dashed line of Figure~\ref{fig:SolSysAtmos_Uranus} right plot). While, the quadrupole S(1) and S(2) narrow line observations expand the sensitive altitude range as high as the upper stratosphere (from $\sim$10$^{-3}$ to $\sim$10$^{-7}$ bar, shown by solid lines in the plot). One of the challenges is to measure the intensity of the H$_2$ S(2) spectrum precisely. This line is located at 12.28~$\mu$m where crowded C$_2$H$_6$ lines also exist. An improved spectral resolution $\lambda$/$\Delta\lambda$ $>$10,000 is required to separate the H$_2$ S(2) line from the interfering C$_2$H$_6$ lines. The past observations by Spitzer/IRS ($\lambda$/$\Delta\lambda \sim$ 600) could not spectrally resolve these lines sufficiently, and the quality of the data analysis had been complicated compared to the other H$_2$ S(1) analysis \citep{Orton14}. 

Furthermore, infrared spectra of Uranus and Neptune have not been often observed since previous ones by the Infrared Space Observatory (ISO) and the Spitzer telescope. As these icy giant planets have long orbital periods, a new additional infrared spectroscopy of these planets in 2030s will provide an important constraint on the seasonal variations of their temperature structure.

\paragraph{(2) Measurements of the abundances of minor species:}
Not only the temperature but the atmospheric compositions of the gaseous and icy giant planets are still not fully understood. Some of the minor species in their atmospheres contribute to the greenhouse effect and/or radiative cooling, which makes the accurate knowledge of their abundances important to constrain the thermal structures.

As shown in Figures~\ref{fig:SolSysAtmos_Titan} and \ref{fig:SolSysAtmos_Uranus}, infrared spectroscopy is the most desired observation technique to enable spectral detection of minor species in the atmosphere (in particular, stratospheric species). Minor species often relate to each other closely via photochemical reactions, therefore, simultaneous measurements of various chemical species are needed for developing an accurate photochemical model. This can be achieved by observing in a wavelength range of $\sim$12--18~$\mu$m, where a particularly dense concentration of multiple molecular lines is seen. For molecules that do not have suitable lines in the infrared region, it is also effective to collaborate with observations by (sub)millimeter telescopes such as ALMA. 
To understand the seasonal variations in Titan’s atmosphere, it is important to continue monitoring observations over as long a period as possible, with a frequency of once every several months. 

Regarding to both of the above mentioned topics (1) and (2), a spatial resolution better than a few arcsec is required. 
This resolution will resolve the latitudinal belt and zone structures (and some local phenomena such as the great red spot) in Jupiter (apparent disk angular diameter of 30--50 arcsec). Saturn (angular diameter of 15--20 arcsec) can also be resolved into the equatorial, mid-latitude, and high-latitude regions. This a few arcsec resolution is comparable with the angular diameters of Uranus and Neptune, therefore it is not sufficient enough to spatially resolve the disk. However, it still has an advantage of detecting the emission from these icy giant planets without diluting the signal against the cold deep sky. 
Titan, which has an angular diameter of about 0.8 arcsec, will be observed as a point source. 
The sensitivity should be better than 10$^{-18}$~W~m$^{-2}$ which is a typical intensity for the weakest target emission lines. 

\subsection{Scientific goals}
The main goal is to obtain the thermal infrared spectra from planetary atmospheres of our solar system at the wavelength range of 12--18~$\mu$m with a high spectral resolution (resolving power $\lambda$/$\Delta\lambda > \sim$10,000).
The vertical structure of the atmospheric temperature and the abundances of minor species including their isotopologues will be derived from the observed spectra. 
A particular interest lies in observing the gaseous and icy giant planets, for which observations have been very limited, and Titan which contains a very complex atmospheric chemistry. 
The data need to be acquired at several epochs so that any temporal (seasonal) variability can be constrained.

\begin{table}[h]
    \label{tab:ssplanet}
    \begin{center}
    \caption{Required observational parameters.}
    \begin{tabular}{|l|p{9cm}|l|}
    \hline
     & Requirement & Remarks \\
    \hline
    Wavelength & 12--18~$\mu$m & $a$ \\
    \hline
    Spatial resolution & $<$a few arcsec & $b$ \\
    \hline
    Wavelength resolution & $\lambda/\Delta \lambda>10,000$ & $c$ \\
    \hline
    \multirow{2}{*}{Sensitivity} & $<$10$^{-8}$~W~m$^{-2}$~sr$^{-1}$ ($5\sigma$, extended-source, 1~hr) & \multirow{2}{*}{$d$} \\
    & $<$10$^{-18}$~W~m$^{-2}$ ($5\sigma$, point-source, 1~hr) & \\
    \hline
    Observing field & N/A & \\
    \hline
    Observing cadence & once every 4--6 months & $e$ \\
    \hline
    \end{tabular}
    \end{center}
    $^a$ To detect the dense spectral lines of hydrocarbons at 12--16~$\mu$m and also H$_2$ S(1) quadrupole line at 17.03~$\mu$m.\\
    $^b$ To spatially resolve Jupiter and Saturn.\\
    $^c$ To spectrally resolve H$_2$ S(2) from surrounding C$_2$H$_6$ lines.\\ 
    $^d$ To achieve a sufficient S/N on H$_2$ S(2) observations. For Titan, a point-source sensitivity is applied. \\
    $^e$ To monitor the seasonal variability of Titan atmosphere.
\end{table}

\clearpage
\section{Icy Small Solar System Bodies}
\label{sec:icysmallbodies}

\noindent
\begin{flushright}
Takafumi Ootsubo$^{1}$, 
Tsuyoshi Terai$^{2}$
\\
$^{1}$ NAOJ
$^{2}$ Subaru Telescope, NAOJ
\end{flushright}
\vspace{0.5cm}

\subsection{Scientific background and motivation}

\subsubsection{Hydrated minerals and H$_2$O ice on asteroids}
Hydrated minerals, any mineral that contains H$_2$O or OH, are formed by aqueous alteration in environments where anhydrous rock and liquid water exist together with a certain pressure and temperature. Hydrated minerals are stable even above the sublimation temperature of water ice. They become an important tracer of water present in the history of the solar system unless they were reset by a temperature change after formation. The study of hydrated minerals is therefore important for understanding of the origin of Earth's water and the earliest thermal processes in the solar system. Since most asteroids have not experienced sufficient thermal evolution to differentiate into layered structures like terrestrial planets since their formation, it is indispensable to investigate the presence of hydrated minerals and water ice on various types of asteroids. 

Hydrated minerals and water ice exhibit diagnostic absorption features in the 3-$\mu$m band (approximately 2.5--3.5~$\mu$m wavelength). Features at around 2.7~$\mu$m are attributed to hydrated minerals and those at around 3.05~$\mu$m to water ice. AKARI, the Japanese infrared satellite, conducted a spectroscopic survey in the near-infrared wavelength region of 2.5--5~$\mu$m for 66 asteroids with a diameter of $>~40$~km. It is found that most C-complex asteroids (17 out of 22) have absorption features around 2.75~$\mu$m in the spectra, which is attributed to hydrated minerals. Some low-albedo X-complex asteroids and one D-complex asteroid have an absorption feature in the 3-$\mu$m band, similar to the C-complex asteroids \citep{2019PASJ...71....1U}.

The result of AKARI survey revealed that many C-type asteroids have hydrated minerals on their surface. However, targets that observed with AKARI are limited with a diameter larger than 40~km, and most of targets are S-, X-, and C-type asteroids. To explore the existence of hydrated minerals and water in the asteroids in more detail, we need much smaller samples and asteroids of different types. GREX-PLUS's near-infrared camera (2.0--4.5~$\mu$m) covers wavelength range of 2.7~$\mu$m and 3.05~$\mu$m features. GREX-PLUS can detect the hydrated minerals on asteroids smaller than 10~km with filters F303. The GREX-PLUS hydrated minerals/water-ice survey with more than 100 asteroids would provide the insight  
into the thermal environments and evolution around the snow line in the early solar system.

\subsubsection{H$_2$O ice abundance on trans-Neptunian objects}
Trans-Neptunian objects (TNOs), a population of small solar system bodies beyond Neptune's orbit, 
are believed to be primitive icy bodies and remnants of planetesimals formed in the early stage of 
our solar system.
They are located far from the sun where it is cold ($\lesssim$~50~K) enough for volatile materials 
to condense and be retained on their surfaces.
Actually, various kinds of ices such as H$_2$O, CH$_4$, CH$_3$OH, C$_2$H$_6$, CO, and NH$_3$, 
were detected from the objects (including outer planet satellites, dwarf planets) in the outer solar 
system \citep[e.g.,][]{2020tnss.book..109B}.
The composition of the surface ices provides essential information about the thermal and chemical 
histories of planetesimals, which would lead us to understanding of the physical and chemical 
conditions in the protoplanetary disk as well as the dynamical evolution of small bodies including 
the radial mixing during planet formation/migration processes.

H$_2$O ice is the most common icy components on small bodies in the outer solar system.
Most of TNOs which are known to be covered by icy surfaces have spectra dominated by H$_2$O ice 
excluding the largest objects such as Pluto, Eris, Makemake, and Sedna that have 
CH$_4$ ice-rich surfaces \citep[e.g.,][]{2013ASSL..356..107D}.
So far, the presence of H$_2$O ice on TNOs was identified by the characteristic absorption bands 
at 1.5~$\mu$m and 2.0~$\mu$m wavelengths in their near-infrared (1--2.5~$\mu$m) reflective spectra.
The depth of these features represents the spectral fraction of H$_2$O ice on the surface.

\begin{figure}[t]
 \begin{minipage}{0.50\hsize}
  \begin{center}
  \includegraphics[width=82mm]{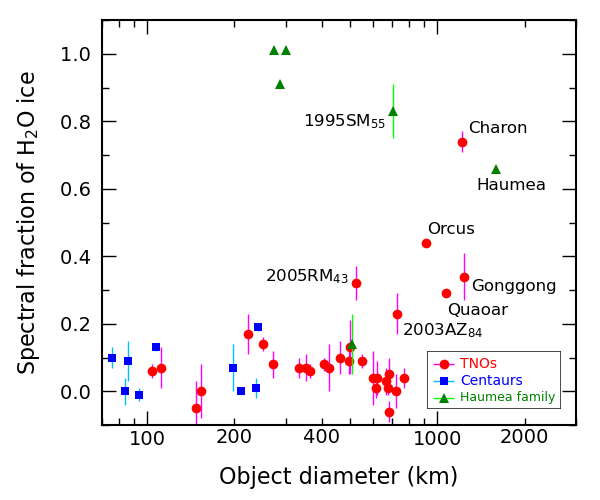}
  \end{center}
  \vspace{-1.0em}
  \caption{
  The plot of spectral fraction of H$_2$O ice \citep{2012AJ....143..146B} vs. body size
  for TNOs (including Charon; red circles), Centaurs (blue squares), and Haumea family members 
  (green triangles).
  }
  \label{fig_terai01}
 \end{minipage}
 \hspace{1.0em}
 \begin{minipage}{0.50\hsize}
  \begin{center}
  \includegraphics[width=82mm]{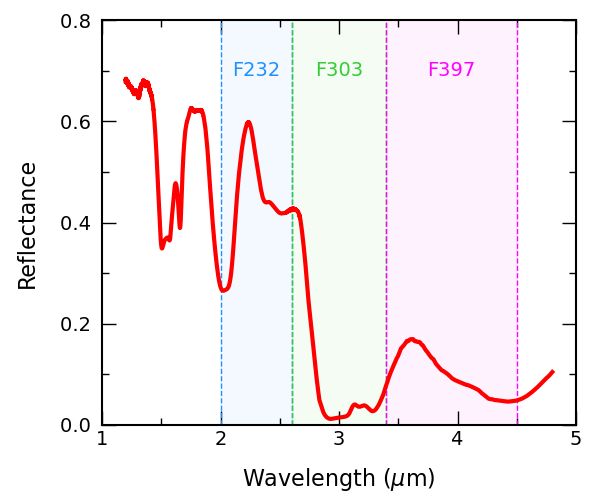}
  \end{center}
  \vspace{-1.0em}
  \caption{
  Near-infrared reflective spectrum of crystalline H$_2$O ice at 60~K \citep{2009ApJ...701.1347M}.
  The shaded areas represent wavelength ranges of the planned filters for GREX-PLUS.
  }
  \label{fig_terai02}
 \end{minipage}
\end{figure}

The abundance of H$_2$O ice has been investigated from a large number of TNOs and Centaurs 
(a small-body population located between Jupiter and Neptune) by several spectroscopic surveys 
performed with 8--10 meter ground-based telescopes
\citep[e.g.,][]{2008AJ....135...55B,2011Icar..214..297B,2012AJ....143..146B}.
These studies revealed that the surface H$_2$O ice fraction has a trend with the body size 
(see Figure~\ref{fig_terai01}): the large objects such as Charon (Pluto's largest moon), Haumea, 
Orcus, Gonggong, and Quaoar have H$_2$O ice-rich surfaces, while none or only small amounts of 
H$_2$O ice were detected on most objects smaller than $\sim$800~km in diameter.
The exception is the Haumea collisional family members that are considered to be originated from 
fragments of Haumea's mantle ejected due to a disruptive impacts and exhibit strong spectral 
features of H$_2$O ice even on small objects of diameter less than 500~km 
\citep{2007Natur.446..294B}.
This fact may indicate that it is not always impossible for small objects to retain H$_2$O ice on 
their surfaces.
In addition, a spectrum with no or very weak features does not necessarily imply an ice-poor surface 
because the absorption signatures are masked if H$_2$O ice contains particulate dark contamination 
\citep{1982Icar...49..244C}.
It is obvious that our understanding is still limited to the H$_2$O ice abundance on TNOs.
Measurements with higher accuracy for H$_2$O ice abundance on a number of small to mid-sized 
TNOs are required to examine whether these objects truly lack H$_2$O ice on their surfaces or not, 
which would be a crucial clue to develop our interpretation to the trend of the ice fraction 
increasing with size, though it is difficult even by using the present largest ground-based 
telescopes due to faintness of these objects.

Observations with GREX-PLUS's near-infrared camera covering the 2.0--4.5~$\mu$m wavelength range 
would provide a breakthrough in this study.
It is well known that H$_2$O ice has strong absorption around 3.0~$\mu$m owing to O-H stretching 
vibrations \citep[see Figure~\ref{fig_terai02}; e.g.,][]{2009ApJ...701.1347M}.
This 3 $\mu$m band is most suitable for precisely measuring the amount of H$_2$O ice and has 
actually been observed on icy satellites of the giant planets and Saturn's ring
\citep[e.g.,][]{1998ASSL..227..579C}.
Wide-field and high-sensitivity imaging with the F303 filter makes it possible to realize 
high-quality and high-efficiency 3 $\mu$m observations for a large number of small/mid-sized TNOs 
and allows us to examine their relationship between the surface H$_2$O ice abundance 
and the body size or another parameter such as the orbits, albedos, visible spectra, and bulk densities in detail.

\subsubsection{Chemical composition of cometary ices: H$_2$O, CO$_2$, and CO}

It is considered that comets are the most pristine objects in the solar system and chemical abundances of the cometary ices can be used to infer the conditions in the early solar nebula. 
Most of the abundant molecular species in the cometary ices are H$_2$O, CO$_2$, and CO. Thus, one of the most important characteristics of cometary ice is the mixing ratio of major volatiles relative to H$_2$O, especially for CO$_2$, CO, and organics. However, because of the severe absorption of telluric CO$_2$ in the atmosphere, it is difficult to access the cometary CO$_2$ with ground-based telescopes. 
In the near-infrared wavelength region, the $\nu_1$ and $\nu_3$ vibrational fundamental bands of H$_2$O are recognized at around 2.7~$\mu$m, while the $\nu_3$ vibrational fundamental band of CO$_2$ around 4.3~$\mu$m and the ro-vibrational fundamental v(1–0) band of CO around 4.7~$\mu$m are also in this spectral region.
The Japanese infrared satellite AKARI has observed more than a dozen comets in the 2.5--5~$\mu$m region \citep{2012ApJ...752...15O}. It is reported that the mixing ratio of CO$_2$ with respect to H$_2$O spans from several to $\sim$30\% among the comets observed within 2.5 au from the Sun. CO was detected only in very few cases. We need a larger comet sample (more than 50) with good CO$_2$ and CO detection for further discussions.

Although the near-infrared spectroscopy is the best way for the study of cometary mixing ratios, multi-band photometry with adequate filters can provide meaningful results. Spitzer/IRAC surveyed 23 comets with 3.6 and 4.5~$\mu$m bands and derived mixing ratios of CO$_2$+CO with respect to H$_2$O (Reach et al. 2013).     
GREX-PLUS's near-infrared camera (2.0--4.5~$\mu$m) covers this wavelength range and can detect H$_2$O, CO$_2$, and CO separately with filters F303, F397, and F520, respectively, while Spitzer/IRAC could not distinguish the emission from CO$_2$ and CO. Wide-field and high-sensitivity imaging with the near- and mid-infrared filters allows us to investigate cometary volatiles in detail.

\subsection{Required observations and expected results}

\subsubsection{Asteroids}
Consecutive imagings with the F232, F303, F397, and F520 filters at a field in a short time ($\ll$~1~hour) are 
required to reduce the effect of brightness variation due to the object's spin. At least one mid-infrared filter is required to estimate a contribution of the thermal emission continuum. More than 100 samples are desirable to cover the diameter range down to smaller than 10 km with various types of asteroids. 

\subsubsection{Trans-Neptunian objects}
Consecutive imagings with the F232, F303, F397 filters at a field in a short time ($\ll$~1~hour) are 
required to reduce the effect of brightness variation due to the object's spin.
Our estimate shows that a 300~sec integration with the F303 filter can observe TNOs as small as
200~km in diameter, which is sufficient for this study.
We also request to perform the Wide survey at low ecliptic latitude (within $\pm$30$^\circ$) regions 
as much as possible so that we can obtain photometric data of many TNOs.
We plan to determine the surface H$_2$O ice abundance from 50--100 TNOs larger than 200~km in 
diameter by combining the data of the Wide survey and pointing observations

\subsubsection{Comets}
Consecutive imaging with 3 filters in near-infrared and more than 1 filter in mid-infrared is required to determine the continuum (mid-infrared) and detect excess emission from the volatiles (near-infrared) in the spectra. Observations for the same comet in short time ($<$~1~day) is desired to reduce the effect of brightness variation due to the comet coma activity.
We plan to observe more than 20 comets and derive the mixing ratios of CO$_2$ and CO with respect to H$_2$O. 

\subsection{Scientific goals}

\subsubsection{Asteroids}
A hydrated mineral and water-ice survey with GREX-PLUS is aiming to reveal whether the existence of water on asteroids is universal or not. Observations with $>~100$ targets with various asteroid types could provide the information of thermal environment and aqueous alternation inside/outside of the snow line in the early solar nebula. Observed samples of wide size range, down to smaller than 10~km, would reveals the collisional history of asteroids and the existence of water inside asteroids.

\subsubsection{Trans-Neptunian objects}
This study is aiming to highly accurately measure the spectral fraction of surface H$_2$O ice for 
50 or more TNOs between $\sim$200--800~km in diameter to investigate the distribution of 
H$_2$O ice abundance on TNOs and to elucidate some fundamental questions such as 
(1)~Are the surfaces of small/mid-sized TNOs really lack of ice or not?
(2)~Why do such objects have (apparently) ice-poor surfaces?
(3)~What factors determine the transitional size ($\sim$800~km in diameter) of H$_2$O ice abundance 
on TNOs?

\subsubsection{Comets}
The goal of the GREX-PLUS observations of comets is the accurate measurement of volatile composition in comet ices. Newly derived CO$_2$, CO, and organics mixing ratios with respect to H$_2$O of $> 20$ comets in addition to AKARI samples will be discussed in the viewpoint of comets' orbits, and thus, their birthplaces in the proto-solar disk.

\begin{table}[bth]
    \label{tab:icybodies}
    \begin{center}
    \caption{Required observational parameters.}
    \begin{tabular}{|l|p{9cm}|l|}
    \hline
     & Requirement & Remarks \\
    \hline
    Wavelength & 2--5 $\mu$m & \\
    \cline{1-2}
    Spatial resolution & $<1$ arcsec & \\
    \hline
    Wavelength resolution & $\lambda/\Delta \lambda>3$ & \\
    \hline
    Field of view & TNOs: 40 degree$^2$, 25 ABmag ($5\sigma$, point-source) & \\
    \hline
    Observing field & TNOs: ecliptic latitude $<$~30$^\circ$ & \\
    \hline
    Observing cadence & TNOs: consecutive imagings with the three NIR filters within $\ll$~1~hour & \\
    \hline
    \end{tabular}
    \end{center}
\end{table}

\clearpage
\section{Star and Planetary Forming Regions}
\label{sec:starformingregions}

\noindent
\begin{flushright}
Chikako Yasui$^{1}$, 
Michihiro Takami$^{2}$
\\
$^{1}$ NAOJ, 
$^{2}$ ASIAA
\end{flushright}
\vspace{0.5cm}

\subsection{Scientific background and motivation}



\subsubsection{Protoplanetary Disks as Sites of Planet Formation}

\begin{figure*}[ht!]
\begin{center}
    \includegraphics[width=8cm]{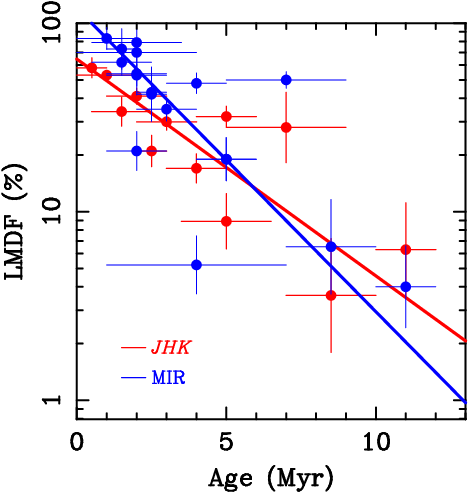}
\end{center}
\caption{Evolution of protoplanetary disks---fraction of stars
with NIR disk excess as a function of age. Figure~5 (left panel) of \citet{Yasui2014}, ``Rapid evolution of the innermost dust disc of protoplanetary discs surrounding intermediate-mass stars''.}
\label{fig:yasui1}
\end{figure*}

Because planets are formed in protoplanetary disks in the process of
star formation, the evolution and dissipation of protoplanetary disks
should be an essential key to determine planet formation.
Observational studies of protoplanetary disks have progressed rapidly
from around 2000.
Figure~\ref{fig:yasui1} is an example of one such result, obtained from
near-infrared observations, where each dot represents an individual
star-forming region, the horizontal axis indicates the age, and the
vertical axis indicates the fraction of stars that still have their
disks within each region.
Using the fact that the stars in each star-forming region are born
almost simultaneously, the lifetime of protoplanetary disks, i.e., the
timescale of planet formation, was estimated to be about 10 Myr, because
almost all stars lose their disks at about 10 Myr.
However, the figure also shows that some young stars have already lost
their disks, while others still retain their disks even at relatively
old ages, indicating that there is a large variation in the timescale
for the disk dispersal of individual stars.
In an attempt to explain the variation, various environmental
dependencies have been proposed:
e.g.,
stellar mass \citep{Ribas2015}, 
cluster density \citep{Fang2013},
and metallicity \citep{Yasui2010}. 
However, the dispersion of the disk frequency is still large, even after considering the dependence of disk lifetime on each parameter.
Therefore, what determines the disk lifetime? is a very important
question in planet formation that has been left unanswered.

Various theoretical models existed prior to about 2000 as mechanisms
that attempted to explain disk dispersal. 
Meanwhile, from around the same period, it has been observationally known
that the disk disappears from inner to outer part of disks almost
simultaneously \citep{Williams2011}. 
The UV-switch model, combination of the mass accretion process from
inner disks onto the central stars and the photoevaporation process of
outer disks was proposed as a possible explanation for disk dispersal
\citep{Clarke2001}, and is now widely accepted.
Mass accretion activities have been confirmed in many objects, mainly by
the observation of hydrogen emission lines,
and the comparison of the line profiles with theoretical models has
revealed the basic properties of mass accretion quantitatively, such as
the mass accretion rate and the geometric structure of accretion
\citep{Hartmann2016}.
Meanwhile, \textit{\textbf{ the photoevaporation process has not been well understood
because there have been very small number of observational detections ($N
\sim 20$), even though it is the main process of disk dissipation.}}
Photoevaporation is a phenomenon in which disk gas is heated by
high-energy photons from the central star or nearby stars and flows out
of the disk.
Although theoretical studies of the complex chemical network have led to
a better understanding of this phenomenon, the very basics, such as what
types of radiation energy (EUV, FUV, or X-rays) primarily induces
photoevaporation, are still poorly understood.


\subsubsection{The Physical Mechanism of High-Mass Star Formation}
The formation of high-mass stars has represented a tricky puzzle both from the theoretical and observational points of view.
After a certain evolutionary phase accretion should be inhibited by radiation pressure, making the formation of stars more massive than $\sim$10 ${\rm M}_\odot$ impossible \citep{Tan14}. This is not consistent with the clear existence of a significant number of more massive stars.
Disk accretion is regarded as the most promising scenario to resolve this issue, as such a geometry could significantly reduce the effects of radiation pressure on the accreting material \citep{Beltran16}. However, there is not yet observational proof that the mass accretion from the innermost region of these disks to the central protostar is ongoing. For instance, millimeter interferometry of thermal dust continuum emission (e.g., using ALMA) allows us to observe disk structures with a spatial resolution as good as ~100 au, with the given angular resolution (typically $\sim$0".03) and distances to nearest high-mass young stars (typically $\gtrsim$2 kpc)\citep{Beltran16}. Such observations do not allow us to investigate to what extent the dust(+gas) remains at $r \ll$100 au against radiation pressure. Moreover, dust continuum observations do not directly provide kinematic information in principle, therefore these do not allow us to investigate whether disk accretion actually occurs. The latter issue could be solved by observing molecular line emission using millimeter interferometry, but only with modest angular/spatial resolutions.

Furthermore, these stars are heavily embedded even when they reach the main sequence so that it is difficult to determine the evolutionary stages of the individual targets. We cannot use optical spectroscopy to determine the stellar properties, and extensive ground-based near-infrared spectroscopy to date has not been able to solve this issue because of a limited number of the observable targets and spectral features \citep[e.g.,][]{Caratti17,Beuther10, Hsieh21}.
There is a proposed evolutionary sequence based on the presence and size of the ionized region \citep[No HII Region $\rightarrow$ Hyper Compact HII Region $\rightarrow$ Ultra Compact HII Region;][]{Beuther05}, which is easily measured in the radio. However, the phase is determined by the amount of stellar ionizing radiation, which is in principle a function of both mass and evolutionary status. The far-IR luminosity is used as an indicator of the mass of the central source \citep[e.g.,][]{Beltran16}, but may also be a function of the mass accretion rate \citep[e.g.,][]{Audard14}.

Breakthroughs on the above issues would be made with space mid-infrared spectroscopy, in particular using JWST and GREX-PLUS. These missions will allow us to observe a variety of spectral lines and features at 3--20 $\mu$m, which are tremendously useful for determining the physical properties for the target protostars and the inner disks. Such lines and features include: Br-$\alpha$ 4.05 $\mu$m, Pf-$\beta$ 4.65 $\mu$m, PAH 3.3-$\mu$m and perhaps He I (e.g., 4.04/4.05 $\mu$m) lines, that are useful for probing the color of UV radiation, allowing the calculation of the temperature of the (proto-)stellar photosphere \citep{Osterbrock89}; and a variety of molecular lines (CO, H$_2$O, CO$_2$, HCN, C$_2$H$_2$ etc.) useful for directly probing accretion and ejection of molecular gas \citep{Elias06} \citep[see also][]{Hsieh21} or the presence of the inner disk ($\ll$100 AU) \citep{Najita03,Pontoppidan10}, thereby allowing us to determine if these protostars are still in an active accretion phase. If the stellar continuum is significantly brighter than the disk emission, CO, SiO and Br-$\alpha$ may be seen in absorption in the stellar photosphere, allowing us to probe stellar evolution at the start of ionizing radiation \citep{Hosokawa10}. Otherwise, if the star is associated with a disk with a very high mass accretion rate, the spectra would be similar to M supergiants with CO and SiO bands \citep{Audard14}. Although many of the above spectral lines/features lie at $\lambda < 5$ $\mu$m, for which the protostars and disks are not directly seen due to extremely large circumstellar extinction, their infrared spectra are probably accessible via {\it bright} infrared reflection nebulae at $\lambda=3-5$ $\mu$m \citep[][]{Takami12}.

\begin{figure*}[ht!]
\begin{center}
    \includegraphics[width=8cm]{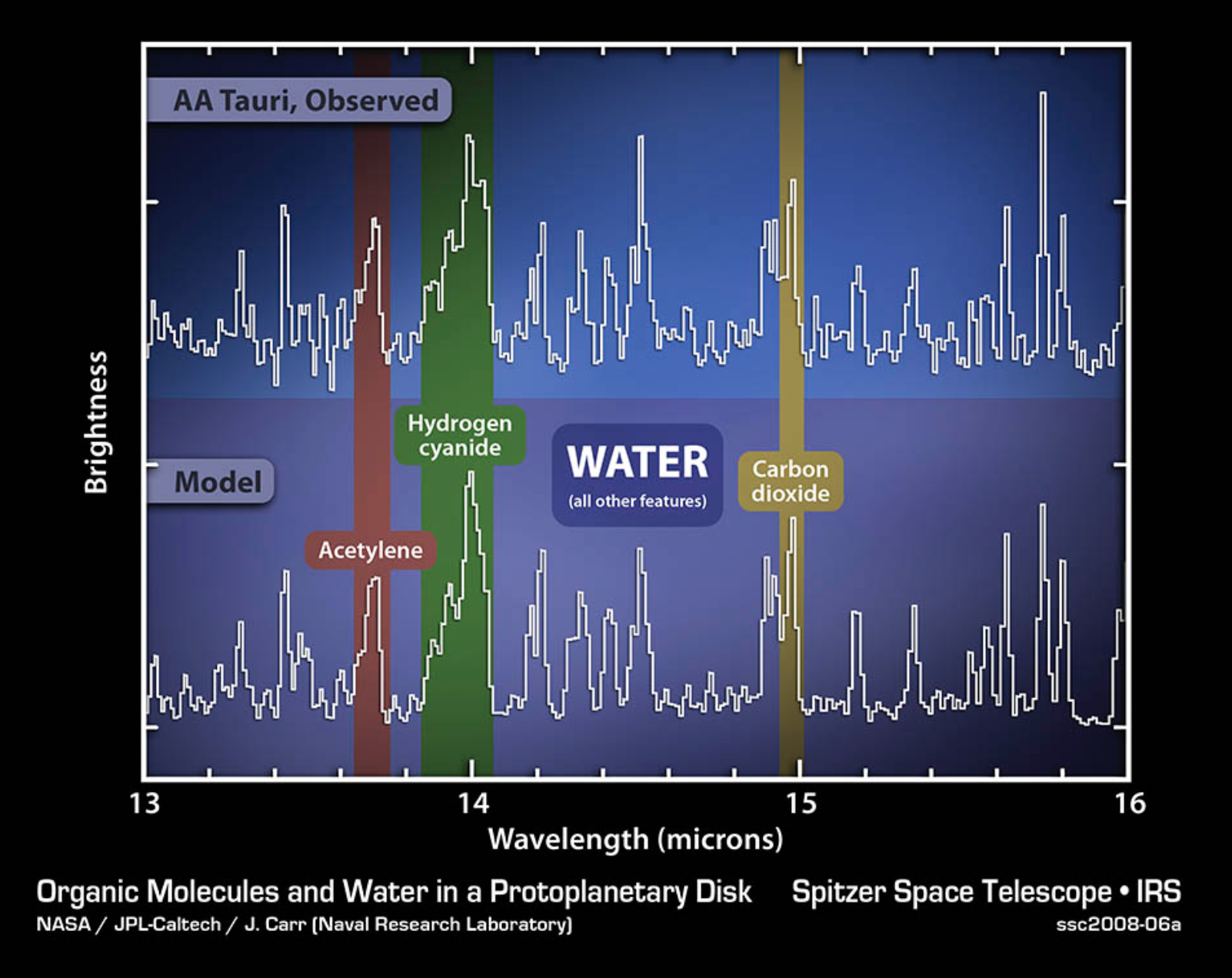}
\end{center}
\caption{A variety of molecular emission lines toward a low-mass young stars observed using Spitzer Space Telescope (Figure courtesy: NASA/JPL/J. Carr). These are considered to be associated with an unresolved circumstellar disk at a few au scales \citep[e.g.,][]{Pontoppidan10}.
\label{fig:takami1}}
\end{figure*}

GREX-PLUS will offer a unique opportunity for high resolution spectroscopy at 12--18 $\mu$m, which covers a number of molecular lines associated with H$_2$O, CO$_2$, HCN, C$_2$H$_2$ (see Figure \ref{fig:takami1}). Although the spectral coverage of its high-resolution (HR) spectrograph is significantly smaller than JWST, its contributions will be powerful and essential as described below.

\subsection{Required observations and expected results}

\subsubsection{Low-Mass Disk Evolution: The First Large Statistical Study with the Mid-Infrared [NeII] Line}

Forbidden lines with low-velocity components, such as [NeII] 12.8 $\mu$m
in the mid-infrared wavelength range and [OI] and [SII] in the optical
wavelength range, are known tracers of photoevaporation that have been
identified so far \citep{Pascucci2022}. 
For [OI] and [SII], although relatively high sensitivities can be
achieved in the optical wavelength region even from the ground-based
observations, the observed lines often show several components other than
photoevaporation, and it is sometimes difficult to separate the
photoevaporation component.
Meanwhile, the line profiles of [NeII] show only the component tracing
photoevaporation (although the profiles for some objects show only the
component tracing jets) and is currently considered a promising tracer
of photoevaporation.
However, since the sensitivity of the mid-infrared wavelength region is low
for ground-based observations, space telescopes are required, but no
previous space telescopes have had high resolution spectrographs.

GREX-PLUS is the first space telescope capable of mid-infrared
high resolution spectroscopy.
High sensitivity observations will dramatically increase the number of
detection samples.
In previous studies \citep{Pascucci2009, Pascucci2020}, sensitivity was
$F_{\rm cont} \le 300$ mJy using VISIR ($R\sim 30$,000) on Melipal/VLT,
and at this sensitivity, the observations were selective for sources 
with strong 12 $\mu$m emissions. 
Meanwhile, GREX-PLUS achieves a sensitivity of $F_{\rm cont} = 5$ mJy at
5$\sigma$ for an integration time of 1 hour, making it possible to observe
objects down to  $F_{\rm cont} = 30$ mJy (1hr, 30$\sigma$).
This sensitivity allows us to detect stars down to $\simeq$0.5 ${\rm M}_\odot$
in the well-known star forming regions, such as Taurus, Cham, Lup, CrA,
etc.
This will be the first comprehensive observation of photoevaporative
activity in nearby star-forming regions.


\subsubsection{High-Mass Star Formation}

We expect that the high resolution spectrometer on GREX-PLUS will yield the powerful and essential progress toward the understanding on the physical mechanism of high-mass star formation using mid-infrared spectroscopy. First, it will allow us to determine the origin of these emission lines, which must originate from either the inner disk region \citep[e.g.,][]{Pontoppidan10}, outflow or inflow.
If it turns out that the emission lines originate from the disk, their line profiles would be useful for investigating the structure of the inner disk regions even without direct spatial information.
We expect that the spectral resolution of the high resolution spectrometer on GREX-PLUS ($R\sim3\times10^4$, $\Delta v=10$ km s$^{-1}$) is optimum for such studies 
\citep{Elias06,Hsieh21}.


We will investigate the following using the spectra obtained using GREX-PLUS: (1) what is responsible for the emission/absorption lines (disk/inflow/outflow) for which stellar masses and evolutionary stages; (2) if the emission/absorption lines are associated with the inner disk, how their physical conditions change with their stellar masses and evolutionary stages, and when and at which conditions those disks start to dissipate (and therefore what actually determines their final stellar masses).

\subsection{Scientific goals}


\subsubsection{Low-Mass Disk Evolution: The Relationship Between Photoevaporation and Other Physical Parameters}

The [NeII] observations in the nearby star-forming regions are expected
to dramatically increase the sample size of observations to up to
$>$300.
The majority of the observed objects have been studied to date to reveal
phenomena related to disk dispersal other than the photoevaporation
process (e.g., mass accretion, disk wind, stellar winds, jets, etc.).
and other parameters besides the age and central stellar mass. 
For the majority of target objects, previous studies have revealed
physical parameters other than age and central stellar mass, parameters
for disk dispersal processes other than photoevaporation (mass
accretion, disk winds, stellar winds, jets, etc.).
We plan to handle data from more than 300 statistically sufficient
objects and discuss the evolution of stars and disks in general, with a
focus on photoevaporation.
This will allow us to clarify under what conditions photoevaporation is
efficient and to answer the fundamental question of what determines the
disk dissipation and thus the planet formation process.



\subsubsection{High-Mass Star Formation}

Using the high resolution spectrometer, we will obtain spectra for a number of high-mass protostars with varieties of masses and evolutionary stages. The required instrument specifications are tabulated below. High-mass star forming regions are very complex so that it is not easy to predict the brightness of the emission features or the depths of the absorption features in advance. Therefore, we propose to organize a pilot survey for a fraction of targets (in particular with the bright ones) to determine the best integration time to study the above emission/absorption lines and features. As our targets are bright, it would be useful for the observations for the system verification of the spectrograph.

\begin{table}[ht]
    \label{tab:takam}
    \begin{center}
    \caption{Required observational parameters.}
    \begin{tabular}{|l|p{9cm}|l|}
    \hline
     & Requirement & Remarks \\
    \hline
    Wavelength & 12--18 $\mu$m &  \\
    \cline{1-2}
    Spatial resolution & No critical requirements & \\
    \hline
    Wavelength resolution & $\lambda/\Delta \lambda = 3 \times 10^4 - 1 \times 10^5$ & \\
    \hline
    Sensitivity & 30 mJy, S/N$\ge$30 &\\
    \hline
    Observing field & None & \\
    \hline
    Observing cadence & N/A & \\
    \hline
    \end{tabular}
    \end{center}
\end{table}

\clearpage
\section{Galaxy Center and Disk}
\label{sec:galaxycenter}

\noindent
\begin{flushright}
Naoteru Gouda$^{1}$, 
\\
$^{1}$ NAOJ, 
\end{flushright}
\vspace{0.5cm}
\subsection{Scientific background and motivation}

{\bf ~~~~~~\normalsize $\sim$ Scientific significance of the Milky Way Galaxy (the Galaxy) $\sim$}

The Galaxy has been considered to have undergone various evolution, including collisions with and merging with several other dwarf galaxies so far. In order to understand the evolution process of our Galaxy (``Galactic Archaeology''), it is very important to understand the dynamical structure and its history in the Galaxy. In addition, the Galaxy is a typical disk galaxy in the present universe, and detailed observations of the Galaxy provide clues to the formation and evolution of many other galaxies in the universe.
In addition, for external galaxies, it is difficult to accurately reveal in detail the physical characteristics of individual stars by observations because of their long distances. On the other hand, the Galaxy where we live is the only one that can reveal its physical characteristics and its spatial distribution and motion in the near future in an accurate and detailed manner for individual stars in the Galaxy. In other words, it is possible to use methods that cannot be adapted to other galaxies in the study of the formation and evolution of galaxies, and the Galaxy is a very good ``test bed'' from this point of view. 
 
Furthermore, stars, planets and life including humanity in the Galaxy are born and raised in the Galaxy, which is a cluster of over 100 billion stars, so they are complexly affected by the Galaxy we live in (the effects depend on the locations where in the Galaxy the stars and planets were born, the orbits the stars 
with planets follow, and the influence from the surrounding space environment, etc.). Therefore, ``knowing'' the Galaxy is also important for solving the mystery of existence of humanity.
So, the exploration of the Galaxy, which is the most familiar galaxy, but still a mystery, is very significant.

\begin{figure}[tbh]
  \begin{center}
   \includegraphics[width=110mm]{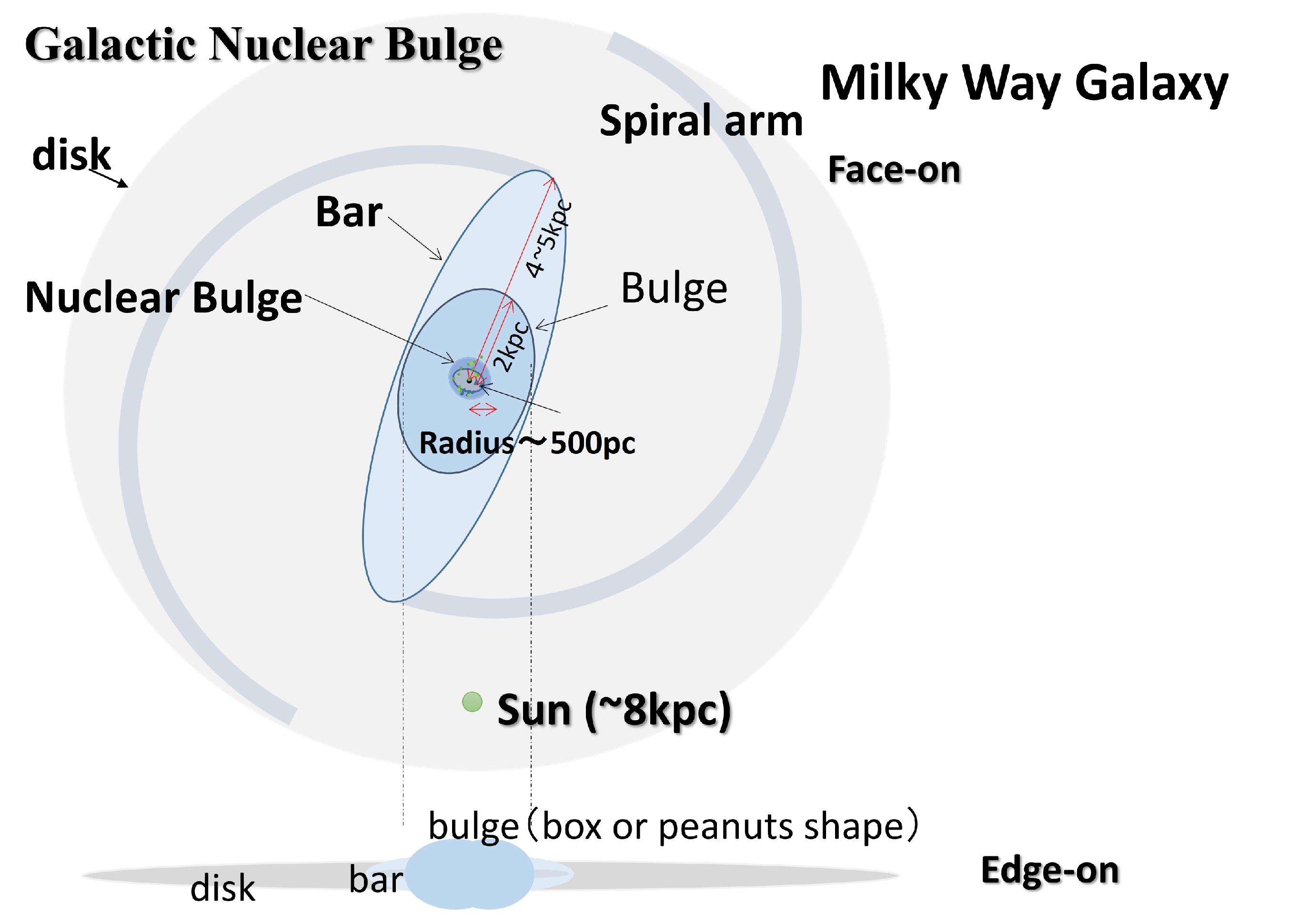}
  \end{center}
  \caption{Imaginary picture of sub-structures of the Galaxy}
  \label{fig-MW}
\end{figure}
 
\subsection{Required observations and expected results}

{\bf ~~~~~~\normalsize $\sim$ Exploration of the Galactic nuclear bulge $\sim$}

The structure of the Galaxy is roughly divided into the disk containing the solar system, the bulge structure at the center of the Galaxy, and a large structure called the halo surrounding the disk and the bulge \citep{2016ARA&A...54..529}. In addition, it is thought that there is a spiral pattern called a spiral arm, and the existence of a bar-like structure (bar). The bulge structure is a bulging structural element (about 2 kpc in size) around the Galactic center, where the stars are densely gathered (see Fig.\ref{fig-MW}). In addition, as a structure in close contact with the bulge, there is a non-axisymmetric structure called a bar structure (the length from the center is about 5 kpc) in the Galaxy, which is considered to be classified as a ``barred spiral galaxy''. Non-axisymmetric structures such as the bar structure strongly influence the dynamical structure of a wide range of the Galaxy including the vicinity of the solar system (such as the orbit and velocity distribution of stars). The velocities and the orbits of stars are strongly influenced by the non-axisymmetric structures. Therefore, for example, only the information on the motions of stars at the disk cannot allow us to understand all the physical information on the disk and the spiral arm. It should be understood together with the physical information of the bulge and bar structure including the formation epoch of the bar structure.
The bulges are roughly classified to two types. One of them is a ``classical bulge'' and another is a ``pseudo-bulge''. The classical bulge exhibits a similar brightness distribution to elliptical galaxies, on the other hand, the pseudo bulge exhibits similar to disk structures, and these are considered to have different origins. Furthermore, the pseudo bulge is categorized to two types, a box/peanut type and a disk-like bulge. There is possibility that the Galaxy has these 3 types of the bulge (\citealt{2016ASSL...418..199G}). The Milky Way has certainly a box/peanut bulge, but furthermore other two types, those are, disk-like bulge and classical bulge might be located in the Galactic nucleus region.
 The Galactic nucleus region whose radius is about 500 pc (or less than 500 pc), is called the Galactic nuclear bulge (see Fig.\ref{fig-MW}) (\citealt{2016ARA&A...54..529}). The nuclear bulge has different physical characters from those of the outer bulge. The nuclear bulge is the place where the material elements of the Galaxy, such as stars, gas, and dark matter, are the most concentrated, and it is the place where the history from the early Galaxy formation to the present is intensively hidden. 
In the Galactic nuclear bulge, stars of various ages have different spatial distributions and motions according to their ages. It is like having different strata in geology. 
Although the Galactic nuclear bulge covers only a small part of the sky, in terms of understanding our Galaxy's structure, and its formation and evolution, it is particularly important region. So, to reveal the Milky Way's central core structure, and its formation history is very interesting and important for understanding the whole part of the Galaxy. We call it ``the Galactic Center Archeology''. It can be said that it is the core for exploring the Galaxy's history. 
In addition, the nuclear bulge region is an important area that links the physical relationship between the outer entire bulge, bar structure and the supermassive black hole at the Galactic center (e.g., \citealt{2017ApJL...844..15D}).

Furthermore, the recent development of powerful, multi- wavelength observational capabilities have allowed for the era of Galactic archaeology (e.g., \citealt{2002ARA&A...40..487}). 
For example, X-ray observations of the Galactic center region have revealed periods of enhanced accretion activity just a few hundred years ago (e.g., \citealt{2013A&A...558A..32C}). On larger scales, ionized gas emission points to a period of AGN activity several
million years ago \citep{2013ApJ...778..58B}. This result is independently supported by X-ray observations of hot gas \citep{2016ApJL...828..12N}. AGN activity on possibly even longer timescales is revealed by gamma ray and ultraviolet spectroscopic studies of large-scale biconical plasma that exists above and below the Galactic plane.
In addition, the galactic central region includes nuclear star clusters. These clusters exhibit ongoing episodes of star formation, resulting in a generally younger stellar population \citep{2016ARA&A...54..529}. As above, the Galactic central region shows very intriguing phenomena.

GREX-PLUS can carry out imaging surveys at the Galactic central region 
in near-infrared and mid-infrared bands. This may allow us to find a lot of intriguing new phenomena and/or celestial objects and clarify them in collaboration
with some X-ray and radio observations.
Furthermore, as described below, GREX-PLUS can provide not only images of celestial objects in the Galactic central region, but also has possibility to provide astrometric information of stars which is very important information for elucidating the Galactic central region such as the Galactic nuclear bulge.

\subsection{Scientific goals}

{\bf ~~~~~~\normalsize $\sim$ Infrared astrometry in the Galactic nuclear bulge $\sim$}

The key to deciphering the history of the central region of the Galaxy, which is about 26 thousand light years ($\sim8$ kpc) away from us, is the positional distribution and movement of stars of various ages that still remain in the nuclear bulge. Stars born at different times in the history of Galaxy formation, like fossils, exist as witnesses of history in the present Galaxy. Information on the current positional distribution and motion of those stars reflects the history of the central region of the Galaxy.
Furthermore the past of the nuclear bulge can lead to the elucidation of the whole Galactic history, such as being able to decide the formation time of the outer long bar structure surrounding the nuclear bulge \citep{2020MN...492..4500}. The history of the formation and evolution of the Galactic center region is clarified by observing the astrometric information such as the positions and velocities of stars in the nuclear bulge precisely to also resolve the outer bar and bulge. 
We are located at the edge of the Galactic disk, and because the light from stars is blocked by the thick gas and dust between us and the Galactic center, the Galactic central region is a place where astronomical observation is difficult. Therefore, infrared astrometric measurements are necessary.
JASMINE is the first satellite mission in the world to perform high precision astrometry from the space with a stable stellar image captured by a space telescope using infrared wavelengths that are transparent to dust and gas, and JASMINE provides the information of the positional distribution and movement of stars in the central region of the Galaxy, which are not yet understood \citep{2021ASPC...528..163G}. 
The mission objective of JASMINE is to use an optical telescope with a primary mirror aperture of around 36 cm to perform infrared astrometric observations (Hw band: 1.0-1.6 $\mu$m). JASMINE can measure, as the highest precision, annual parallaxes at a precision of less than or equal to 25 $\mu$as and proper motions, or transverse angular velocities across the celestial sphere, at a precision of less than or equal to 25 $\mu$as/year in the direction of an area of a few square degrees of the Galactic nuclear bulge in order to create a catalogue of the positions and movements of stars within the region. This mission will help to achieve revolutionary breakthroughs in astronomy, including the formation history of the Galactic nuclear bulge (Galactic Center Archeology); Galacto-seismology; the supermassive black hole at the Galactic Center; the gravitational field in the Galactic nuclear bulge; the activity around the Galactic Center; formation of star clusters; the orbital elements of X-ray binary stars and the identification of the compact object in an X-ray binary; the physics of fixed stars; star formation; planetary systems; and gravitational lensing. Such data will allow for the compilation of a more meaningful catalog when combined with data from ground-based observations of the line-of-sight velocities and chemical compositions of stars in the bulge.
However, relatively small telescope provides astrometric information only for bright stars (below Hw$\sim$14.5magnitude).
On the other hand, GREX-PLUS may provide astrometric information for stars including dark stars while the precisions of the astrometric measurements may not be so good if we use only data provided by GREX-PLUS\footnote{If GREX-PLUS satisfies the conditions of specifications of the instruments which are suggested by Table~\ref{tb:astrometry}, and also if it can carry out the survey of stellar images at the Galactic nuclear bulge region in the near-infrared band, then GREX-PLUS may have possibility to provide highly precise ($\sim$ 25 $\mu$as) astrometric measurements for stars brighter than K$\sim$14 magnitude.}. However, for stars observed by both JASMINE and GREX-PLUS, if JASMINE can provide highly accurate astrometric information of these stars, GREX-PLUS can use the astrometric information of these stars to get precise astrometric information even for darker stars which JASMINE cannot measure. This strengthens the scientific performance
of kinematical analysis at the Galactic nuclear bulge, in particular,
to lead investigations of new objects and new phenomena, such as
the discovery of intermediate massive black holes and clarification of physical characters of ultra-light dark matter, etc..
In this way, still many objects and phenomena are left to perform astrometric observations that are scientifically interesting and important in the Galactic nuclear bulge and GREX-PLUS has possibility to contribute to this exploration.

\begin{table}
    \begin{center}
        \caption{Required observational parameters.}
        \label{tb:astrometry}
        \begin{tabular}{|l|l|l|} \hline
        &Requirments&Remarks \\ \hline
        Wavelength& $\sim 2.2~\mu {\rm m}$&a \\ \hline
        Observation region& $-1.5^{\circ}<{\ell}<1.5^{\circ}, -0.3^{\circ}<b<0.3^{\circ}$ &b \\ \hline
        Observation periods& 7 days in spring, 7 days in autumn every year for 5 years&c \\ \hline
        Ratio of PSF size to pixel size& $(\lambda/D)\cdot (f/w) \sim 1.5-2.0$ & d \\ \hline
        Pointing stability& $<100~{\rm mas}/20~{\rm seconds}$ & e \\ \hline
        Thermal~stability of instruments& $<1~{\rm nm}/5~{\rm hours}$& f \\ \hline
        \end{tabular}
    \end{center}
    ${}^a$ K-band is suitable for astrometric measurements of stars in the Galactic central region.\\
    ${}^b$ This region corresponds to the Galactic nuclear stellar disk region in the Galactic bulge.\\
    ${}^c$ For 3 years operation instead of 5 years operation, 12 days in spring, and 12 days in autumn every year are necessary.\\
    ${}^d$ Required ratio of the PSF size to the pixel size is suitable for the precise estimation of the centroids of stellar images $(\lambda:{\rm wavelength}, D:{\rm the~diameter~of~the~primary~mirror}, f:{\rm focal~length}, w:{\rm pixel ~size})$ The present proposed specification of GREX-PLUS for this ratio is 2 and so this requirement is satisfied.\\
    ${}^e$ The present proposed specification of the pointing stability is 100 mas/300 seconds and so this requirement is satisfied.\\
    ${}^f$ This condition is for the time variation of the centroids of stellar images on the focal plane.
\end{table}

\chapter{Synergy with Other Projects}
\label{chap:synergywithotherprojects}

In this Chapter, we will review expected synergies between GREX-PLUS and other telescope projects.


\paragraph{Subaru Telescope}

Optical wide and deep imaging survey fields produced with the Hyper Suprime-Cam (HSC; \citealt{2018PASJ...70S...1M}) on the Subaru Telescope are natural target fields for any infrared wide-field imaging surveys including GREX-PLUS.
This is a clear synergy case.
In the middle of the 2020s, the Subaru Telescope will have a next-generation adaptive optics (AO) system, Ground-Layer AO (GLAO), that enables to have near diffraction-limited imaging quality across a wide-area up to $K$-band (i.e. wavelength 2.2 $\mu$m).
This will be achieved by the instrument called, ULTIMATE-Subaru \citep{2022SPIE12185E..21M}.
The expected angular resolution of $0.2''$ in $K$-band is several times better than that of GREX-PLUS.
For the scientific theme of EGS2, galaxy mass assembly, this higher spatial resolution is very much useful to resolve the galaxy internal structure and examine the stellar mass assembly process.
Therefore, it is another excellent synergy case to make follow-up observations of galaxies detected in GREX-PLUS imaging surveys by using ULTIMATE-Subaru to spatially resolve their internal structure.

\paragraph{Rubin Observatory}

The Legacy Survey of Space and Time (LSST) aims to start in January, 2025 and will obtain ``movie'' of about 20,000 square degrees in the southern sky over 10 year.
The imaging depth will be as deep as 27.5 AB to 24.9 AB in $u,g,r,i,z,y$ after the 10-year survey.
These optical wide area and deep images will be the best fields for near-infrared imaging surveys including GREX-PLUS.
This is a natural synergistic point between Rubin Observatory and GREX-PLUS.

\paragraph{Extremely Large Telescopes}

The next generation ground-based 30--40 m extremely large telescopes (ELTs) will have unprecedented sensitivity and angular resolution in near-infrared wavelengths with an AO system.
On the other hand, the field-of-view is too small to make any wide-field survey and ELTs themselves can not find interesting objects to be observed in great details.
Therefore, ELTs inevitably need other telescopes that provide excellent targets to be followed-up with ELTs.
GREX-PLUS is one of such target providers for ELTs.
Since GREX-PLUS wide-field imaging surveys are only $>1$ degree-scale surveys in the wavelength 2--8 $\mu$m in coming decades, the unique role of GREX-PLUS as a target providers for ELTs stands unrivaled.
The spectroscopic capability of ELTs is also extremely high and can have spectral resolution $>100,000$.
However, the atmospheric absorption is unavoidable and the sensitivity and wavelength coverage is limited in wavelengths longer than 2.5 $\mu$m.
Therefore, the GREX-PLUS high resolution spectcroscopic capability is very unique even compared to ELTs.

\paragraph{JWST}

James Webb Space Telescope (JWST) was highly successfully launched on December 25th 2021 and has started its science observations in 2022.
Thanks to the successful launch and cruise to the Sun-Earth L2, the fuel consumption was minimal and the fuel is expected to last over 20 years.
JWST observes in the 0.6-28 $\mu$m wavelength range, which includes that of GREX-PLUS.
The primary mirror aperture of 6.5 m is overwhelmingly large as a space telescope, yielding ultra-high sensitivity and high spatial resolution.
On the other hand, although its field-of-view is larger than that of the Hubble Space Telescope, it is limited to less than 10 arcmin$^2$.
Hence, it is difficult to conduct square-degree-scale imaging surveys.
Since GREX-PLUS will achieve such ultra wide-field imaing surveys, they are complementary each other.
Follow-up observations by using JWST for unique and interesting objects detected in GREX-PLUS surveys provide straightforward synergy.
The wavelength resolution of the JWST spectrographs is limited to 3,000 at maximum, that is 10 times coarser than that of GREX-PLUS.
For example, JWST can not resolve the Keperlian motion of water molecules around the ``snowline'', while it can detect the water 18 $\mu$m line in protoplanetary disks very easily.
Therefore, we can select the disks to be targeted with GREX-PLUS after the JWST's water line detections, which constitutes another very nice synergy case between GREX-PLUS and JWST.

\paragraph{Euclid}

Euclid is a wide-field survey space telescope developed by the European Space Agency (ESA) to be launched in 2023.
The primary mirror aperture is 1.2 m, the same as that of GREX-PLUS.
It has a field-of-view of approximately 2,000 arcmin$^2$, which is even wider than that of GREX-PLUS.
The observed wavelength range is from optical to near-infrared 2 $\mu$m, and there is no overlap with GREX-PLUS. 
It also has a wide-field spectroscopic capability, but with a low wavelength resolution of 250.
The main scientific goal is to resolve the origin of the accelerated expansion of the Universe by measuring gravitational weak-lensing effects and baryonic acoustic oscillations, and to conduct an ultra-wide-field survey of up to 15,000 square degrees.
The ultra-wide imaging survey will be shallow (24 AB) and it will also conduct a narrower survey (40 square degrees) to reach a deeper depth (26 AB).
Compared to GREX-PLUS, there is a difference in the combination of the width and depth of the surveys, in addition to the difference in the wavelength band.
As synergy between GREX-PLUS and Euclid, GREX-PLUS survey fields will be selected among Euclid survey fields, especially, from its Deep surveys, to have the imaging data in the wavelength shorter than 2 $\mu$m.
The GREX-PLUS imaging data in the wavelength longer than 2 $\mu$m are also extremely useful to trace the longer wavelength radiation from objects detected in the Euclid imaging data.

\paragraph{Roman}

Nancy Grace Roman Space Telescope (hereafter Roman) is the NASA's next-generation flagship space telescope, scheduled for launch in 2026. 
The primary mirror aperture is 2.4 m, twice larger than that of GREX-PLUS, and its observational wavelength range is 0.5-2.3 $\mu$m. 
GREX-PLUS covers the wavelength range of 2 $\mu$m or longer, which are complementary to each other. 
In particular, the High Latitude Survey conducted by Roman provides a deep 26.7 AB survey in the wavelength less than 2 $\mu$m over 1,700 square degrees, and is expected to have the best synergy in depth and coverage with the GREX-PLUS survey in the wavelength longer than 2 $\mu$m. 
Roman has a prism and grism spectrograph. 
The wavelength resolution is low resolution, which is completely different from the GREX-PLUS high resolution spectrometer in terms of both wavelength band and wavelength resolution.
Roman will also conduct microlensing surveys near the Galactic center, that is very powerful to identify exoplanets with the orbital radius of 1 au or larger.
Roman also has Coronagraph Instrument to realize direct observations of exoplanets, including polarimetric spectroscopy as well as imaging.

Extremely high synergy between Roman and GREX-PLUS is expected in a wide range of scientific themes.
In fact, most of science themes with the GREX-PLUS wide-field camera assumes wide-field deep imaging data in the wavelength less than 2 $\mu$m by Roman.
Therefore, GREX-PLUS imaging surveys must be conducted in the Roman imaging survey fields.
There is also potential synergy in observations of the Galaxy center (Section~\ref{sec:galaxycenter}) between Roman microlensing surveys and GREX-PLUS.
Synergy between Roman Coronagraph Instrument of direct observations of exoplanets and GREX-PLUS high resolution spectrometer observations is also possible and under investigation.

\paragraph{JASMINE}

Japan Astrometry Satellite Mission for INfrared Exploration (JASMINE; \citealt{2021ASPC...528..163G}) is scheduled for launch in 2028 by an Epsilon Launch Vehicle of ISAS/JAXA.
With the primary mirror of 36 cm, JASMINE will conduct astrometric observations in the near-infrared wavelength of 1.0-1.6 $\mu$m.
The target field is the Galactic nuclear region to understand the nuclear bulge formation history in the Milky Way.
JASMINE will also conduct exoplanet transit observations when the Galaxy center is not visible from the orbit.
As described in Section~\ref{sec:galaxycenter}, GREX-PLUS can realize the Galaxy center observation in wavelengths longer than JASMINE.
Since the dust extinction is less severe in the wavelengths of GREX-PLUS than in those of JASMINE, the observations will be highly complementary.
For the astrometric measurements in GREX-PLUS imaging data, it will be required to calibrate the astrometry with JASMINE data which have higher astrometric accuracy.
There is another synergy in exoplanet observations.
JASMINE will discover many transit exoplanet systems, especially, Earth-type planets in the habitable zone around M-type dwarf stars.
These exoplanets will be excellent follow-up targets for GREX-PLUS high resolution spectrometer to examine rare molecules in the planetary atmosphere.

\paragraph{HiZ-GUNDAM}

High-z Gamma-ray bursts for Unraveling the Dark Ages Mission (HiZ-GUNDAM; \citealt{2020SPIE11444E..2ZY}) is a mission candidate for launch around 2030 by an Epsilon Launch Vehicle of ISAS/JAXA.
The scientific aim is to explore the very early Universe by using gamma-ray bursts (GRBs).
To do so, HiZ-GUNDAM will carry wide-field gamma-ray detectors and an optical-to-near-infrared small telescope for the immediate follow-up of afterglows of GRBs.
The infrared wavelength coverage of the telescope is up to 2.5 $\mu$m.
The primary mirror aperture is 30 cm.
While the scientific theme is partly overlapping with those of the wide-field camera of GREX-PLUS, the sensitivity (i.e. mirror aperture) and the wavelength coverage are very different.
Therefore, HiZ-GUNDAM can not achieve the scientific goals of GREX-PLUS.
At the same time, GREX-PLUS can not replace HiZ-GUNDAM because GREX-PLUS does not carry any gamma-ray detectors.
On the other hand, if HiZ-GUNDAM and GREX-PLUS are in orbit simultaneously, the most distant GRBs found by HiZ-GUNDAM will be excellent follow-up targets for GREX-PLUS.




%

\paragraph{ALMA}

GREX-PLUS also has very high synergy with world-leading radio telescopes, for example, 
the Atacama Large Millimeter/submillimeter Array (ALMA).
It has unprecedented high sensitivity, high angular resolution, and high spectral resolution in a wide wavelength range from 300 $\mu$m to 3.6 mm.
Since its field-of-view is very tiny, however, ALMA generally needs targets to observe {\it a priori}.
GREX-PLUS wide-field surveys will supply a number of excellent targets to ALMA.
Especially, ALMA follow-up observations of [O~{\sc iii}] 88 $\mu$m and [C~{\sc ii}] 158 $\mu$m lines for very high-$z$ galaxy candidates detected with GREX-PLUS are natural synergy.
Another excellent synergy is dust continuum observations of these high-$z$ galaxies with ALMA.
GREX-PLUS high resolution spectroscopy has also high synergy with ALMA in molecular chemistry in the interstellar medium, star forming regions, and protoplanetary disks.

By 2030, ALMA will enhance its capability to broaden the receiver bandwidth by a factor of two.
This development will also enhance the synergy with GREX-PLUS very much.
For example, the speed of the line search of distant galaxy candidates simply becomes a factor of two faster as well as an improvement of the continuum sensitivity.



\bibliography{main_all}


\end{document}